# Second Order Accurate Schemes for Magnetohydrodynamics With Divergence-Free Reconstruction

By
**Dinshaw S. Balsara**

Physics Department, University of Notre Dame

(dbalsara@nd.edu)

**Mailing Address:**

Physics Department

College of Science

University of Notre Dame

225 Nieuwland Science Hall

Notre Dame, IN 46556

**Phone :** (574) 631-9639

**Fax :** (574) 631-5952

**Abstract**

While working on an adaptive mesh refinement (AMR) scheme for divergence-free magnetohydrodynamics (MHD) Balsara (2001, J. Comput. Phys., 174, 614) discovered a unique strategy for the reconstruction of divergence-free vector fields. Balsara also showed that for one-dimensional variations in flow and field quantities the reconstruction reduces exactly to the total variation diminishing (TVD) reconstruction. In Balsara (2001) the innovations were put to use in studying AMR-MHD. While the other consequences of the invention especially as they pertain to numerical scheme design were mentioned, they were not explored in any detail. In this paper we begin such an exploration. We study the problem of divergence-free numerical MHD and show that the work done so far still has four key unresolved issues. We resolve those issues in this paper. It is shown that the magnetic field can be updated in divergence-free fashion with a formulation that is better than the one in Balsara and Spicer (1999, J. Comput. Phys., 149, 270). The problem of reconstructing MHD flow variables with spatially second order accuracy is also studied. Some ideas from WENO reconstruction, as they apply to numerical MHD, are developed. Genuinely multidimensional reconstruction strategies for numerical MHD are also explored. The other goal of this paper is to show that the same well-designed second order accurate schemes can be formulated for more complex geometries such as cylindrical and spherical geometry. Being able to do divergence-free reconstruction in those geometries also resolves the problem of doing AMR in those geometries; the appendices contain detailed formulae for the same. The resulting MHD scheme has been implemented in Balsara's RIEMANN framework for parallel, self-adaptive computational astrophysics. The present work also shows that divergence-free reconstruction and the divergence-free time-update can be done for numerical MHD on unstructured meshes. As a result, we establish important analogies between MHD on structured meshes and MHD on unstructured meshes because such analogies can guide the design of MHD schemes and AMR-MHD techniques on unstructured meshes. The present paper also lays out the roadmap for designing MHD schemes for structured and unstructured meshes that have better than second order accuracy in space and time. All the schemes designed here are shown to be second order accurate. We also show that the accuracy does not depend on the quality of the Riemann solver. We have compared the numerical dissipation of the unsplit MHD schemes presented here to the dimensionally split MHD schemes that have been used in the past and found the former to be superior. The dissipation does depend on the Riemann solver but the dependence becomes weaker as the quality of the interpolation is improved. Several stringent test problems are presented to show that the methods work, including problems involving high velocity flows in low plasma-$\beta$ magnetospheric environments. Similar advances can be made in other fields, such as electromagnetics, radiation MHD and incompressible flow, that rely on a Stokes-law type of update strategy.

**I) Introduction**

Recent years have seen a flowering of interest in higher order Godunov schemes for MHD. This stems from the fact that the MHD equations are useful in several fields of space physics, astrophysics and engineering. The robust and accurate performance of higher order Godunov schemes has, therefore, made them attractive to practitioners in those fields. Even in the field of astrophysics, where there has been an excessive dependence on the ZEUS code of Stone and Norman (1992), a need for change is evident because of the recent paper by Falle (2002). Early development of higher order Godunov schemes for MHD focused on interpreting the MHD equations as a simple system of conservation laws. This was done by Brio and Wu (1988), Zachary, Malagoli and Colella (1994), Powell (1994), Dai and Woodward (1994), Ryu and Jones (1995), Roe and Balsara (1996), Balsara (1998a,b) and Falle, Komissarov and Joarder (1998). More recent efforts have focused on understanding the structure of the induction equation:

$$\frac{\partial \mathbf{B}}{\partial t} + c \, \nabla \times \mathbf{E} = 0 \qquad (1.1)$$

and the divergence-free evolution that it implies for the magnetic field. In eqn. (1.1), $\mathbf{B}$ is the magnetic field, $\mathbf{E}$ is the electric field and c is the speed of light. Eqn. (1.1) ensures that the magnetic field will remain divergence-free forever if it is divergence-free to begin with. Other important systems of equations that satisfy an update equation of this form include the equations of incompressible flow, radiation MHD and relativistic MHD. Relativistic MHD has been treated in Balsara (2001a) and radiation MHD in Balsara (1999a,b). In the specific cases of ideal MHD, radiation MHD and relativistic MHD the electric field is given by

$$\mathbf{E} = -\frac{1}{c} \mathbf{v} \times \mathbf{B} \qquad (1.2)$$

where $\mathbf{v}$ is the fluid velocity. Brackbill and Barnes (1980) and Brackbill (1985) have shown that violating the $\nabla \cdot \mathbf{B} = 0$ constraint leads to unphysical plasma transport orthogonal to the magnetic field. This comes about because violating the constraint results in the addition of extra source terms in the momentum and energy equations. Yee (1966) was the first to formulate divergence-free schemes for electromagnetism. Brecht et al (1981) and DeVore (1991) did the same for FCT-based MHD. Dai and Woodward (1998), Ryu et al (1998), Balsara and Spicer (1999b) and Londrillo and DelZanna (2000) showed that simple extensions of higher order Godunov schemes permit one to formulate divergence-free time-update strategies for the magnetic field. Toth (2000) has made a comparative study of such schemes and found the scheme of Balsara and Spicer (1999b) to be one of the most accurate second order schemes that he tested. Balsara and Kim (2003) have intercompared divergence-cleaning and divergence-free schemes for numerical MHD. They find that the schemes that are based on divergence-cleaning show significant inadequacies when used for astrophysical applications.

Despite these advances, several issues in the design of higher order Godunov schemes for numerical MHD with have so far remained unresolved. They are:
**Issue # 1)** In divergence-free formulations for numerical MHD the magnetic field components have to be specified at the zone faces and yet one has to use zone-centered magnetic fields to obtain the pressure from the energy density. How should the zone-centered magnetic fields be obtained? Most practitioners have used arithmetic averaging.

But there is nothing in the theory that suggests that arithmetic averaging is optimal or that it produces the unique volume-averaged field value of choice.

**Issue # 2)** In divergence-free formulations for numerical MHD the zone-centered magnetic fields are limited in TVD fashion and yet the face-centered variables are held to be the fundamental variables that have to be evolved. Why doesn't one limit the fundamental variables, i.e. the face-centered magnetic field components, in the scheme? Furthermore, how should the magnetic fields be limited? Does TVD limiting find a natural extension when limiting face-centered variables?

**Issue # 3)** In zone-centered formulations of numerical MHD the normal component of the magnetic field at the zone face has a jump in it. Powell (1994) has shown that the seven wave model for the Riemann solver is ill-defined when the normal component of the magnetic field at the zone boundary has a jump in it. Yet Powell's own resolution of that situation consisted of changing the MHD equations so as to introduce an eighth wave. The introduction of the eighth wave also introduces magnetic monopoles in the magnetic field. Some of the most thorough searches for magnetic monopoles have not revealed a single convincing detection in nature of a magnetic monopole, see Abbott et al (1998). Is there any way of reconciling numerical MHD with the factual evidence? Such a reconciliation requires that we find a scheme which naturally preserves the $C^0$ continuity of the normal component of the magnetic field at zone faces. Such a scheme would eliminate the jump in the normal component at the zone faces, thereby eliminating the need for an eight wave model.

**Issue # 4)** The divergence-free schemes of Dai and Woodward (1998), Ryu et al (1998) and Balsara and Spicer (1999b) all had to obtain the electric field at the zone edges via an averaging procedure. Yet it would seem natural that the dualism between electric field and flux components that was found in Balsara and Spicer (1999b) should be used to obtain the electric fields directly at the zone edges where they are needed. Is there any way to do that?

The issues in the previous paragraph come into sharp focus once one realizes that van Leer (1979) offered a well-defined strategy for the reconstruction of zone-centered variables. Yet a reconstruction strategy for face-centered, divergence-free vector fields had not been invented till the work of Balsara (2001b). While working on an adaptive mesh refinement scheme for the divergence-free evolution of vector fields, Balsara showed that there is a unique strategy for the reconstruction of divergence-free vector fields. Balsara also showed that for one-dimensional variations in flow and field quantities the reconstruction reduces exactly to the total variation diminishing (TVD) reconstruction of van Leer (1979) and Harten (1983). This is a very useful demonstration because it shows that higher order Godunov schemes for MHD which use this reconstruction will benefit from the same stability properties that Tadmor (1984) and Sweby (1984) have demonstrated for TVD schemes for hydrodynamics. Balsara (2001b) recognized the potential of his work for scheme design on structured as well as unstructured meshes. Balsara (2001b) also realized that his work makes it possible to design schemes for numerical MHD that are better than second order accurate. Balsara (2001b) also understood that the issues enumerated in the previous paragraph could be resolved via his work. The present paper explores some of the consequences of Balsara's work that pertain to the design of higher order schemes for numerical MHD on structured meshes. One of the goals of this paper is to present a genuinely well-designed second order accurate scheme so that all the issues presented in the previous paragraph are properly resolved. Some ideas from WENO reconstruction are used but that is only in the sprit of trying to ameliorate the tendency of TVD limiters to chop off extrema. The other goal of this paper is to show that the same well-designed second order schemes can be formulated for more complex geometries such as cylindrical and spherical geometry. The resulting MHD scheme has been implemented in Balsara's RIEMANN framework for parallel, self-adaptive computational astrophysics. The present work also shows that a similar resolution exists for MHD on unstructured meshes and some of that work is also catalogued here. As a result, yet another goal of this work is to establish important analogies between MHD on structured meshes and MHD on unstructured meshes because such analogies can guide the design of MHD schemes on unstructured meshes. The present paper also lays out the roadmap for designing MHD schemes for structured and unstructured meshes that have better than second order accuracy. Similar advances can be made in other fields, such as electromagnetics, radiation MHD, incompressible flow and all-speed flow that rely on a Stokes-law type of update strategy.

In Section II we provide a new staggered mesh time update scheme that makes it possible to have temporally second and third order accurate time-integration of divergence-free vector fields. The new scheme also resolves issue # 4 above. In Section III we explain the reconstruction strategy for divergence-free vector fields and how it permits one to resolve issues # 1, 2 and 3 above. In Section IV we extend the reconstruction strategy of Balsara (2001b) to cylindrical and spherical structured meshes and to tetrahedral meshes. In Section V we present a pointwise description of the numerical MHD scheme presented here. Section VI presents an accuracy analysis for the schemes designed here. In Section VII we show that the method works well on several stringent test problems. In Section VIII we offer some conclusions.

**II) A New Staggered Mesh Time Update Scheme**

Balsara and Spicer (1999b) constructed a divergence-free staggered mesh scheme for ideal MHD that was, at least formally, spatially and temporally second order accurate. Their scheme did, however, use an averaging procedure to obtain the electric field at the zone edges. As a result it did not resolve issue # 4 in the previous section. In this section we present a new staggered mesh time update scheme that makes it possible to have temporally second and third order accurate time-integration of divergence-free vector fields. Compared to previous MHD time-stepping strategies the new scheme offers greater temporal accuracy and freedom from dimensional sweeping. The new scheme also resolves issue # 4 above. The equations of ideal MHD can be cast in a conservative form that is suited for the design of higher order Godunov schemes. In that form they become:

$$\frac{\partial \mathbf{U}}{\partial t} + \frac{\partial \mathbf{F}}{\partial x} + \frac{\partial \mathbf{G}}{\partial y} + \frac{\partial \mathbf{H}}{\partial z} = 0 \tag{2.1}$$

where **F**, **G** and **H** are the ideal fluxes. Written out explicitly, eqn. (2.1) becomes :

$$\frac{\partial}{\partial t}\begin{pmatrix} \rho \\ \rho v_x \\ \rho v_y \\ \rho v_z \\ \varepsilon \\ B_x \\ B_y \\ B_z \end{pmatrix} + \frac{\partial}{\partial x}\begin{pmatrix} \rho v_x \\ \rho v_x^2 + P + \mathbf{B}^2/8\pi - B_x^2/4\pi \\ \rho v_x v_y - B_x B_y/4\pi \\ \rho v_x v_z - B_x B_z/4\pi \\ (\varepsilon + P + \mathbf{B}^2/8\pi)v_x - B_x(\mathbf{v}\cdot\mathbf{B})/4\pi \\ 0 \\ (v_x B_y - v_y B_x) \\ -(v_z B_x - v_x B_z) \end{pmatrix}$$

$$+ \frac{\partial}{\partial y}\begin{pmatrix} \rho v_y \\ \rho v_x v_y - B_x B_y/4\pi \\ \rho v_y^2 + P + \mathbf{B}^2/8\pi - B_y^2/4\pi \\ \rho v_y v_z - B_y B_z/4\pi \\ (\varepsilon + P + \mathbf{B}^2/8\pi)v_y - B_y(\mathbf{v}\cdot\mathbf{B})/4\pi \\ -(v_x B_y - v_y B_x) \\ 0 \\ (v_y B_z - v_z B_y) \end{pmatrix} + \frac{\partial}{\partial z}\begin{pmatrix} \rho v_z \\ \rho v_x v_z - B_x B_z/4\pi \\ \rho v_y v_z - B_y B_z/4\pi \\ \rho v_z^2 + P + \mathbf{B}^2/8\pi - B_z^2/4\pi \\ (\varepsilon + P + \mathbf{B}^2/8\pi)v_z - B_z(\mathbf{v}\cdot\mathbf{B})/4\pi \\ (v_z B_x - v_x B_z) \\ -(v_y B_z - v_z B_y) \\ 0 \end{pmatrix} = 0$$

(2.2)

where $\varepsilon = \rho v^2/2 + P/(\gamma-1) + \mathbf{B}^2/8\pi$ is the total energy. We see that the flux components of ideal MHD obey the following symmetries:

$$F_7 = -G_6, \quad F_8 = -H_6, \quad G_8 = -H_7,$$
$$F_{r,7} = -G_{r,6}, \quad F_{r,8} = -H_{r,6}, \quad G_{r,8} = -H_{r,7}$$

(2.3)

The Balsara and Spicer (1999b) scheme was based on realizing that there is a dualism between the fluxes that are produced by a higher order Godunov scheme and the electric fields that were needed in eqn. (1.1). The last three components of the **F**, **G** and **H** fluxes could also be reinterpreted as electric fields in our dual approach. The electric fields are needed at the edge centers as shown in Figure 1a and are to be used to update the face-centered magnetic fields. The Balsara and Spicer (1999b) scheme drew on the dualism to produce the edge-centered electric fields by averaging components of the face-centered fluxes. However, the averaging of flux components that are evaluated at different locations has several undesirable consequences. First, it restricts the method to be second order accurate in space and time and obscures the natural pathway by which the spatial and temporal accuracy of the scheme may be improved. Second, the spatial averaging of the fluxes to produce electric fields may introduce undesirable phase errors into the scheme.

Shu and Osher (1988) constructed a set of multi-stage Runge-Kutta time-stepping schemes that preserved the TVD property and also permitted second, third and even higher order of temporal accuracy to be achieved. Balsara (2001b) showed that his divergence-free reconstruction of vector fields satisfies a TVD property. As a result such Runge-Kutta time-stepping schemes also permit us to obtain a TVD-preserving update strategy for the magnetic field. Shu and Osher (1988, 1989), Liu, Osher and Chan (1994), Jiang and Shu (1996) and Balsara and Shu (2000) demonstrated the great utility of the multi-stage Runge-Kutta time-stepping schemes in designing schemes that have third and higher order of spatial accuracy. In practice only the second and third order accurate schemes of Shu and Osher (1988) are used because the other schemes require smaller Courant numbers and are rather difficult to implement. Because it is one of the design goals of this sequence of papers to arrive at second and higher order accurate schemes for numerical MHD, we catalogue the multi-stage Runge-Kutta time-stepping schemes in this section as they would apply to the divergence-free update of vector fields.

The Runge-Kutta time-stepping schemes consist of writing eqn. (2.1) in the form

$$\frac{d\mathbf{U}}{dt} = L(\mathbf{U}) \tag{2.4}$$

Where $L(\mathbf{U})$ is a discretization of the spatial operator. The second order TVD Runge-Kutta scheme is simply the Heun scheme:

$$\begin{aligned}\mathbf{U}^{(1)} &= \mathbf{U}^n + \frac{1}{2}\Delta t\, L(\mathbf{U}^n) \\ \mathbf{U}^{n+1} &= \mathbf{U}^n + \Delta t\, L(\mathbf{U}^{(1)})\end{aligned} \tag{2.5}$$

The third order TVD Runge-Kutta scheme is given by:

$$\begin{aligned}\mathbf{U}^{(1)} &= \mathbf{U}^n + \Delta t\, L(\mathbf{U}^n) \\ \mathbf{U}^{(2)} &= \frac{3}{4}\mathbf{U}^n + \frac{1}{4}\mathbf{U}^{(1)} + \frac{1}{4}\Delta t\, L(\mathbf{U}^{(1)}) \\ \mathbf{U}^{n+1} &= \frac{1}{3}\mathbf{U}^n + \frac{2}{3}\mathbf{U}^{(2)} + \frac{2}{3}\Delta t\, L(\mathbf{U}^{(2)})\end{aligned} \tag{2.6}$$

The schemes, as presented so far, might look much like their zone-centered variants. However, one should realize that the last three components of the solution vector $\mathbf{U}$ are not the zone-centered variables but rather the face-centered magnetic field components. The magnetic field components have to be updated using appropriately defined quadrature points at the zone's edges. For example, to obtain a spatially second order accurate update a single quadrature point at the center of each edge will suffice. For a spatially second order accurate scheme the quadrature points associated with each of the edges are shown in Figure 1a with solid dots. Using the first stage in eqn. (2.5) to illustrate the magnetic field update we get (eqns. (2.7) to (2.9) should not be viewed as matrix equations):

$$B^{(1)}_{x, i+1/2,j,k} = B^n_{x, i+1/2,j,k} - \frac{c \Delta t}{2\Delta y \Delta z}\begin{pmatrix} \Delta z\, E_{z, i+1/2,j+1/2,k}(\mathbf{U}^n) - \Delta z\, E_{z, i+1/2,j-1/2,k}(\mathbf{U}^n) \\ + \Delta y\, E_{y, i+1/2,j,k-1/2}(\mathbf{U}^n) - \Delta y\, E_{y, i+1/2,j,k+1/2}(\mathbf{U}^n) \end{pmatrix} \quad (2.7)$$

$$B^{(1)}_{y, i,j-1/2,k} = B^n_{y, i,j-1/2,k} - \frac{c \Delta t}{2\Delta x \Delta z}\begin{pmatrix} \Delta x\, E_{x, i,j-1/2,k+1/2}(\mathbf{U}^n) - \Delta x\, E_{x, i,j-1/2,k-1/2}(\mathbf{U}^n) \\ + \Delta z\, E_{z, i-1/2,j-1/2,k}(\mathbf{U}^n) - \Delta z\, E_{z, i+1/2,j-1/2,k}(\mathbf{U}^n) \end{pmatrix} \quad (2.8)$$

$$B^{(1)}_{z, i,j,k+1/2} = B^n_{z, i,j,k+1/2} - \frac{c \Delta t}{2\Delta x \Delta y}\begin{pmatrix} \Delta x\, E_{x, i,j-1/2,k+1/2}(\mathbf{U}^n) - \Delta x\, E_{x, i,j+1/2,k+1/2}(\mathbf{U}^n) \\ + \Delta y\, E_{y, i+1/2,j,k+1/2}(\mathbf{U}^n) - \Delta y\, E_{y, i-1/2,j,k+1/2}(\mathbf{U}^n) \end{pmatrix} \quad (2.9)$$

The difference between the present scheme and the one in Balsara and Spicer (1999b) stems from the fact that the fluxes are now evaluated at the facial quadrature points as shown by the dots in Figure 1b. The facial quadrature points should not be confused with the quadrature points on each edge. A facial quadrature point does coincide in physical space with the quadrature point on the edge that the face abuts. However, the facial quadrature point associated with a given face is taken to be a one-sided limit (taken from within the face) as the facial quadrature point approaches the edge's quadrature point. The dualism between flux and electric field is again utilized. However, the spatial averaging is avoided as follows (eqns. (2.10) to (2.12) should not be viewed as matrix equations) :

$$E_{x, i,j+1/2,k+1/2}(\mathbf{U}^n) = \frac{1}{4c}\begin{pmatrix} H^+_{7, i,j+1/2,k+1/2}(\mathbf{U}^n) + H^-_{7, i,j+1/2,k+1/2}(\mathbf{U}^n) \\ - G^+_{8, i,j+1/2,k+1/2}(\mathbf{U}^n) - G^-_{8, i,j+1/2,k+1/2}(\mathbf{U}^n) \end{pmatrix} \quad (2.10)$$

$$E_{y, i+1/2,j,k+1/2}(\mathbf{U}^n) = \frac{1}{4c}\begin{pmatrix} F^+_{8, i+1/2,j,k+1/2}(\mathbf{U}^n) + F^-_{8, i+1/2,j,k+1/2}(\mathbf{U}^n) \\ - H^+_{6, i+1/2,j,k+1/2}(\mathbf{U}^n) - H^-_{6, i+1/2,j,k+1/2}(\mathbf{U}^n) \end{pmatrix} \quad (2.11)$$

$$E_{z, i+1/2,j+1/2,k}(\mathbf{U}^n) = \frac{1}{4c}\begin{pmatrix} G^+_{6, i+1/2,j+1/2,k}(\mathbf{U}^n) + G^-_{6, i+1/2,j+1/2,k}(\mathbf{U}^n) \\ - F^+_{7, i+1/2,j+1/2,k}(\mathbf{U}^n) - F^-_{7, i+1/2,j+1/2,k}(\mathbf{U}^n) \end{pmatrix} \quad (2.12)$$

In the above equations as well as in Figure 1 we have assumed a Cartesian mesh with edges of size $\Delta x$, $\Delta y$ and $\Delta z$. The reader is urged to look at Figs. 1a and 1b to realize that eqns. (2.10) to (2.12) are not spatial averages. Balsara and Spicer (1999b) do treat more general geometries for the ideal MHD case and the results developed in this section can also be easily extended to general geometries such as cylindrical or spherical geometries. Notice that in Figure 1b the reconstructed variables are interpolated to either side of each of the four facial quadrature points associated with any given face. These interpolated variables provide the left and right states associated with the Riemann problems at those points. Four Riemann problems are solved at each of those four facial quadrature points. As a result, the quadrature point on each edge (i.e. see Fig. 1a and equns. (2.10) – (2.12)) draws on the flux components that are evaluated at each of the four facial quadrature points that coincide (via the one-sided limiting procedure shown in Fig. 1b) with the edge quadrature point.

The zone-centered variables, i.e. the first five components of the vector $\mathbf{U}$ in eqn. (2.1), may be updated in the normal fashion in which conserved variables are routinely updated in a higher order Godunov scheme. However, notice that the flux vectors at each of the four facial quadrature points associated with a face can also be used to update $\mathbf{U}$. We illustrate this for the first five components of eqn. (2.2) using the first stage in eqn. (2.5) as an example:

$$
\begin{aligned}
\mathbf{U}^{(1)}_{1,\ldots,5,i,j,k} = \mathbf{U}^{n}_{1,\ldots,5,i,j,k} \\
- \frac{1}{2}\frac{\Delta t}{\Delta x} \Bigg[ & \frac{1}{4}\Big(F^{-}_{1,\ldots,5,\,i+1/2,j,k+1/2}(\mathbf{U}^n) + F^{+}_{1,\ldots,5,\,i+1/2,j,k-1/2}(\mathbf{U}^n) + F^{-}_{1,\ldots,5,\,i+1/2,j+1/2,k}(\mathbf{U}^n) + F^{+}_{1,\ldots,5,\,i+1/2,j-1/2,k}(\mathbf{U}^n)\Big) \\
& - \frac{1}{4}\Big(F^{-}_{1,\ldots,5,\,i-1/2,j,k+1/2}(\mathbf{U}^n) + F^{+}_{1,\ldots,5,\,i-1/2,j,k-1/2}(\mathbf{U}^n) + F^{-}_{1,\ldots,5,\,i-1/2,j+1/2,k}(\mathbf{U}^n) + F^{+}_{1,\ldots,5,\,i-1/2,j-1/2,k}(\mathbf{U}^n)\Big) \Bigg] \\
- \frac{1}{2}\frac{\Delta t}{\Delta y} \Bigg[ & \frac{1}{4}\Big(G^{-}_{1,\ldots,5,\,i+1/2,j+1/2,k}(\mathbf{U}^n) + G^{+}_{1,\ldots,5,\,i-1/2,j+1/2,k}(\mathbf{U}^n) + G^{-}_{1,\ldots,5,\,i,j+1/2,k+1/2}(\mathbf{U}^n) + G^{+}_{1,\ldots,5,\,i,j+1/2,k-1/2}(\mathbf{U}^n)\Big) \\
& - \frac{1}{4}\Big(G^{-}_{1,\ldots,5,\,i+1/2,j-1/2,k}(\mathbf{U}^n) + G^{+}_{1,\ldots,5,\,i-1/2,j-1/2,k}(\mathbf{U}^n) + G^{-}_{1,\ldots,5,\,i,j-1/2,k+1/2}(\mathbf{U}^n) + G^{+}_{1,\ldots,5,\,i,j-1/2,k-1/2}(\mathbf{U}^n)\Big) \Bigg] \\
- \frac{1}{2}\frac{\Delta t}{\Delta z} \Bigg[ & \frac{1}{4}\Big(H^{-}_{1,\ldots,5,\,i+1/2,j,k+1/2}(\mathbf{U}^n) + H^{+}_{1,\ldots,5,\,i-1/2,j,k+1/2}(\mathbf{U}^n) + H^{-}_{1,\ldots,5,\,i,j+1/2,k+1/2}(\mathbf{U}^n) + H^{+}_{1,\ldots,5,\,i,j-1/2,k+1/2}(\mathbf{U}^n)\Big) \\
& - \frac{1}{4}\Big(H^{-}_{1,\ldots,5,\,i+1/2,j,k-1/2}(\mathbf{U}^n) + H^{+}_{1,\ldots,5,\,i-1/2,j,k-1/2}(\mathbf{U}^n) + H^{-}_{1,\ldots,5,\,i,j+1/2,k-1/2}(\mathbf{U}^n) + H^{+}_{1,\ldots,5,\,i,j-1/2,k-1/2}(\mathbf{U}^n)\Big) \Bigg]
\end{aligned}
$$

(2.13)

This completes our description of the time update strategy.

We offer the following observations:
1) The scheme we have presented does not resort to spatial averaging for the electric field and, therefore, resolves issue # 4 in Section I.
2) The update in eqn. (2.13) is locally conservative of the mass, momentum and energy densities. Likewise, eqns. (2.7) to (2.9) show that the x, y and z-components of the magnetic field (viewed as area-weighted averages) are indeed updated in a locally conservative fashion in the x=constant, y=constant and z=constant planes. This resolves Falle's (2002) misconception about the conservation properties of divergence-free schemes.
3) The way is shown for including more quadrature points. As a result, if the spatial reconstruction is better than second order then the use of more quadrature points on the edges permits us to obtain an update scheme for the magnetic field that is better than second order accurate. The use of equns. (2.10) – (2.12) to go from facial quadrature points to edge quadrature points extends naturally to higher orders. The better than second order update of the conserved variables will also require flux evaluation at the facial quadrature points. The edge quadrature points and the facial quadrature points can be judiciously chosen so as to minimize flux evaluations. While the goal of this paper is to achieve a well-designed second order scheme for MHD the present innovations will be used in subsequent papers to demonstrate even higher order accurate schemes.

4) The use of four Riemann problems instead of one at each zone face does add to the cost of the scheme. One may in many situations ameliorate most of that additional cost by using the HLLE type Riemann solver which is indeed very inexpensive. The use of a low cost and low quality Riemann solver is justified because we are also simultaneously improving the quality of the interpolation. We cite the work of Cockburn and Shu (1998) who have shown that as the quality of the interpolation is improved the HLLE-type Riemann solver works almost as well as the Roe-type Riemann solver. The reason is that the higher quality interpolation reduces most of the difference between the left and right states of the Riemann problem. As a result, the extra dissipation introduced by the HLLE Riemann solver is minimized. In the present work we have used both the HLLE and Roe-type Riemann solvers.

5) Brio, Zakharian and Webb (2001) and Wendroff (1999) have designed two dimensional Riemann solvers for hydrodynamics. Because of the way facial quadrature points contribute to the quadrature points on each edge, similarly constructed Riemann solvers for MHD may prove valuable in numerical MHD.

6) Runge-Kutta Discontinuous Galerkin (RKDG) schemes have been designed for hydrodynamics by Cockburn and Shu (1998). The ideas about quadrature points suggest that it is possible to make a natural extension of RKDG schemes to MHD. This point was first made in Balsara (2001b) and has been amplified here. It will be pursued in detail in subsequent papers.

7) The scheme in eqn. (2.6) can alternatively be written as:-

$$\mathbf{U}^{(1)} = \mathbf{U}^n + \Delta t\, L(\mathbf{U}^n)$$

$$\mathbf{U}^{(2)} = \mathbf{U}^n + \frac{1}{4} \Delta t\, L(\mathbf{U}^n) + \frac{1}{4} \Delta t\, L(\mathbf{U}^{(1)}) \qquad (2.14)$$

$$\mathbf{U}^{n+1} = \mathbf{U}^n + \frac{1}{6} \Delta t\, L(\mathbf{U}^n) + \frac{1}{6} \Delta t\, L(\mathbf{U}^{(1)}) + \frac{2}{3} \Delta t\, L(\mathbf{U}^{(2)})$$

In that form it provides the ingredients for doing flux and field corrections for AMR-MHD in a fashion that is third order accurate in time. This becomes an especially powerful idea if it is used together with RKDG formulations that are spatially third or fourth order accurate. The result is an AMR-MHD strategy that breaks free of the second order restrictions in Berger and Colella (1989) and Balsara (2001b). This work will be reported in a subsequent paper.

8) The ideas developed here extend naturally to unstructured meshes in general and tetrahedral meshes in particular, as pointed out by Balsara (2001b). We illustrate this point in Fig. 2a which shows a general tetrahedron. Magnetic field components along the normals to each of the faces are now taken to be the fundamental magnetic field quantities that are to be evolved. To instantiate this for face 1, which opposes vertex 1, we show a locally orthogonal coordinate system $\left( \hat{\xi}_1, \hat{\phi}_1, \hat{\eta}_1 \right)$ where $\hat{\xi}_1$ and $\hat{\phi}_1$ lie within face 1 and $\hat{\eta}_1$ is orthogonal to it. The magnetic field component $B_{\eta_1} = \vec{B} \bullet \hat{\eta}_1$ at the centroid of face 1 is the fundamental magnetic field quantity that needs to be evolved on face 1. For a second order accurate scheme the magnetic field components can be evolved using electric field components that are projected along each of the edges and collocated at the edge centers. The quadrature points for the electric fields are shown by solid dots in Fig. 2a. Again instantiating this for face 1, the temporal update of $B_{\eta_1}$ depends only on $E_{42} = \vec{E} \bullet \hat{r}_{4,2}$, $E_{23} = \vec{E} \bullet \hat{r}_{2,3}$ and $E_{34} = \vec{E} \bullet \hat{r}_{3,4}$ where $\hat{r}_{4,2}$ is a unit vector joining vertex 4 to vertex 2 and the other unit vectors are defined similarly. The update strategy for $B_{\eta_1}$ would be analogous to that in eqn. (2.7) and the multi-stage schemes in eqns. (2.5) and (2.6) can be used to obtain higher order temporal accuracy. The electric field components at the edge-centered quadrature points in Fig. 2a can in turn be obtained by using the components of the flux vector evaluated at the facial quadrature points. The corresponding facial quadrature points are shown in Fig. 2b where the red points lie in face 1, the green points in face 2, the blue points in face 3 and the black points in face 4. Taking the red facial quadrature point that abuts edge 4,2 as an example, we realize that the solution of the Riemann problem at that point will produce electric field components along $\hat{\xi}_1$ and $\hat{\phi}_1$ which can in turn be projected along $\hat{r}_{4,2}$ and thus be made to contribute to $E_{42}$ in a fashion that is very similar to eqn. (2.10). The analogy between Figs. 1a, b and 2a, b is striking and will be developed further in a subsequent paper. We also point out that the present discussion of tetrahedral meshes also shows that issue # 4 from Section I can be resolved on unstructured meshes in a fashion that is entirely analogous to its resolution on structured meshes.

### III) Divergence-Free TVD Reconstruction of Magnetic Fields

In sub-section III.1 we very briefly recapitulate the divergence-free reconstruction on Cartesian meshes and cast it in a format that is suitable for extension to different limiting strategies and different geometries. For a thorough discussion we refer the reader to Balsara (2001b). In sub-section III.2 we describe a fast TVD limiting strategy for the primitive variables which has been found to work very well for MHD. In sub-section III.3 we discuss characteristically designed TVD and r=2 WENO limiting as it applies to MHD and in sub-section III.4 we discuss r=3 WENO limiting as it applies to MHD. The r=2 and 3 WENO schemes for hydrodynamics have been catalogued in Jiang and Shu (1996). In sub-section III.5 we extend the genuinely multidimensional interpolation ideas of Barth (1995) to divergence-free MHD. For a detailed justification of the different reconstruction strategies in sub-sections III.3 to III.5 we point the reader to Balsara (1998b), Jiang and Shu (1996), Balsara and Shu (2000) and Barth (1995). We only focus on the problem of reconstructing MHD variables as it plays out within the context of these different reconstruction strategies. The overarching goal of this section is to show that the full gamut of reconstruction strategies that are available for finite volume reconstruction in CFD are also available to the divergence-free formulations for MHD.

**III.1) Divergence-free Reconstruction of Magnetic Fields on 3d Cartesian Meshes**

To establish some notation we refer the reader to Figure 1. We establish a short-form notation that is given by

$$B_x^\pm \equiv B_{x, i\pm 1/2, j, k} \; ; \; B_y^\pm \equiv B_{y, i, j\pm 1/2, k} \; ; \; B_z^\pm \equiv B_{z, i, j, k\pm 1/2} \tag{3.1}$$

TVD and WENO limiting procedures for the magnetic field components in eqn. (3.1) will be catalogued in subsequent sub-sections. For now let us just specify that the undivided difference of the x-component of the field $B_x^\pm$ in the transverse, i.e. y and z, directions by $\Delta_y B_x^\pm$ and $\Delta_z B_x^\pm$. Similar variations are set up for the other two components in their transverse directions. We consider a zone extending from $(x_1, y_1, z_1)$ as its lower coordinate limits to $(x_2, y_2, z_2)$ as its upper coordinate limits. To simplify the notation, we set the origin at the zone center so that we make the transcription $x \to x - (x_1 + x_2)/2$ and similarly for y and z. The zone has edges of size $\Delta x$, $\Delta y$, and $\Delta z$. The piecewise linear variation of $B_x$, $B_y$ and $B_z$ in the x, y and z zone faces are, therefore, given by:

$$B_x\left(x = \pm\Delta x/2, y, z\right) = B_x^\pm + \frac{\Delta_y B_x^\pm}{\Delta y} y + \frac{\Delta_z B_x^\pm}{\Delta z} z \tag{3.2}$$

$$B_y\left(x, y = \pm\Delta y/2, z\right) = B_y^\pm + \frac{\Delta_x B_y^\pm}{\Delta x} x + \frac{\Delta_z B_y^\pm}{\Delta z} z \tag{3.3}$$

$$B_z\left(x, y, z = \pm\Delta z/2\right) = B_z^\pm + \frac{\Delta_x B_z^\pm}{\Delta x} x + \frac{\Delta_y B_z^\pm}{\Delta y} y \tag{3.4}$$

The reconstructed fields in the interior of the zone which match the linear variation of the fields on the zone faces can be written as:

$$B_x(x, y, z) = a_0 + a_x x + a_y y + a_z z + a_{xx} x^2 + a_{xy} x y + a_{xz} x z \tag{3.5}$$

$$B_y(x, y, z) = b_0 + b_x x + b_y y + b_z z + b_{xy} x y + b_{yy} y^2 + b_{yz} y z \tag{3.6}$$

$$B_z(x, y, z) = c_0 + c_x x + c_y y + c_z z + c_{xz} x z + c_{yz} y z + c_{zz} z^2 \tag{3.7}$$

For the vector field to be divergence-free we need to satisfy

$$\frac{\partial B_x(x, y, z)}{\partial x} + \frac{\partial B_y(x, y, z)}{\partial y} + \frac{\partial B_z(x, y, z)}{\partial z} = 0 \tag{3.8}$$

at all points in the zone's interior. Imposing the divergence-free condition, i.e. eqn. (3.8), in a continuous sense gives four further constraints on the coefficients of the three polynomials given by eqns. (3.5) to (3.7). The constraints are:

$$a_x + b_y + c_z = 0 \; ; \; 2 a_{xx} + b_{xy} + c_{xz} = 0 \; ; \; a_{xy} + 2 b_{yy} + c_{yz} = 0 \; ;$$
$$a_{xz} + b_{yz} + 2 c_{zz} = 0 \tag{3.9}$$

The coefficients in eqns. (3.5) to (3.7) are obtained by matching one field component along with its two transverse variations on each of the six faces of the zone, see eqns. (3.2) to (3.4). On carrying out the algebra we get:

$$a_x = \frac{\left(B_x^+ - B_x^-\right)}{\Delta x} \tag{3.10}$$

$$a_y = \frac{1}{2} \left( \frac{\Delta_y B_x^+}{\Delta y} + \frac{\Delta_y B_x^-}{\Delta y} \right) \tag{3.11}$$

$$a_z = \frac{1}{2} \left( \frac{\Delta_z B_x^+}{\Delta z} + \frac{\Delta_z B_x^-}{\Delta z} \right) \tag{3.12}$$

$$a_{xy} = \frac{1}{\Delta x} \left( \frac{\Delta_y B_x^+}{\Delta y} - \frac{\Delta_y B_x^-}{\Delta y} \right) \tag{3.13}$$

$$a_{xz} = \frac{1}{\Delta x} \left( \frac{\Delta_z B_x^+}{\Delta z} - \frac{\Delta_z B_x^-}{\Delta z} \right) \tag{3.14}$$

$$a_{xx} = -\frac{1}{2} \left( b_{xy} + c_{xz} \right) \tag{3.15}$$

$$a_0 = \frac{\left(B_x^+ + B_x^-\right)}{2} - a_{xx} \frac{\Delta x^2}{4} \tag{3.16}$$

To obtain the formulae for the coefficients in eqn. (3.6), make the following replacements a → b, b → c, c → a, x → y, y → z and z → x in eqns. (3.10) to (3.16) above. Similarly, to obtain the formulae for the coefficients in eqn. (3.7), make the replacements a → c, b → a, c → b, x → z, y → x and z → y in eqns. (3.10) to (3.16) above. Allow self-evident permutations of the form $c_{zx} \equiv c_{xz}$.

We now make several very important points, the first three of which address the issues that were raised in Section I :

1) The volume-averaged magnetic field component in the x-direction in the zone being considered is given by

$$\langle B_x \rangle_{vol-avg} = a_0 + a_{xx} \frac{\Delta x^2}{12} = \frac{(B_x^+ + B_x^-)}{2} - a_{xx} \frac{\Delta x^2}{6} \tag{3.17}$$

Due to the presence of the $a_{xx}$ term we see that eqn. (3.17) is not exactly the mean x-component of the field evaluated at the zone center. However, it is the unique value available from reconstruction. While arithmetic averaging may seem to be acceptable because the second term in (3.17) may be deemed to be a small effect, the second term does not necessarily remain small in situations when the field undergoes substantial variation in a few zones. A strong shock moving obliquely to the computational mesh might exemplify such situations. The problem is especially acute for strong shocks moving through low-$\beta$ plasmas where using the value from eqn. (3.17) has been found to provide some unique advantages to the evaluation of the pressure as will be shown in one of the test problems in Section VII. We see from eqn. (3.17) that a study of the reconstruction problem yields a unique resolution of issue # 1 in Section I.

2) Notice too that we limit the normal components of the magnetic field that are stored at each face. This can be done using the faces that adjoin the face being considered as will be shown in sub-sections III.b, III.c, III.d and III.e. As a result, the scheme presented here applies limiters directly to the fundamental variables thus resolving issue # 2 in Section I.

3) The reconstruction presented here is designed so that the polynomial function for $B_x(x, y, z)$ in eqn. (3.5) naturally matches $B_x(x = \pm \Delta x/2, y, z)$ in eqn. (3.2). As a result the normal component of the magnetic field is $C^0$ continuous at the x-boundary by construction. A similar result holds at the y and z boundaries showing that issue # 3 in Section I is resolved.

4) Those who have experience in higher order Godunov methods might be perplexed by the fact that $B_x$ is allowed to be continuous at the x-boundary. However, note that $B_y$ and $B_z$ are indeed discontinuous at the x-boundary. This allows us to use a Riemann solver to mediate in the proper upwinding of $B_y$ and $B_z$. The continuity of $B_x$ at the x-boundary is essential for the physical consistency of the MHD Riemann problem. Thus the method meets the dual requirements of upwinding and physical consistency of the Riemann solver.

5) The vector field that results from our divergence-free reconstruction is indeed divergence-free when integrated over any closed surface in the computational domain. The closed surface does not have to conform to the zone boundaries.

6) When the results of this section are taken together with the work presented in the previous section we see that a Discontinuous Galerkin formulation for MHD would be especially beneficial where one would evolve not just the magnetic field component at each zone face but also its first ( or higher) moments in the plane of that face. This will be presented in a subsequent paper.

7) There are similar problems with preserving the divergence of the magnetic field in numerical electrodynamics, see Yee (1966); incompressible flow, see Harlow and Welch (1996); relativistic MHD, see Balsara (2001a) and radiation MHD, see Balsara [6,7]. Hence those application areas would also benefit from the strategies developed here.

8) The volume-averaged magnetic field can be evaluated once in each fractional time-step and stored in an auxiliary variable. This ameliorates the extra cost associated with evaluating eqn. (3.17) whenever the zone-centered magnetic fields are needed. As a result one can track the temperature evolution of low-$\beta$ astrophysical flows with fidelity without incurring an extra computational cost.

**III.2) Fast TVD Limiting of the Primitive Variables**

In this sub-section we describe a fast TVD limiting strategy for the primitive variables (i.e. zone-centered density, pressure and velocities and face-centered magnetic fields) which has been found to work very well for MHD. We begin by limiting the face-centered x-component of the magnetic field in the transverse directions as follows:

$$\Delta_y B_x^\pm = \text{VL\_Limiter} \left( B_{x, i\pm 1/2, j+1, k} - B_{x, i\pm 1/2, j, k} , B_{x, i\pm 1/2, j, k} - B_{x, i\pm 1/2, j-1, k} \right) \qquad (3.18)$$

$$\Delta_z B_x^\pm = \text{VL\_Limiter} \left( B_{x, i\pm 1/2, j, k+1} - B_{x, i\pm 1/2, j, k} , B_{x, i\pm 1/2, j, k} - B_{x, i\pm 1/2, j, k-1} \right) \qquad (3.19)$$

where VL_Limiter stands for the van Leer limiter. The other components of the magnetic field can be similarly limited. The density is also limited in zone-centered fashion using the van Leer limiter. The pressure and velocities are, however, limited in zone-centered fashion using the minmod limiter. This mix of limiters seems to be very well-suited for MHD because it seems to be robust enough to handle all the problems that we have used it on and yet refined enough to produce rather crisp representation of linearly degenerate wave fields. While the justification for this choice of limiters is indeed entirely empirical we do, nevertheless, offer an intuitive justification of sorts for this choice. It is intuitively justified by pointing out that the MHD equations have four genuinely non-linear magnetosonic wave families. A compressive limiter is not needed for those wave families because they are genuinely non-linear waves and self-steepen of their own accord. The MHD equations also have three linearly degenerate wave families consisting of two Alfven waves and an entropy wave. Those wave families might benefit from the steepening provided by a compressive limiter such as the van Leer limiter. Since much of the variation in an Alfven wave usually consists of the variation in the magnetic field and since much of the variation in an entropy wave usually consists of the variation in the density we apply the van Leer limiter to those variables.

It is also useful to explain how the pressure is obtained from the total energy density. The limiters in eqns. (3.18) to (3.19) are applied to the face-centered fields. (The limiters described in the next three sub-sections produce even better results.) The undivided differences are then used in eqns. (3.10) to (3.16) to obtain the coefficients of the reconstructed field. Eqn. (3.17) is then used to obtain the volume-averaged magnetic field components and they are used, along with the velocities, to obtain the zone-centered pressure from the total energy. Because eqn. (3.17) includes quadratic variation in the field quantities it is a better representation of the volume-averaged magnetic field and can be especially helpful in some situations involving low-β plasmas. The quadratic variation of the reconstructed magnetic field on structured meshes is an inevitable consequence of its multidimensional structure and the divergence-free constraint that it has to satisfy. In the past, treating low-β plasmas required using corrective techniques of the type described in Balsara and Spicer (1999b) and Janhunen (2000). Such techniques may still be needed for extremely low-β plasmas but the range of β's that can be treated without resort to such techniques is expanded by using the reconstruction described here.

In the next three sub-sections we will describe more sophisticated algorithms for obtaining the slopes of the face-centered magnetic field components. In some of those algorithms we will need the pressure to be interpolated to the zone-boundaries from the two adjoining zone-centers. In all such situations, the zone-centered pressure is always obtained by using the total energy and the best volume-averaged magnetic field that we can form at any stage in the computational method.

### III.3) Characteristically Designed TVD and r=2 WENO Limiting of Magnetic Fields

Characteristic limiting has been known to produce better entropy enforcement than limiting on the primitives. This has been shown for hydrodynamics by Harten (1983) and for MHD by Balsara (1998b). We, therefore, describe characteristic limiting in this sub-section. The limiting of the zone-centered variables is done exactly as in the hydrodynamic case and so we do not describe that any further. In this paper we focus on the characteristic limiting of the face-centered magnetic field components. Say we want to limit the variation of the y-component of the magnetic field along the x-direction. The y-component of the magnetic field is defined at the y = constant face of a zone. We thus define a short form notation as:

$$B_{y,i} \equiv B_{y,i,j+1/2,k} \qquad (3.20)$$

Other flow variables such as the density, pressure, velocities as well as the other field components can be assigned at the face being considered. Again using a short form notation we illustrate this process for the density:

$$\rho_i = \frac{1}{2}\left(\rho_{i,j,k} + \rho_{i,j+1,k}\right) \qquad (3.21)$$

As a result, we can define a vector of primitive variables $V_i = \left(\rho_i, v_{x,i}, v_{y,i}, v_{z,i}, P_i, B_{x,i}, B_{y,i}, B_{z,i}\right)^T$ for a row of y = constant zone faces along the x-direction. The subscript "i" tags the y = constant zone faces in the x-direction. At each face in that row we can define the wave speed $\lambda_i^\alpha$ for the $\alpha^{th}$ wave that propagates in the x-direction. Here $\alpha = 1,\ldots,7$. Associated with each wave speed in each zone we can also evaluate the left and right eigenvectors in the primitive variables that are denoted by $l^{\alpha}(V_i)$ and $r^{\alpha}(V_i)$. For each characteristic field α in zone "i" we define the weight associated with that characteristic field as:

$$w_{i,m}^{\alpha} = l^{\alpha}(V_i) \bullet V_{i+m} \qquad \text{where m} = -1, 0, 1 \qquad (3.22)$$

Characteristically designed TVD limiting consists of applying the limiter to the differences of the characteristic weights as follows:

$$\Delta w_i^{\alpha} = \text{Limiter}\,(\,w_{i,+1}^{\alpha} - w_{i,0}^{\alpha},\, w_{i,0}^{\alpha} - w_{i,-1}^{\alpha}) \qquad (3.23)$$

where different limiters can be applied to different characteristic fields. For example, one may apply a compressive limiter, like the van Leer limiter, to the linearly degenerate fields and a minmod limiter to the genuinely non-linear fields. The characteristically designed TVD slopes in the primitive variables are given by:

$$\Delta V_i = \sum_{\alpha} \Delta w_i^{\alpha}\, r^{\alpha}(V_i) \qquad (3.24)$$

The $B_y$ component of $\Delta V_i$ yields the characteristically limited value for the y-component of the magnetic field in the x-direction.

We now turn to a description of the r=2 WENO algorithm which uses the same characteristic weights as in eqn. (3.22) but uses a smoother representation of the interpolation in order to avoid chopping off extrema. In doing so one also obtains one further order of spatial accuracy. Since the present scheme is designed to be a finite volume scheme, that further gain in order of accuracy will only be obtained when the variation in the flow is restricted to one dimension. Much of the algorithm presented here was first described by Liu, Osher and Chan (1994) and we refer the interested reader to that paper for terminology and details. However, we do present here a variation of their algorithm which gives numerical MHD the extra stability that it seems to need. The algorithm relies on examining the direction of wave propagation and the structure of the characteristic variables. Using the weights in eqn. (3.22) it is possible to identify two possible stencils $\{w_{i,-1}^{\alpha}, w_{i,0}^{\alpha}\}$ and $\{w_{i,0}^{\alpha}, w_{i,+1}^{\alpha}\}$ which could both potentially provide linear interpolation of the characteristic variable in the zone of interest. The WENO algorithm relies on examining the structure of the characteristic fields in order to decide on the amount of importance that is to be attributed to either of the two possible stencils. MHD calculations seem to require more limiting than the corresponding hydrodynamic ones. In view of that fact we make a modification of the basic algorithm for obtaining the optimal weights for the two possible stencils as follows:

if ( $\lambda_{i-1}^{\alpha} < 0$ and $\lambda_{i}^{\alpha} < 0$) then

$C_{i,-}^{\alpha} = 1.0$ ; $C_{i,+}^{\alpha} = 0.5$

if ( $\lambda_{i}^{\alpha} > 0$ and $\lambda_{i+1}^{\alpha} > 0$) then

$C_{i,-}^{\alpha} = 0.5$ ; $C_{i,+}^{\alpha} = 1.0$ (3.25)

else

$C_{i,-}^{\alpha} = 1.0$ ; $C_{i,+}^{\alpha} = 1.0$

end if

The above modification is applied to the WENO interpolation of the face-centered variables and also to the WENO interpolation of the zone-centered variables. The normalized weights associated with the characteristic slopes from the two possible stencils are given by:

$$\omega_{i,-}^{\alpha} = \frac{\beta_{i,-}^{\alpha}}{\beta_{i,-}^{\alpha} + \beta_{i,+}^{\alpha}} \; ; \; \omega_{i,+}^{\alpha} = \frac{\beta_{i,+}^{\alpha}}{\beta_{i,-}^{\alpha} + \beta_{i,+}^{\alpha}} \quad (3.26)$$

where $\beta_{i,-}^{\alpha}$ and $\beta_{i,+}^{\alpha}$ are the un-normalized weights and are given by:

$$\beta_{i,-}^{\alpha} = \frac{C_{i,-}^{\alpha}}{\left(w_{i,0}^{\alpha} - w_{i,-1}^{\alpha}\right)^2 + \varepsilon} \; ; \; \beta_{i,+}^{\alpha} = \frac{C_{i,+}^{\alpha}}{\left(w_{i,1}^{\alpha} - w_{i,0}^{\alpha}\right)^2 + \varepsilon} \quad (3.27)$$

Here $\varepsilon$ is a small positive number that is introduced to prevent the denominator from becoming zero. Typically, $\varepsilon \sim 10^{-12}$ for the calculations presented here. The WENO limited slope of the characteristic weight, i.e. $\Delta w_i^{\alpha}$ from the previous paragraph, is given by

$$\Delta w_i^{\alpha} = \omega_{i,-}^{\alpha} \left(w_{i,0}^{\alpha} - w_{i,-1}^{\alpha}\right) + \omega_{i,+}^{\alpha} \left(w_{i,+1}^{\alpha} - w_{i,0}^{\alpha}\right) \quad (3.28)$$

and application of eqn. (3.24) yields the characteristically designed WENO slopes for the primitive variables. The $B_y$ component of $\Delta V_i$ yields the characteristically limited value for the y-component of the magnetic field in the x-direction.

Three important points are worth making:
1) The characteristic steepening algorithms of Yang (1990) also work very well with the TVD/WENO algorithms presented here as shown by Balsara (1998b) and Balsara and Shu (2000).
2) Eqn. (3.21) does produce a slight averaging though second order accuracy and stability are retained. However, a complete resolution of this issue will be presented in our subsequent work on RKDG schemes.

3) The r=2 WENO and the r=3 WENO schemes, as described in Jiang and Shu (1996), are said to be third and fifth order accurate when used in a zone-centered, pointwise WENO scheme. We are not designing pointwise WENO schemes in this paper. Hence our goal is only to use the slopes that are available from these WENO schemes as a way of producing a better quality slope in what is basically a second order accurate, finite-volume scheme for MHD.

**III.4) Characteristically Designed r=3 WENO Limiting of Magnetic Fields**

We now describe a modification of the r=3 WENO algorithm of Jiang and Shu (1996) and we refer the interested reader to that paper for terminology and details. The present interpolation uses a larger stencil than the r=2 WENO algorithm and, therefore, produces an even smoother representation of the interpolation. We now use the characteristic weights:

$$w_{i,m}^{\alpha} = l^{\alpha}(V_i) \bullet V_{i+m} \qquad \text{where } m = -2, -1, 0, 1, 2 \qquad (3.29)$$

Using the weights in eqn. (3.22) it is possible to identify three possible stencils $\{w_{i,-2}^{\alpha}, w_{i,-1}^{\alpha}, w_{i,0}^{\alpha}\}$, $\{w_{i,-1}^{\alpha}, w_{i,0}^{\alpha}, w_{i,+1}^{\alpha}\}$ and $\{w_{i,0}^{\alpha}, w_{i,+1}^{\alpha}, w_{i,+2}^{\alpha}\}$, all three of which could potentially interpolate the characteristic variable by a quadratic interpolant in the zone of interest. We again modify the interpolation as follows:

if ( $\lambda_{i-1}^{\alpha} < 0$ and $\lambda_i^{\alpha} < 0$ ) then
$C_{i,-1}^{\alpha} = 0.3$ ; $C_{i,0}^{\alpha} = 0.6$ ; $C_{i,+1}^{\alpha} = 0.1$
if ( $\lambda_i^{\alpha} > 0$ and $\lambda_{i+1}^{\alpha} > 0$ ) then
$C_{i,-1}^{\alpha} = 0.1$ ; $C_{i,0}^{\alpha} = 0.6$ ; $C_{i,+1}^{\alpha} = 0.3$ \qquad (3.30)
else
$C_{i,-1}^{\alpha} = 1.0$ ; $C_{i,0}^{\alpha} = 1.5$ ; $C_{i,+1}^{\alpha} = 1.0$
end if

The above modification is applied to the WENO interpolation of the face-centered variables and also to the WENO interpolation of the zone-centered variables. The normalized weights that are ascribed to the characteristic slopes associated with each of the three stencils are given by:

$$\omega_{i,-1}^{\alpha} = \frac{\beta_{i,-1}^{\alpha}}{\beta_{i,-1}^{\alpha} + \beta_{i,0}^{\alpha} + \beta_{i,+1}^{\alpha}} \; ; \; \omega_{i,0}^{\alpha} = \frac{\beta_{i,0}^{\alpha}}{\beta_{i,-1}^{\alpha} + \beta_{i,0}^{\alpha} + \beta_{i,+1}^{\alpha}} \; ; \; \omega_{i,+1}^{\alpha} = \frac{\beta_{i,+1}^{\alpha}}{\beta_{i,-1}^{\alpha} + \beta_{i,0}^{\alpha} + \beta_{i,+1}^{\alpha}} \quad (3.31)$$

where $\beta_{i,-1}^{\alpha}$, $\beta_{i,0}^{\alpha}$ and $\beta_{i,+1}^{\alpha}$ are the un-normalized weights and are given by:

$$\beta_{i,-1}^{\alpha} = \frac{C_{i,-1}^{\alpha}}{IS_{i,-1} + \varepsilon} \; ; \beta_{i,0}^{\alpha} = \frac{C_{i,0}^{\alpha}}{IS_{i,0} + \varepsilon} \; ; \beta_{i,+1}^{\alpha} = \frac{C_{i,+1}^{\alpha}}{IS_{i,+1} + \varepsilon} \tag{3.32}$$

The smoothness measures $IS_{i,-1}$, $IS_{i,0}$ and $IS_{i,+1}$ in eqn. (3.32) evaluate the extent of the variation in each of the three stencils and are given by:

$$IS_{i,-1} = \frac{13}{12}\left(w_{i,-2}^{\alpha} - 2 w_{i,-1}^{\alpha} + w_{i,0}^{\alpha}\right)^2 + \frac{1}{4}\left(w_{i,-2}^{\alpha} - 4 w_{i,-1}^{\alpha} + 3 w_{i,0}^{\alpha}\right)^2 \tag{3.33}$$

$$IS_{i,0} = \frac{13}{12}\left(w_{i,-1}^{\alpha} - 2 w_{i,0}^{\alpha} + w_{i,+1}^{\alpha}\right)^2 + \frac{1}{4}\left(w_{i,-1}^{\alpha} - w_{i,+1}^{\alpha}\right)^2 \tag{3.34}$$

$$IS_{i,+1} = \frac{13}{12}\left(w_{i,0}^{\alpha} - 2 w_{i,+1}^{\alpha} + w_{i,+2}^{\alpha}\right)^2 + \frac{1}{4}\left(3 w_{i,0}^{\alpha} - 4 w_{i,+1}^{\alpha} + w_{i,+2}^{\alpha}\right)^2 \tag{3.35}$$

As a result, the WENO limited slope of the characteristic weight, i.e. $\Delta w_i^{\alpha}$ from the previous sub-section, is given by:

$$\Delta w_i^{\alpha} = \omega_{i,-1}^{\alpha} \frac{1}{2}\left(w_{i,-2}^{\alpha} - 4 w_{i,-1}^{\alpha} + 3 w_{i,0}^{\alpha}\right) + \omega_{i,0}^{\alpha} \frac{1}{2}\left(w_{i,+1}^{\alpha} - w_{i,-1}^{\alpha}\right) \\ + \omega_{i,+1}^{\alpha} \frac{1}{2}\left(-3 w_{i,0}^{\alpha} + 4 w_{i,+1}^{\alpha} - w_{i,+2}^{\alpha}\right) \tag{3.36}$$

Application of eqn. (3.24) yields the characteristically designed WENO slopes for the primitive variables. When doing some stringent MHD test problems we found the r=3 WENO interpolation to be too compressive for the genuinely non-linear fields when strong shocks are present in the problem. The r=3 WENO interpolation tries to fit the steepest possible slopes that are consistent with enforcing non-oscillatory behavior in the characteristic variables. In doing that it fails to take into account the fact that certain variables, such as density and pressure, need to remain bounded in physical space because of positivity considerations. There are two easy ways to resolve this situation: 1) Take the slopes provided by the r=3 WENO scheme and apply a further step of multidimensional limiting of the physical variables as shown in the next section. 2) Apply the r=3 WENO only to the linearly degenerate characteristic fields. Use r=2 WENO for the genuinely non-linear fields. The latter option is preferred for its simplicity.

There is also a very refined way to resolve this situation. It consists of realizing that the r=3 WENO interpolation works very well in all smooth regions of the flow. It also works very well for discontinuities of modest strength. It only produces TVD-violating solutions at strong magnetosonic shocks. Balsara (1998b) had shown that the characteristic variables can be used to analyze the structure of MHD flows and selectively introduce extra dissipation and stability where needed. We use a similar philosophy here in designing blended-WENO interpolation. Blended-WENO interpolation uses the more accurate r=3 WENO interpolation in most portions of the flow including those regions where the TVD condition is weakly violated. In regions where it is strongly violated it reverts to the stabler r=2 WENO interpolation. Thus a high-accuracy WENO interpolation scheme is blended with a lower-accuracy WENO interpolation which lends it more stability. Notice that well-resolved local extrema will almost always be treated with the high-accuracy WENO interpolation. Even marginally-resolved extrema will not be clipped entirely because they will be treated by the r=2 WENO interpolation which preserves extrema. Genuinely non-linear characteristic fields can be treated differently from linearly degenerate characteristic fields so that the blended-WENO scheme can bring out the best solution in each wave family. The coefficients that control this blending can be tailored to suit the specific needs of the hyperbolic system being considered. Blended-WENO consists of constructing the following ratios in each zone "i".

$$r_{i,0} = \frac{w^{\alpha}_{i,+1} - w^{\alpha}_{i,0}}{w^{\alpha}_{i,0} - w^{\alpha}_{i,-1}} \quad ; r_{i,+1} = \frac{w^{\alpha}_{i,+2} - w^{\alpha}_{i,+1}}{w^{\alpha}_{i,+1} - w^{\alpha}_{i,0}} \quad ; r_{i,-1} = \frac{w^{\alpha}_{i,0} - w^{\alpha}_{i,-1}}{w^{\alpha}_{i,-1} - w^{\alpha}_{i,-2}} \quad ; \tag{3.37}$$

We then interpolate the genuinely non-linear characteristic variables within zone "i" as follows:

$$\begin{aligned}
&\textit{if } \left\{ \left( r_{i,0} \geq 0 \right) \textit{or} \left[ \left( \frac{1}{\beta_{nl}} \leq \left| r_{i,0} \right| \right) \textit{and} \left( \left| r_{i,0} \right| \leq \beta_{nl} \right) \right] \right\} \textit{and} \\
&\quad \left\{ \left( r_{i,+1} \geq 0 \right) \textit{or} \left[ \left( \frac{1}{\beta_{nl}} \leq \left| r_{i,+1} \right| \right) \textit{and} \left( \left| r_{i,+1} \right| \leq \beta_{nl} \right) \right] \right\} \textit{and} \\
&\quad \left\{ \left( r_{i,-1} \geq 0 \right) \textit{or} \left[ \left( \frac{1}{\beta_{nl}} \leq \left| r_{i,-1} \right| \right) \textit{and} \left( \left| r_{i,-1} \right| \leq \beta_{nl} \right) \right] \right\} \textit{ then}
\end{aligned} \tag{3.38}$$

  Use r=3 WENO in zone "i"

*else*

  Use r=2 WENO in zone "i"

*endif*

The linearly-degenerate characteristic variables within zone "i" are interpolated as follows:

$$\text{if } \left\{ \left( r_{i,0} \geq 0 \right) or \left[ \left( \frac{1}{\beta_{ld}} \leq |r_{i,0}| \right) and \left( |r_{i,0}| \leq \beta_{ld} \right) \right] \right\} \text{ then}$$

  Use r=3 WENO in zone "i"

*else*                                 (3.39)

  Use r=2 WENO in zone "i"

*endif*

Extensive testing has shown that the blending coefficients $\beta_{nl} = 1.5$ and $\beta_{ld} = 2.5$ yield a very pleasing and stable scheme for MHD which is robust, has high resolution and retains accuracy over a wide spectrum of problems. Several points about blended-WENO follow:

1) The algorithm is scale-free. When $r_{i,0} < 0$ the restriction $\frac{1}{\beta_{nl}} \leq |r_{i,0}| \leq \beta_{nl}$ keeps the interpolation symmetrical. The use of three such conditionals in equn. (3.38) ensures that small scale fluctuations that are on the scale of two or three mesh points do not develop in the non-linear waves.

2) Ideas that are somewhat similar in spirit have been tried in Gozalo and Abgrall (2001). Unlike their schemes, the blended-WENO interpolation uses the high-accuracy r=3 WENO interpolation even when the TVD condition, i.e. $r_{i,0} \geq 0$, is violated, as long as it is not violated by too much.

3) The present scheme produces very nice results even with an HLLE Riemann solver. The blended-WENO interpolation is so good that even with an HLLE Riemann solver contact discontinuities and torsional Alfven waves are represented crisply. When a linearized, Roe-type Riemann solver is used a characteristically-designed viscosity of the form given in Balsara (1998b) should be included at strong shocks. It is worth pointing out that when large numbers of chemical species are evolved along with the MHD equations it is quite advantageous to use the high-quality blended-WENO interpolation and a low-quality HLLE Riemann solver instead of using a lower quality interpolation and a high-quality Roe-type Riemann solver. The reason is that the blended-WENO interpolation can be done on the species fractions, which results in the reconstruction of each specie being treated like a scalar interpolation step. The computational complexity of the HLLE Riemann solver increases linearly with the number of species used, a property which is not trivially shared by the Roe-type Riemann solver.

4) As shown in Balsara (1998b) an ACM steepener can be used to improve the representation of linearly degenerate characteristic fields. Owing to the good resolving abilities of the blended-WENO interpolation this too may be an unnecessarily expensive choice for most problems.

5) The blended-WENO algorithm can also be used as an intelligent limiter for RKDG schemes. In that role it can detect problematic regions and relinquish formal accuracy in a very controlled fashion. The blending coefficients $\beta_{nl}$ and $\beta_{ld}$ can take on much higher values for RKDG schemes. We can reinterpret the blended-WENO concept so that we choose the lower order, stabilizing scheme to be the highest order WENO scheme that can stably integrate the hyperbolic system by itself. Thus for hydrodynamical problems the r=3 WENO interpolation can be used as the lower order scheme instead of the r=2 WENO interpolation.

6) In Balsara and Shu (2000) we had shown that higher order finite-difference WENO schemes have several advantages provided their monotonicity properties can be controlled. For hydrodynamical problems the r=3 WENO interpolation is almost always a stable choice and can be blended in with the r=5 WENO interpolation in Balsara and Shu (2000). This can be made to yield a superb scheme that is minimally fifth order accurate and usually has the excellent resolution and advection properties of the r=5 WENO scheme. For MHD problems the r=2 WENO is the stable choice and can be blended in with the $r \geq 3$ WENO schemes in Balsara and Shu (2000) using the methods given here.

**III.5) Genuinely Multidimensional Limiting**

The discussion in sub-section III.1 shows us that the divergence-free reconstruction problem couples all three dimensions. We realize, therefore, that it is advantageous to seek genuinely multidimensional strategies for limiting the fluid and magnetic field variables. We explore only the least expensive forms of such limiting strategies here. Such a strategy was presented by Barth (1995). We first catalogue the scheme as it applies to zone-centered variables for two reasons. First, the article by Barth may not be accessible to readers of archival journals. Second, we point out that there is a degree of freedom in the limiter that allows us to make contact with compressive and non-compressive limiters in one dimension. We then extend the strategy to face-centered magnetic fields.

Barth's scheme is based on restricting the interpolated variable to lie within the bounds provided by all the neighboring zones that abut it. Thus a zone (i,j,k) in three dimensions will have 26 other zones that abut it. Consider a zone-centered variable such as the density, $\rho_{i,j,k}$, in zone (i,j,k). Within each zone, evaluate the slopes. This can be done using central differences such as:

$$\Delta_x \rho_{i,j,k} = 0.5\,(\rho_{i+1,j,k} - \rho_{i-1,j,k})\ ;\ \Delta_y \rho_{i,j,k} = 0.5\,(\rho_{i,j+1,k} - \rho_{i,j-1,k})\ ;\ \Delta_z \rho_{i,j,k} = 0.5\,(\rho_{i,j,k+1} - \rho_{i,j,k-1})$$
(3.40)

It can also be done by using the using the r=2 and r=3 WENO slopes provided in sub-sections III.3 and III.4 respectively. The reconstructed density in zone (i,j,k) before any multidimensional limiting is applied is given by:

$$\rho(x,y,z) = \rho_{i,j,k} + \frac{\Delta_x \rho_{i,j,k}}{\Delta x} x + \frac{\Delta_y \rho_{i,j,k}}{\Delta y} y + \frac{\Delta_z \rho_{i,j,k}}{\Delta z} z \tag{3.41}$$

Multidimensional limiting relies on finding a scalar limiter function $\tau_{i,j,k}$ in zone (i,j,k) so that the multidimensionally limited density can be written as:

$$\rho(x,y,z) = \rho_{i,j,k} + \tau_{i,j,k} \left( \frac{\Delta_x \rho_{i,j,k}}{\Delta x} x + \frac{\Delta_y \rho_{i,j,k}}{\Delta y} y + \frac{\Delta_z \rho_{i,j,k}}{\Delta z} z \right) \tag{3.42}$$

The scalar limiter function $\tau_{i,j,k}$ is chosen so as to keep the function in eqn. (3.42) within the bounds provided by the neighboring zones. This is done as follows: Let $\rho_{i,j,k}^{max}$ and $\rho_{i,j,k}^{min}$ denote the maximum and minimum values of eqn. (3.41) in zone (i,j,k). Let $\hat{\rho}_{i,j,k}^{max}$ and $\hat{\rho}_{i,j,k}^{min}$ denote the maximum and minimum values of the zone-averaged density in the 27 zones whose indices range from (i-1,j-1,k-1) to (i+1,j+1,k+1). The limiter function is now given by

$$\tau_{i,j,k} = \min\left( 1, \ \psi \min\left( \frac{\hat{\rho}_{i,j,k}^{max} - \rho_{i,j,k}}{\rho_{i,j,k}^{max} - \rho_{i,j,k}}, \frac{\rho_{i,j,k} - \hat{\rho}_{i,j,k}^{min}}{\rho_{i,j,k} - \rho_{i,j,k}^{min}} \right) \right) \tag{3.43}$$

Setting $\psi$ to any value between 0.5 and 1.0 gives us second order accuracy. When restricted to one dimensional variations it is easy to verify that $\psi = 0.5$ yields a non-compressive minmod limiter while $\psi = 1.0$ yields a compressive limiter. Consistent with the suggestion in sub-section III.2 setting $\psi = 1.0$ for all the MHD variables would yield oscillatory solutions for strong shock problems in numerical MHD. Setting $\psi = 1.0$ for the density and magnetic fields while using $\psi = 0.5$ for the pressure and velocities yields much better results.

The process of multidimensionally limiting the magnetic field can be illustrated by focusing on the x-component of the magnetic field, $B_{x,\ i+1/2,j,k}$. The limiting relies on using the x-components of the magnetic field in the 8 neighboring zone faces that abut the face being considered. Within the x=constant face, evaluate the slopes in the two transverse directions. This can be done using central differences such as:

$$\Delta_y B_{x,\ i+1/2,j,k} = 0.5\ (B_{x,\ i+1/2,j+1,k} - B_{x,\ i+1/2,j-1,k})\ ;\ \Delta_z B_{x,\ i+1/2,j,k} = 0.5\ (B_{x,\ i+1/2,j,k+1} - B_{x,\ i+1/2,j,k-1}) \tag{3.44}$$

It can also be done by using the r=3 WENO slopes provided in sub-section III.4. The reconstructed x-component of the magnetic field in face (i+1/2,j,k), before any multidimensional limiting is applied, is given by:

$$B(y,z) = B_{x,\,i+1/2,j,k} + \frac{\Delta_y B_{x,\,i+1/2,j,k}}{\Delta y}\, y + \frac{\Delta_z B_{x,\,i+1/2,j,k}}{\Delta z}\, z \tag{3.45}$$

Multidimensional limiting relies on finding a scalar limiter function $\tau_{i+1/2,j,k}$ in face $(i+1/2,j,k)$ so that the multidimensionally limited magnetic field component can be written as:

$$B(y,z) = B_{x,\,i+1/2,j,k} + \tau_{i+1/2,j,k} \left( \frac{\Delta_y B_{x,\,i+1/2,j,k}}{\Delta y}\, y + \frac{\Delta_z B_{x,\,i+1/2,j,k}}{\Delta z}\, z \right) \tag{3.46}$$

As before, let $B^{max}_{x,\,i+1/2,j,k}$ and $B^{min}_{x,\,i+1/2,j,k}$ denote the maximum and minimum values of eqn. (3.45) in face $(i+1/2,j,k)$. Let $\hat{B}^{max}_{x,\,i+1/2,j,k}$ and $\hat{B}^{min}_{x,\,i+1/2,j,k}$ denote the maximum and minimum values of the x-component of the face-averaged magnetic field in the 9 faces whose indices range from $(i+1/2,j-1,k-1)$ to $(i+1/2,j+1,k+1)$. The limiter function is now given by

$$\tau_{i+1/2,j,k} = \min\left(1,\ \psi\ \min\left( \frac{\hat{B}^{max}_{x,\,i+1/2,j,k} - B_{x,\,i+1/2,j,k}}{B^{max}_{x,\,i+1/2,j,k} - B_{x,\,i+1/2,j,k}}\ ,\ \frac{B_{x,\,i+1/2,j,k} - \hat{B}^{min}_{x,\,i+1/2,j,k}}{B_{x,\,i+1/2,j,k} - B^{min}_{x,\,i+1/2,j,k}} \right)\right) \tag{3.47}$$

Several important points are worth making:
1) When comparing the limiting in this sub-section to the fast TVD limiting in sub-section III.2 we realize that for one-dimensional variations the present reconstruction strategy does reduce to something that is quite similar to the fast TVD reconstruction strategy. However, in multiple dimensions, they are different. In three dimensions, the dimension-by-dimension TVD limiting applies three limiters to each of the directions individually. The present limiter applies only one limiter to all the directions viewed collectively. Thus the multidimensional limiter permits greater latitude of variation to the slopes than any limiter that is based on a dimension-by-dimension analysis.
2) When the slopes from the r=3 WENO interpolation in sub-section III.4 are used in place of eqns. (3.40) and (3.44) a very nice scheme results. The reason is that the r=3 WENO does a very nice job of constructing the slopes based on a detailed examination of the upwind direction. The multidimensional limiting described in this sub-section is very unobtrusive even as it nudges the r=3 WENO interpolation to stay within physical bounds. This may be a way to provide a modicum of stability to the r=3 WENO slopes on very stringent MHD problems.
3) Even when used with the slopes in eqns. (3.40) and (3.44) the present scheme can produce rather pleasing results even on very stringent problems as will be shown in the Results section, i.e. Section VII.
4) Barth (1995) originally designed the multidimensional limiter described here for use on unstructured meshes. Everything that has been described here goes over cleanly for unstructured meshes. This shows that divergence-free schemes for numerical MHD that use genuinely multidimensional, divergence-free reconstruction on unstructured meshes can be designed using the ideas developed here. We will explore that in a subsequent paper.

5) When $\psi = 1.0$ is used for all MHD variables a scheme that has superb second order convergence properties on relatively smooth problems results. Such a scheme also works well on problems with moderately strong shocks. However, such a scheme has some difficulties on extremely stringent MHD problems such as the blast problem in sub-section VII.2. Using $\psi = 0.5$ instead for the pressure and velocities yields a very robust scheme which, despite being second order accurate, has somewhat poorer second order convergence properties. The convergence properties of either scheme will be catalogued in Section VI. For that reason we draw on an idea in Balsara and Spicer (1999a) where it was suggested that it is possible to detect the presence of strong MHD shocks by using a pair of switches. Eqns. (9) to (12) from Balsara and Spicer (1999a) provide the details. The results in this paper have been produced by using $\tilde{\beta} = 0.3$ and $\delta = 0.5$ in eqns. (9) and (10) of Balsara and Spicer (1999a) though substantially larger values, such as $\tilde{\beta} = 1.0$ and $\delta = 1.0$, can also be used. In regions of the flow that are free of strong shocks we use $\psi = 1.0$ for all MHD variables. When shocks are detected in a zone we revert to using $\psi = 0.5$ locally for the pressure and velocities. For the rest of this paper, whenever we refer to the genuinely multidimensional limiter from this section without being more specific we will be referring to the limiting technique described in this point.

**IV) Extending the Divergence-Free Reconstruction Strategy to General Geometries**

In sub-section IV.1 we extend the divergence-free reconstruction strategy to cylindrical geometry. In sub-section IV.2 we extend it to spherical structured meshes and in sub-section IV.3 we extend it to tetrahedral meshes. In doing so we show that the ideas for divergence-free reconstruction of vector fields that are being explored here are very general and extend naturally to complex geometries on which physical problems are done.

**IV.1) Cylindrical Geometry**

Consider a zone in cylindrical geometry extending from $( r_1, \theta_1, z_1 )$ at its lower coordinate limits to $( r_2, \theta_2, z_2 )$ at its upper coordinate limits. Here "r" is the radial coordinate, "θ" is the toroidal coordinate and "z" is measured along the z-axis. The divergence-free condition is given by:

$$\frac{1}{r} \left[ \frac{\partial \, (r \, B_r)}{\partial \, r} + \frac{\partial \, (B_\theta)}{\partial \, \theta} + \frac{\partial \, (r \, B_z)}{\partial \, z} \right] = 0 \tag{4.1}$$

First introduce volumetric coordinates

$$x = \frac{r^2}{2} \; ; \; y = \theta \; ; \; z = z \tag{4.2}$$

with the volumetric centroid given by

$$x_c = \frac{r_c^2}{2} = \frac{1}{2}\left(\frac{r_1^2}{2} + \frac{r_2^2}{2}\right) \; ; \; y_c = \frac{1}{2}(\theta_1 + \theta_2) \; ; \; z_c = \frac{1}{2}(z_1 + z_2) \tag{4.3}$$

Then set the origin at the zone center so that we make the transcription $x \to x - x_c$ and similarly for y and z. Eqn. (4.1) can be made to look like eqn. (3.8) by setting

$$B_x = r\, B_r \; ; \; B_y = \frac{B_\theta}{r} \; ; \; B_z = B_z \tag{4.4}$$

Hence the expressions for $B_x(x, y, z)$, $B_y(x, y, z)$ and $B_z(x, y, z)$ should be identical to eqns. (3.5) to (3.7) in sub-section III.1. As a result, the coefficients of the reconstructed polynomials are given by eqns. (3.10) to (3.16). For evaluating those coefficients we will need expressions for terms like $\Delta_x B_y^\pm$. The limiting procedure only gives $\Delta_r B_\theta^\pm$. We obtain $\Delta_x B_y^\pm$ from $\Delta_r B_\theta^\pm$ as follows:

$$\Delta_x B_y^\pm = \Delta_r B_\theta^\pm \frac{1}{r_c} + \langle B_\theta^\pm \rangle \frac{r_1 - r_2}{r_c^2} \tag{4.5}$$

The magnetic fields in eqn. (4.4) are collocated at the centroids of the faces. The volume averaged $\langle B_z \rangle_{vol-avg}$ is still given by an expression that is similar to eqn. (3.17). However, because of the inclusion of factors of "r" in eqn. (4.4), our expressions for $\langle B_r \rangle_{vol-avg}$ and $\langle B_\theta \rangle_{vol-avg}$ are given by:

$$\begin{aligned}
\langle B_r \rangle_{vol-avg} &= \frac{2\, a_0}{(r_1 + r_2)} + \frac{a_x}{(r_2^2 - r_1^2)}\left[\frac{1}{3}(r_2^3 - r_1^3) - r_c^2(r_2 - r_1)\right] \\
&+ \frac{a_{xx}}{(r_2^2 - r_1^2)}\left[\frac{1}{10}(r_2^5 - r_1^5) - \frac{r_c^2}{3}(r_2^3 - r_1^3) + \frac{r_c^4}{2}(r_2 - r_1)\right]
\end{aligned} \tag{4.6}$$

$$\begin{aligned}
\langle B_\theta \rangle_{vol-avg} &= \left[b_0 + b_{yy}\frac{\Delta\theta^2}{12}\right]\frac{2}{3}\frac{(r_2^3 - r_1^3)}{(r_2^2 - r_1^2)} \\
&+ \frac{b_x}{(r_2^2 - r_1^2)}\left[\frac{1}{5}(r_2^5 - r_1^5) - \frac{r_c^2}{3}(r_2^3 - r_1^3)\right]
\end{aligned} \tag{4.7}$$

For the sake of completeness we also write out the r, θ and z-components of the momentum equations. In doing so, we wish to show that higher order Godunov schemes for MHD can be written in a form that fully conserves angular momentum. The r-component of the momentum equation becomes:

$$\frac{\partial(\rho v_r)}{\partial t} + \frac{1}{r}\frac{\partial}{\partial r}\left[r\left(\rho v_r^2 - \frac{B_r^2}{4\pi}\right)\right] + \frac{1}{r}\frac{\partial}{\partial \theta}\left[\rho v_r v_\theta - \frac{B_r B_\theta}{4\pi}\right] + \frac{\partial}{\partial z}\left[\rho v_r v_z - \frac{B_r B_z}{4\pi}\right]$$
$$+ \frac{\partial}{\partial r}\left[P + \frac{\mathbf{B}^2}{8\pi}\right] = \frac{\rho v_\theta^2}{r} - \frac{B_\theta^2}{4\pi r}$$

(4.8)

The θ-component of the momentum equation, when written out in an angular momentum conserving form, becomes:

$$\frac{\partial(r\rho v_\theta)}{\partial t} + \frac{1}{r}\frac{\partial}{\partial r}\left[r^2\left(\rho v_r v_\theta - \frac{B_r B_\theta}{4\pi}\right)\right] + \frac{1}{r}\frac{\partial}{\partial \theta}\left[r\left(\rho v_\theta^2 - \frac{B_\theta^2}{4\pi} + P + \frac{\mathbf{B}^2}{8\pi}\right)\right]$$
$$+ \frac{\partial}{\partial z}\left[r\left(\rho v_\theta v_z - \frac{B_\theta B_z}{4\pi}\right)\right] = 0$$

(4.9)

The z-component of the momentum equation becomes:

$$\frac{\partial(\rho v_z)}{\partial t} + \frac{1}{r}\frac{\partial}{\partial r}\left[r\left(\rho v_r v_z - \frac{B_r B_z}{4\pi}\right)\right] + \frac{1}{r}\frac{\partial}{\partial \theta}\left[\rho v_\theta v_z - \frac{B_\theta B_z}{4\pi}\right]$$
$$+ \frac{\partial}{\partial z}\left[\rho v_z^2 - \frac{B_z^2}{4\pi} + P + \frac{\mathbf{B}^2}{8\pi}\right] = 0$$

(4.10)

The continuity and induction equations take their standard forms. Since those equations are catalogued in several textbooks, they are not repeated here.

Several useful points are worth making:

1) In situations where we have $r \sim O(\Delta r)$ eqns. (4.6) and (4.7) can differ considerably from the Cartesian case.

2) In geometries with more complicated metrics the volume averaging does not have to be exact and can, instead, be done via numerical quadratures. This might be relevant in more complicated geometries such as Schwarzschild and Kerr geometries. This will be shown in a later paper.

3) The present sub-section clearly shows that divergence-free AMR-MHD can also be done in 3d geometries with more complicated metrics. The only extension needed by the Balsara (2001b) algorithm is the divergence-free reconstruction in those more complicated metrics, which has been provided here. This point is reinforced by the next sub-section.

4) A precise interpretation of the mathematics presented here would show that there are special points where the face-centered magnetic fields should be collocated. However, for second order schemes the precise location of the collocation points should not become an issue. Appendix A gives a discussion on collocation points in cylindrical geometry. These are also the locations where the zone-faces should be cut when carrying out adaptive mesh refinement. Appendix A also provides closed-form formulae for use in cylindrical AMR.

**IV.2) Spherical Geometry**

Consider a zone in spherical geometry extending from $(r_1, \theta_1, \phi_1)$ at its lower coordinate limits to $(r_2, \theta_2, \phi_2)$ at its upper coordinate limits. Here "r" is the radial coordinate, "$\theta$" is the azimuthal coordinate and "$\phi$" is the toroidal coordinate. The divergence-free condition is given by:

$$\frac{1}{r^2 \sin\theta} \left[ \frac{\partial \left(r^2 \sin\theta \, B_r\right)}{\partial r} + \frac{\partial \left(r \sin\theta \, B_\theta\right)}{\partial \theta} + \frac{\partial \left(r B_\phi\right)}{\partial \phi} \right] = 0 \qquad (4.11)$$

First introduce volumetric coordinates

$$x = \frac{r^3}{3} \; ; \; y = 1 - \cos\theta \; ; \; z = \phi \qquad (4.12)$$

Then set the origin at the zone's centroid so that we make the transcription $x \rightarrow x - (x_1 + x_2)/2$ and similarly for y and z. Eqn. (4.11) can be made to look like eqn. (3.8) by setting

$$B_x = r^2 B_r \; ; \; B_y = \frac{\sin\theta \, B_\theta}{r} \; ; \; B_z = \frac{B_\phi}{r \sin\theta} \qquad (4.13)$$

Let $(r_c, \theta_c, \phi_c)$ be the centroid of the zone being considered. The magnetic fields in eqn. (4.13) are collocated at the centroids of the faces. $\langle B_r \rangle_{vol-avg}$ is given by:

$$\begin{aligned}\langle B_r \rangle_{vol-avg} &= \frac{3 a_0 (r_2 - r_1)}{(r_2^3 - r_1^3)} + \frac{a_x}{(r_2^3 - r_1^3)} \left[ \frac{1}{4}(r_2^4 - r_1^4) - r_c^3 (r_2 - r_1) \right] \\ &+ \frac{a_{xx}}{(r_2^3 - r_1^3)} \left[ \frac{1}{21}(r_2^7 - r_1^7) - \frac{r_c^3}{6}(r_2^4 - r_1^4) + \frac{r_c^6}{3}(r_2 - r_1) \right]\end{aligned} \qquad (4.14)$$

$\langle B_\theta \rangle_{vol-avg}$ is given by:

$$V_0 \langle B_\theta \rangle_{vol-avg} = b_0 \left[ \frac{1}{4}(r_2^4 - r_1^4) \right][\theta_2 - \theta_1] + b_x \left[ \frac{1}{21}(r_2^7 - r_1^7) - \frac{r_c^3}{12}(r_2^4 - r_1^4) \right][\theta_2 - \theta_1]$$

$$+ b_y \left[ \frac{1}{4}(r_2^4 - r_1^4) \right] \left[ \cos\theta_c (\theta_2 - \theta_1) - (\sin\theta_2 - \sin\theta_1) \right]$$

$$+ b_{xy} \left[ \frac{1}{21}(r_2^7 - r_1^7) - \frac{r_c^3}{12}(r_2^4 - r_1^4) \right] \left[ \cos\theta_c (\theta_2 - \theta_1) - (\sin\theta_2 - \sin\theta_1) \right]$$

$$+ b_{yy} \left[ \frac{1}{4}(r_2^4 - r_1^4) \right] \left[ \frac{1}{4}(\sin 2\theta_2 - \sin 2\theta_1) - 2\cos\theta_c (\sin\theta_2 - \sin\theta_1) + \left( \cos^2\theta_c + \frac{1}{2} \right)(\theta_2 - \theta_1) \right]$$

where $V_0 = \frac{1}{3}(r_2^3 - r_1^3)(\cos\theta_1 - \cos\theta_2)$

(4.15)

$\langle B_\phi \rangle_{vol-avg}$ is given by:

$$V_0 \langle B_\phi \rangle_{vol-avg} = \left( c_0 + c_{zz} \frac{\Delta\phi^2}{12} \right) \left[ \frac{1}{4}(r_2^4 - r_1^4) \right] \left[ -\frac{1}{4}(\sin 2\theta_2 - \sin 2\theta_1) + \frac{1}{2}(\theta_2 - \theta_1) \right]$$

$$+ c_x \left[ \frac{1}{21}(r_2^7 - r_1^7) - \frac{r_c^3}{12}(r_2^4 - r_1^4) \right] \left[ -\frac{1}{4}(\sin 2\theta_2 - \sin 2\theta_1) + \frac{1}{2}(\theta_2 - \theta_1) \right]$$

$$+ c_y \left[ \frac{1}{4}(r_2^4 - r_1^4) \right] \left[ -\frac{1}{3}(\sin^3\theta_2 - \sin^3\theta_1) - \frac{\cos\theta_c}{4}(\sin 2\theta_2 - \sin 2\theta_1) + \frac{\cos\theta_c}{2}(\theta_2 - \theta_1) \right]$$

where $V_0 = \frac{1}{3}(r_2^3 - r_1^3)(\cos\theta_1 - \cos\theta_2)$

(4.16)

For the sake of completeness we also write out the r, θ and φ-components of the momentum equations. As before, we wish to show that higher order Godunov schemes can be written in a format that fully conserves angular momentum. The r-component of the momentum equation becomes:

$$\frac{\partial(\rho v_r)}{\partial t} + \frac{1}{r^2} \frac{\partial}{\partial r}\left[ r^2 \left( \rho v_r^2 - \frac{B_r^2}{4\pi} \right) \right] + \frac{1}{r \sin\theta} \frac{\partial}{\partial \theta}\left[ \sin\theta \left( \rho v_r v_\theta - \frac{B_r B_\theta}{4\pi} \right) \right]$$

$$+ \frac{1}{r \sin\theta} \frac{\partial}{\partial \phi}\left[ \rho v_r v_\phi - \frac{B_r B_\phi}{4\pi} \right] + \frac{\partial}{\partial r}\left[ P + \frac{\mathbf{B}^2}{8\pi} \right] = \frac{\rho (v_\theta^2 + v_\phi^2)}{r} - \frac{(B_\theta^2 + B_\phi^2)}{4\pi r}$$

(4.17)

The θ-component of the momentum equation becomes:

$$\frac{\partial(\rho v_\theta)}{\partial t} + \frac{1}{r^2} \frac{\partial}{\partial r}\left[r^2\left(\rho v_r v_\theta - \frac{B_r B_\theta}{4\pi}\right)\right] + \frac{1}{r \sin\theta} \frac{\partial}{\partial \theta}\left[\sin\theta\left(\rho v_\theta^2 - \frac{B_\theta^2}{4\pi}\right)\right]$$

$$+ \frac{1}{r \sin\theta} \frac{\partial}{\partial \phi}\left[\rho v_\theta v_\phi - \frac{B_\theta B_\phi}{4\pi}\right] + \frac{1}{r} \frac{\partial}{\partial \theta}\left[P + \frac{\mathbf{B}^2}{8\pi}\right] =$$

$$- \frac{\rho\left(v_r v_\theta - v_\phi^2 \cot\theta\right)}{r} + \frac{\left(B_r B_\theta - B_\phi^2 \cot\theta\right)}{4\pi r}$$

(4.18)

The φ-component of the momentum equation, when written out in an angular momentum conserving form, becomes:

$$\frac{\partial(r \sin\theta \, \rho v_\phi)}{\partial t} + \frac{1}{r^2} \frac{\partial}{\partial r}\left[r^3 \sin\theta\left(\rho v_r v_\phi - \frac{B_r B_\phi}{4\pi}\right)\right] + \frac{1}{r \sin\theta} \frac{\partial}{\partial \theta}\left[r \sin^2\theta\left(\rho v_\theta v_\phi - \frac{B_\theta B_\phi}{4\pi}\right)\right]$$

$$+ \frac{1}{r \sin\theta} \frac{\partial}{\partial \phi}\left[r \sin\theta\left(\rho v_\phi^2 - \frac{B_\phi^2}{4\pi} + P + \frac{\mathbf{B}^2}{8\pi}\right)\right] = 0$$

(4.19)

The continuity and induction equations take their standard forms and are not repeated here. Appendix B gives a discussion on collocation points in spherical geometry. These are also the locations where the zone-faces should be cut when carrying out adaptive mesh refinement. Appendix B also provides closed-form formulae for use in spherical AMR.

### IV.3) Tetrahedral Meshes

The divergence-free reconstruction of the magnetic field on tetrahedral meshes is even simpler than the reconstruction on logically rectilinear meshes. The reconstructed magnetic field in the interior of the tetrahedron can be written as:

$$B_x(x, y, z) = a_0 + a_x x + a_y y + a_z z \tag{4.20}$$

$$B_y(x, y, z) = b_0 + b_x x + b_y y + b_z z \tag{4.21}$$

$$B_z(x, y, z) = c_0 + c_x x + c_y y + c_z z \tag{4.22}$$

The divergence-free constraint gives:

$$a_x + b_y + c_z = 0 \tag{4.23}$$

thereby showing that eqns. (4.20) to (4.22) have eleven independent coefficients. Specifying the normal component of the magnetic field on each face of the tetrahedron as well as its linear variation in the two transverse directions that lie within that face constitutes the specification of three pieces of information on each of the four faces of the tetrahedron, see Fig. 2a. However, not all pieces of information are independent since the normal components of the magnetic field at the faces satisfy an integral version of the divergence-free condition. Writing $A_1$ as the area of face 1 and assuming that the normals to the faces of the tetrahedron being considered point outward, the divergence-free condition can be written as:

$$A_1 \, B_{\eta_1} + A_2 \, B_{\eta_2} + A_3 \, B_{\eta_3} + A_4 \, B_{\eta_4} = 0 \tag{4.24}$$

We, therefore, see that exactly eleven independent pieces of independent information are specified at the faces of the tetrahedron which is consistent with the eleven independent coefficients that are used to reconstruct the magnetic field in the interior of the tetrahedron.

The most efficient way of obtaining the coefficients of eqns. (4.20) to (4.22) is explained in pointwise form below:

1) Consider the given tetrahedron shown on the left side of Fig. 3. First realize that for each face of the tetrahedron, the magnetic field component (along the unit normal to that face) and its linear variation within that face are given to us. Thus for any face we can obtain the value of its magnetic field component at each of that face's vertices. Taking face 1 in Fig. 3 as an example, we know $B_{\eta_1} = \vec{B} \bullet \hat{\eta}_1$ at the centroid of face 1 as well as its linear variation along $\hat{\xi}_1$ and $\hat{\phi}_1$. Knowledge of the linear variation of $B_{\eta_1}$ in face 1 allows us to obtain $\vec{B}_2 \bullet \hat{\eta}_1$, $\vec{B}_3 \bullet \hat{\eta}_1$ and $\vec{B}_4 \bullet \hat{\eta}_1$ at the vertices 2, 3 and 4 respectively.

2) Specification of the magnetic field's components along the normals of the three faces that join to a vertex allows us to obtain the Cartesian components of the three dimensional magnetic field vector at each vertex of the tetrahedron. Taking vertex 2 as an example, the specification of $\vec{B}_2 \bullet \hat{\eta}_1$, $\vec{B}_2 \bullet \hat{\eta}_3$ and $\vec{B}_2 \bullet \hat{\eta}_4$ allows us to obtain $\vec{B}_2$ at vertex 2. Do this for each of the vertices of the tetrahedron. This is shown on the left side of Fig. 3 where the magnetic field vectors at the vertices are shown as thick dashed vectors.

3) Then map the given tetrahedron to the reference tetrahedron using a linear transformation. The reference tetrahedron is shown to the right of Fig. 3. Notice, as shown in Fig. 3, that the *same* magnetic field vector that was evaluated at each vertex of the given tetrahedron is now assigned to the corresponding vertex of the reference tetrahedron.

4) Specify three piecewise linear polynomials in the reference tetrahedron. (Do not confuse these polynomials with eqns. (4.20) to (4.22).) Find the coefficients of the three piecewise linear polynomials in the reference tetrahedron so that they match up to the three Cartesian components of the magnetic field vector at each of the four vertices of the reference tetrahedron. In the reference tetrahedron the algebra for obtaining the coefficients is very simple and is explicitly given in Appendix C. ( It should be pointed out that the condition in eqn. (4.23) does not hold for the coefficients of the three piecewise linear polynomials in the reference tetrahedron.)

5) Transform the three piecewise linear polynomials that were found in the reference tetrahedron back to the given tetrahedron using the inverse linear transformation. This yields the polynomials specified in eqns. (4.20) to (4.22) thus completing the reconstruction procedure.

Several points are worth making:
1) The analogy between the present sub-section and sub-section III.1 is very strong. The reconstruction procedure described in this subsection resolves issues # 1 to 3 in Section I such as they pertain to tetrahedral meshes. The volume-averaged magnetic field vector within a tetrahedron is given by evaluating eqns. (4.20) to (4.22) at the centroid of the tetrahedron. As in sub-section III.1 this yields a unique volume-averaged magnetic field vector which allows us to obtain a zone-centered pressure.
2) The reconstruction described in this sub-section can be extended to third and higher orders. Its utilization in TVD, WENO and RKDG schemes for MHD on unstructured meshes will be detailed in subsequent papers.
3) In this sub-section we have shown that divergence-free reconstruction can also be done on tetrahedral meshes. It follows that divergence-free prolongation of fields can be done from a tetrahedral mesh to a more enriched (adaptively refined) tetrahedral mesh. As a result, solution adaptive techniques for MHD can also be designed on tetrahedral meshes and the details will be presented in a later paper.
4) Cut cell approaches have also been used for adaptively refining at boundaries, see Aftosmis, Berger and Melton (1997, 1998). There are two ways of dealing with cells that are partially cut by a physical boundary, both of which have been described in Coirier and Powell (1993). One strategy consists of treating the cut parts as tetrahedra and sub-cycling them in time. Such a strategy would immediately benefit from the ideas presented here. The other strategy consists of cell-merging and can be treated by a slight extension of the ideas in Balsara (2001b). As a result cut cell formulations for AMR-MHD also become feasible because of the ideas presented in Balsara (2001b) and this paper.

## V) Pointwise Description of the Divergence-Free MHD Scheme

In this section we give a pointwise description of a single stage in the multi-stage schemes described by eqns. (2.5) or (2.6) in Section II. The procedure goes as follows:
1) Use the reconstruction described in Sub-section III.2 (or III.5 if genuinely multidimensional limiting is to be used) to obtain the slopes in the face-centered magnetic field components. Use the results from Sub-section III.1 to obtain the volume-averaged magnetic fields, see eqn. (3.17). If one is dealing with a non-Cartesian geometry then use the results from Section IV to obtain the volume-averaged magnetic fields.
2) This point splits into three different choices. The choice described in 2a obtains when the limiting is applied to the primitive variables in order to obtain a very inexpensive scheme. The choice described in 2b obtains when the best limiting is to be done on the characteristic variables. It yields a more expensive scheme. The choice 2c also describes a form of characteristic limiting but uses averaging to obtain the slopes of the magnetic field at the zone boundaries. It is less expensive than choice 2b. The three choices are as follows:

2.a) (Primitive limiting for: Fast TVD limiting or genuinely multidimensional limiting) Use the volume-averaged magnetic fields from point 1 above to obtain the pressure. Limit the slopes of the zone-centered variables (density, pressure and velocities) using the results in sub-section III.2 or III.5.

2.b) (Characteristic limiting) Use the volume-averaged magnetic fields from point 1) above to obtain the pressure. Limit the the face-centered magnetic fields using the results from Sub-sections III.3 or III.4. As a result, the slopes of the face-centered magnetic fields have been changed slightly from their values in point 1) above. Re-evaluate the volume-averaged magnetic fields and the zone-centered pressure. ( As before, if one is dealing with a non-Cartesian geometry then use the results from Section IV to obtain the volume-averaged magnetic fields.) Now limit the zone-centered variables using the appropriate characteristic limiting described in Sub-sections III.3 or III.4.

2.c) (Inexpensive characteristic limiting) Use the volume-averaged magnetic fields from point 1) above to obtain the pressure. Then limit the zone-centered variables using the appropriate characteristic limiting described in Sub-sections III.3 or III.4. This also yields limited slopes for the zone-centered magnetic field. To obtain the limited slopes of the transverse magnetic field at each face, average the limited slopes for the zone-centered magnetic field on either side of that face. This reduces the number of evaluations of the characteristic variables making 2.c much less expensive than 2.b.

3) For each facial quadrature point on each face we now need to solve the Riemann problem. As a result, we need to obtain the left and right states for the Riemann problem at those quadrature points. Use the zone-centered variables with their reconstructed slopes and the face-centered magnetic fields with their reconstructed polynomials to obtain the desired left and right states for the Riemann problem. Once the left and right states are obtained, solve the Riemann problem.

4) Use eqns. (2.10) to (2.12) to obtain the electric field components at the edge-centered quadrature points.

5) Use eqns. (2.7) to (2.9) to update the face-centered magnetic fields and eqn. (2.13) to update the conserved variables.

6) The fractional time step is done so return to the beginning of the next fractional time step.

**VI) Accuracy Analysis**

Several tests can be designed to demonstrate the accuracy of MHD schemes in one dimension. It is also possible to make multidimensional versions of such tests by running such tests in an oblique direction on a two-dimensional mesh. Running one-dimensional problems in an oblique direction can sometimes mask the true error by introducing a fictitious cancellation of the leading error terms. We seek to avoid that here. In this section we present a genuinely two-dimensional MHD problem. In Balsara and Shu (2000) we presented a genuinely two-dimensional Euler problem associated with a fluid vortex that was made to propagate at $45^0$ to the computational mesh. In this paper we extend the test problem to MHD. It is especially good for accuracy testing because it consists of a smoothly-varying and dynamically stable configuration that carries out non-trivial motion in the computational domain. The problem is set up on a two-dimensional domain given by [-5,5]X[-5,5]. The domain is periodic in both directions. An unperturbed magnetohydrodynamic flow with ($\rho$, P, $v_x$, $v_y$, $B_x$, $B_y$) = (1, 1, 1, 1, 0, 0) is initialized on the computational domain. The ratio of specific heats is given by $\gamma = 5/3$. The vortex is initialized at the center of the computational domain by way of fluctuations in the velocity and magnetic fields given by

$$\left( \delta v_x, \delta v_y \right) = \frac{\kappa}{2\pi} e^{0.5(1-r^2)} \left( -y, x \right) \tag{6.1}$$

$$\left( \delta B_x, \delta B_y \right) = \frac{\mu}{2\pi} e^{0.5(1-r^2)} \left( -y, x \right) \tag{6.2}$$

The magnetic vector potential in the z-direction associated with the magnetic field in eqn. (6.2) is given by

$$\delta A_z = \frac{\mu}{2\pi} e^{0.5(1-r^2)} \tag{6.3}$$

The magnetic vector potential plays an important role in the divergence-free initialization of the magnetic field on the computational domain. The circular motion of the vortex produces a centrifugal force. The tension in the magnetic field lines provides a centripetal force. The magnetic pressure also contributes to the dynamical balance in addition to the gas pressure. The condition for dynamical balance is given by

$$\frac{\partial P}{\partial r} = \left[ \rho \left( \frac{\kappa}{2\pi} \right)^2 - \frac{1}{2\pi} \left( \frac{\mu}{2\pi} \right)^2 \right] r\, e^{(1-r^2)} + \frac{1}{4\pi} \left( \frac{\mu}{2\pi} \right)^2 r^3\, e^{(1-r^2)} \tag{6.4}$$

For the fluid case, Balsara and Shu (2000) provide an isentropic solution for the above equation. For the MHD case it is simplest to set the density to unity and solve the above equation for the pressure. The fluctuation in the pressure is then given by

$$\delta P = \frac{1}{8\pi} \left( \frac{\mu}{2\pi} \right)^2 \left( 1 - r^2 \right) e^{(1-r^2)} - \frac{1}{2} \left( \frac{\kappa}{2\pi} \right)^2 e^{(1-r^2)} \tag{6.5}$$

As a result all aspects of the flow field are available in analytical form for all time which makes this problem very useful for accuracy analysis. The vortex can be set up with any strength because it is an exact solution of the MHD equations. It is worth pointing out that this test problem is easily extended to three dimensions by having a non-zero value for the z-component of the magnetic field. The simplest extension consists of giving the magnetic field a constant pitch angle with respect to the z-axis.

The MHD vortex problem was set up with $\kappa = 1$ and $\mu = \sqrt{4\pi}$ which makes the Alfven speed of the vortex equal to its rotational speed. The temporally third order accurate time-stepping scheme from Section II was used for all the tests in this section. This ensures that the present accuracy analysis will reveal the spatial accuracy of the algorithm without any influence from the time-stepping scheme. The use of temporally second order accurate time-stepping does not cause a significant change in the results presented here. Meshes with 50X50, 100X100, 200X200 and 400X400 zones were used, with 800X800 zone meshes becoming necessary only in some occasions. A Roe-type linearized Riemann solver was also used for the first two tests in this section while the remaining tests used an HLLE-type Riemann solver. The problem was run for a time of 10 units by which time the vortex returns to its initial location. As explained by Balsara and Shu (2000) this test problem is very important in calibrating codes that are used in turbulence studies for two important reasons. First, the dominant structures in turbulence are vortex tubes and the present test problem calibrates the code with flows that are morphologically closest to the ones that develop in actual turbulence simulations. Second, schemes that are less than second order accurate are known to be useless for turbulence simulations and the present test problem helps winnow out undesirable schemes.

Table I shows the error measured in the $L_1$ norm and the order of accuracy for the momentum density vector and the magnetic field vector. The error in Table I measures the sum of the errors in each of the three components, as measured in the $L_1$ norm. The algorithm described in Section V was used with r=2 WENO limiting described in Sub-section III.3. Table II shows the convergence history when the slopes from the $r = 3$ WENO limiting described in Sub-section III.4 were used for *all* the characteristic fields in the problem. While this is not a good idea for problems involving very strong shocks, it is acceptable for the present problem. The Roe-type Riemann solver was used in both these tests. We see from Table I that the $r = 2$ WENO-based scheme does reach its design accuracy of being second order accurate. However, it reaches this accuracy only on very large meshes. Furthermore, that order of accuracy is reached from below. From Table II we see that the more refined slope reconstruction that is available in the $r = 3$ WENO-based scheme causes it to reach the design accuracy much faster and on smaller meshes. Based on Tables I and II it is fair to say that if the scheme used for Table I produces a desired accuracy then the scheme used for Table II will produce the same accuracy with only a little more than half as many zones in each direction. Both the schemes, as designed here, are limited to second order accuracy because we do not use the full WENO interpolation. Instead, we use just the average slope from the higher quality interpolation in a finite volume scheme. Tables I and II taken together do, however, make a compelling case for improving the quality of the interpolation. We also mention that both the 200X200 and 400X400 zone simulations with $r = 3$ WENO limiting show no sign of mesh imprinting. On the other hand, all the simulations that were done with $r = 2$ WENO limiting, except for the 400X400 zone simulation, showed some evidence of mesh imprinting. We conclude, therefore, that the improved interpolation also makes for a more visually pleasing result.

In Section II we noted that the use of four Riemann solvers per zone face did make the scheme expensive. It was suggested there that replacing the Roe type Riemann solver by an HLLE Riemann solver provides a way out. The HLLE Riemann solver also carries the bonus that it is free of the carbuncle instability, see Pandolfini and D'Ambrosio (2001) and references therein. Mirroring an observation by Cockburn and Shu (1998), we had hypothesized in Section II that when a good enough interpolation strategy is used the accuracy of the scheme does not depend on the quality of the Riemann solver. In this paragraph we put that hypothesis to the test. Table III shows the error associated with the present algorithm using r=2 WENO limiting and an HLLE Riemann solver (instead of a Roe type Riemann solver). We see that the errors in Table III are entirely comparable to those in Table I. This is a quantitative demonstration that the reconstruction strategies designed in this paper are sophisticated enough that the accuracy of the resulting scheme is approximately independent of the type of Riemann solver used. We have carried out numerous one dimensional tests with either Riemann solver and found that the Roe type Riemann solver offers a noticeable advantage in resolution over the HLLE Riemann solver when a mid-grade interpolation strategy such as the r=2 WENO is used. However, when a high-quality interpolation such as the blended-WENO is used the difference is diminished. Furthermore, based on the many multidimensional tests we have carried out, we have also found that the advantage of the Roe-type Riemann solver is diminished in multiple dimensions. It is also worth pointing out that the scheme described in this paragraph has half the computational complexity as the scheme used for Table I. The scheme described in this paragraph has the same computational complexity as the dimensionally swept algorithm described in Balsara and Spicer (1999b).

Table IV shows the error associated with the blended-WENO interpolation when used along with the HLLE Riemann solver. We see that the convergence to the design accuracy is not as good as the r=3 WENO results shown in Table II. However, the scheme does reach its designed accuracy and has the dual benefits of having high resolution and being robust enough to handle strong shock problems.

In the previous paragraphs we have examined the second order convergence properties of schemes that rely on reconstruction of the characteristic variables. The construction of eigenvectors and the projection of variables on to the space of those eigenvectors makes such schemes computationally costly. In sub-sections III.2 and III.5 we showed that it is possible to limit the primitive variables instead of the characteristic variables, yielding more economical schemes. It is worth asking whether such schemes provide a real advantage in terms of accuracy gained per unit of CPU usage? When all schemes are run with an HLLE Riemann solver, the schemes in sub-sections III.2 and III.5 update twice as many zones per second as the r = 3 WENO-based scheme in Sub-section III.4. Table V shows the convergence of the fast TVD scheme from sub-section III.2 when used along with an HLLE Riemann solver. We see clearly that the scheme only barely approaches its design accuracy on an 800X800 mesh. Comparing Table V to Table II we see that the fast TVD scheme on a specified mesh gives the same accuracy as the r = 3 WENO-based interpolation strategy on a mesh that has half as many zones in each direction. Thus it pays to use a costlier scheme on a smaller mesh in this instance. Table VI shows the convergence of the genuinely multidimensional scheme from sub-section III.5 with $\psi =1.0$ used for all the MHD variables. We see that the scheme quickly reaches its design accuracy on very small meshes, just like the r = 3 WENO-based scheme. Comparing Table VI to Table II we see that we get the same accuracy from either scheme on meshes of the same size. However, the scheme in Table VI runs twice as fast. The genuinely multidimensional scheme which uses $\psi = 1.0$ for all the MHD variables also shows no evidence of mesh imprinting on a 200X200 zone mesh. Table VII shows the convergence of the genuinely multidimensional scheme from sub-section III.5 with $\psi = 1.0$ used for the density and magnetic field while $\psi = 0.5$ was used for all the pressures and velocities. Comparing Table VII to Table V we see that we have a second order scheme whose accuracy and convergence are comparable to the fast TVD scheme from sub-sections III.2. The genuinely multidimensional scheme shown in Table VII is quite robust even for problems with quite strong shocks though it does not have superlative second order convergence properties. The genuinely multidimensional scheme shown in Table VI has superlative second order convergence though it loses robustness for the most stringent problems. The last point in sub-section III.5 shows us a way to get the best of both worlds by detecting strong shocks and switching to the more robust scheme in their vicinity.

In summary we conclude that when accuracy and robustness are taken together schemes that rely on characteristic interpolation have an edge over schemes that interpolate over the primitive variables. Computationally inexpensive Riemann solvers, such as the HLLE limiter, do not degrade the accuracy of these schemes. We also point out in passing that dimensionally swept MHD schemes do not display good accuracy on multidimensional problems even if they have been optimized for superlative one-dimensional performance. The analogous result for the Euler equations has been shown in Balsara and Shu (2000).

## VII) Test Problems

In this section we present several stringent test problems. Since the reader's attention is likely to be focused on the newer fast TVD, r=3 WENO and genuinely multidimensional schemes, we will highlight those schemes in this section.

## VII.a) Numerical Dissipation and Long-Term Decay of Alfven Waves

In several problems, such as the turbulence problems that are of interest in astrophysics, one is interested in the evolution of waves. It is easy to show that the MHD wave families can propagate with minimal loss at $45^0$ to the mesh. As shown by Balsara and Spicer (1999b) it is a lot harder to achieve good propagation of waves that are required to propagate at a small angle to one of the mesh lines. Such issues are very relevant in turbulence studies because if some wave families propagate with high dissipation in certain directions to the mesh then the mode mixing, the formation of a turbulent spectrum and the decay of turbulence can all be negatively impacted.

Building on Balsara and Spicer (1999b) we construct a test problem which examines the dissipation of torsional Alfven waves that are made to propagate at a small angle to the mesh. We use a 120X120 zone mesh with uniform zones. The computational domain spans [-r/2, r/2]X[-r/2, r/2] in the xy-plane with r = 6. A Courant number of 0.4 was used in all the calculations in this sub-section. A uniform density, $\rho_0 = 1$, and pressure, $P_0 = 1$, are initialized on the mesh. The unperturbed velocity, $v_0 = 0$, and the unperturbed magnetic field $B_0 = 1$. The amplitude of the Alfven wave fluctuation is parametrized in terms of the velocity fluctuation which has a value of $\varepsilon = 0.2$ in this problem. Choosing different values for these parameters yields different test problems, including test problems with standing Alfven waves. The Alfven wave is made to propagate at an angle of $\tan^{-1}(1/r) = \tan^{-1}(1/6) = 9.462^0$ with respect to the y-axis. The direction of wave propagation is along the unit vector:

$$\hat{n} = n_x \hat{i} + n_y \hat{j} = \frac{1}{\sqrt{r^2+1}} \hat{i} + \frac{1}{\sqrt{r^2+1}} \hat{j} \tag{7.1}$$

The phase of the wave is taken to be

$$\phi = \frac{2\pi}{n_y}(n_x x + n_y y - V_A t) \text{ where } V_A = \frac{B_0}{\sqrt{4\pi\rho_0}} \tag{7.2}$$

The velocity is given by

$$\vec{v} = (v_0 n_x - \varepsilon n_y \cos\phi)\hat{i} + (v_0 n_y + \varepsilon n_x \cos\phi)\hat{j} + \varepsilon \sin\phi \hat{k} \tag{7.3}$$

The magnetic field is given by

$$\vec{B} = (B_0 n_x + \varepsilon n_y \sqrt{4\pi\rho_0} \cos\phi)\hat{i} + (B_0 n_y - \varepsilon n_x \sqrt{4\pi\rho_0} \cos\phi)\hat{j} \\ - \varepsilon \sqrt{4\pi\rho_0} \sin\phi \hat{k} \tag{7.4}$$

The corresponding magnetic vector potential is given by

$$\vec{A} = -\frac{\varepsilon\sqrt{4\pi\rho_0}}{2\pi}\cos\phi \hat{i} + (-B_0 n_y x + B_0 n_x y + \frac{\varepsilon n_y \sqrt{4\pi\rho_0}}{2\pi}\sin\phi)\hat{k} \tag{7.5}$$

and may be used for initializing the magnetic field. The problem is run to a time of 129 time units by which time the Alfven waves have crossed the computational domain several times. The decay of the maximum values of the z-component of the velocity and magnetic field provides a good measure of the dissipation in the numerical scheme. As a result, we keep track of those quantities at every time step and plot them out on log-linear plots. The rms values of the z-component of the velocity and magnetic field decay in a fashion that is entirely similar to the maximal values of the same quantities. As a result, they are not shown here. As shown by Kim et al (1999) such plots provide a good qualitative understanding of the numerical viscosities and resistivities in the schemes being considered.

Several schemes were tried using the above test problem. The schemes we tested were:

a) The dimensionally swept scheme from Balsara and Spicer (1999b) which used TVD interpolation of the characteristic variables, see Balsara (1998b), and a linearized Riemann solver. This scheme used eqns. (7) to (10) from Balsara and Spicer (1999b) to obtain the electric field instead of eqns. (2.10) to (2.12) from this paper. The former involve spatial averaging while the latter do not.
b) The scheme that uses the same TVD interpolation and linearized Riemann solver as in point a) but uses those building blocks within the context of the unsplit algorithms developed here. This scheme, therefore, provides a fair comparison with the dimensionally swept scheme in point a).
c) The scheme that uses the r=2 WENO interpolation described in sub-section III.3 along with a linearized Riemann solver.
d) The scheme that uses the r=3 WENO interpolation for all the characteristic fields as described in sub-section III.4 along with a linearized Riemann solver.
e) The same as d) but with an HLLE-type Riemann solver.
f) The scheme that uses the blended WENO interpolation described in sub-section III.4 along with a linearized Riemann solver.
g) The same as f) but with an HLLE-type Riemann solver.

Fig. 4.a shows the evolution of the maximal value of the z-velocity for all the schemes as a function of time. Fig. 4.b shows the evolution of the maximal value of the z-component of the magnetic field for the same schemes. We see clearly that the dimensionally swept scheme is the most dissipative. The scheme that uses the same interpolation and Riemann solver within the context of the unsplit algorithms developed here is a good bit less dissipative. The scheme that uses the r=2 WENO interpolation is less dissipative than the previous two schemes. The schemes that use the r=3 WENO interpolation are substantially less dissipative than the schemes that use the poorer TVD or r=2 WENO interpolation. This statement holds true regardless of the Riemann solver. Thus the results of using the r=3 WENO scheme with an HLLE-type Riemann solver are almost as good as the results with a Roe-type linearized Riemann solver. We see, therefore, that as the quality of the interpolation improves the Riemann solver becomes less important to the overall quality of the result. Thus a modest increment in the computational cost associated with using a better quality interpolation is amply offset by the substantial reduction in computational cost associated with using an extremely inexpensive Riemann solver. Lastly, we show that the blended WENO interpolation yields results that are almost as good as the r=3 WENO interpolation. The r=3 WENO is to be preferred for problems without very strong shocks. The blended WENO scheme is a little more dissipative but can handle problems with extremely strong shocks. It should also be pointed out that the directionally split algorithms are unsuitable for AMR applications. In such applications the fluxes and electric fields that are needed for flux correction or electric field correction steps should contain all the contributions from all the waves that arrive at the boundaries or the edges, see Balsara (2001). The fluxes and fields that are obtained from directionally split algorithms do not have this attribute while the unsplit algorithms designed here do have this attribute. The algorithms presented in this paper, therefore, have the twin benefits of being less dissipative than the dimensionally split algorithms and also being more suitable for use in AMR applications. The unsplit algorithms that use the HLLE-type Riemann solver have a computational cost that is competitive with dimensionally split algorithms.

The present results have significance for simulations of astrophysical turbulence. Lee et al (2003) used a dimensionally split TVD algorithm to conclude that one requires a minimum of 1024 zoned in each direction to carry out a turbulence simulation with a clearly demarcated inertial range that is distinct from the dissipation range. This makes such calculations excessively costly. However, notice that a scheme which reduces the dissipation on smaller scales will also make it possible to have a smaller dissipation range. This makes it possible to obtain the same length of inertial range using a much smaller computational domain.

**VII.b) The Rotor Problem**

This test problem was first presented in Balsara and Spicer (1999b). We present a version of the rotor test problem here. The problem is set up on a two dimensional unit square having 200X200 zones. It consists of having a dense, rapidly spinning cylinder, in the center of an initially stationary, light ambient fluid. The two fluids are threaded by a magnetic field that is uniform to begin with and has a value of 2.5 units. The pressure is set to unity in both fluids. The ambient fluid has unit density. The rotor has a constant density of 10 units out to a radius of 0.1. Between a radius of 0.1 and 0.13 a linear taper is applied to the density so that the density in the cylinder linearly joins the density in the ambient. The taper is, therefore, spread out over six computational zones and it is a good idea to keep that number fixed as the resolution is increased or decreased. The ambient fluid is initially static. The rotor rotates with a uniform angular velocity that extends out to a radius of 0.1. At a radius of 0.1 it has a toroidal velocity of one unit. Between a radius of 0.1 and 0.13 the rotor's toroidal velocity drops linearly in the radial velocity from one unit to zero so that at a radius of 0.13 the velocity blends in with that of the ambient fluid. The ratio of specific heats is taken to be 5/3. The RIEMANN framework for computational astrophysics was applied to this problem. Figs 5a, 5b, 5c and 5d show the density, pressure, Mach number and the magnitude of the magnetic field respectively at a time of 0.29. The second order accurate Runge-Kutta time-stepping scheme described in Section II was used with a Courant number of 0.4. The spatial interpolation was carried out by using the genuinely multidimensional limiting described in sub-section III.5. The volume-averaged magnetic fields were obtained by using eqn. (3.17) from Sub-section III.1. An HLLE Riemann solver was used. Divergence-free reconstruction of the magnetic field, as described in Sub-section III.1, was used to obtain the left and right states of the magnetic field for the Riemann solver. Balsara and Spicer (1999b) provide a detailed physical description of this problem. We see that the results presented in Fig. 5 are entirely consistent with the description in Balsara and Spicer (1999b) thereby showing that the genuinely multidimensional limiting that is presented in this paper is indeed very valuable for numerical MHD.

**VII.c) The Blast Problem**

This test problem was first presented in Balsara and Spicer (1999b). In this case the details of setting up the problem are exactly as described in Balsara and Spicer (1999b). The problem was run on a mesh of 200X200 zones. The RIEMANN framework for computational astrophysics was applied to this problem. Figs 6a, 6b, 6c and 6d show the logarithm (base 10) of the density, the logarithm (base 10) of the pressure, the magnitude of the velocity and the magnitude of the magnetic field respectively at a time of 0.01. The second order accurate Runge-Kutta time-stepping scheme described in Section II was used with a Courant number of 0.4. The spatial interpolation was carried out by using the fast TVD limiting described in Sub-section III.2. The volume-averaged magnetic fields were obtained by using eqn. (3.17) from Sub-section III.1. An HLLE Riemann solver was used. Divergence-free reconstruction of the magnetic field, as described in sub-section III.1, was used to obtain the left and right states of the magnetic field for the Riemann solver. Balsara and Spicer (1999b) provide a detailed physical description of this problem. We see that the results presented in Figs. 6a, 6b, 6c and 6d are consistent with that description. Notice that the plasma-$\beta$ in the ambient medium is 0.000251. The fastest wave structure in the problem consists of an almost spherical fast magnetosonic shock which propagates through this low-$\beta$ ambient plasma. From Fig. 6b we see that there are regions where the strong shock propagates obliquely to the mesh without causing any problems in maintaining the positivity of the pressure variable. This is a direct consequence of using the divergence-free reconstruction to obtain the volume-averaged magnetic fields at the zone centers. Figure 6e shows the logarithm (base 10) of the pressure variable when r=3 WENO slopes from Sub-section III.4 were kept within physical bounds by using the multidimensional limiter in Sub-section III.5. (The r=3 WENO limiting was applied to *all* the characteristic fields.) We see that the results are practically as good as those from the fast TVD scheme. Figure 6f shows the logarithm (base 10) of the pressure variable when the slopes from the genuinely multidimensional limiter were used by themselves. Here we do see a small unphysical drop in pressure in regions where the outer fast shock is propagating obliquely to the mesh which probably owes to the fact that the genuinely multidimensional limiter applies the minimal amount of limiting to the problem. The genuinely multidimensional limiter does, nevertheless, show itself to be a very worthy performer on this extremely stringent problem. Toth (2000) has reported that a very large and unphysical drop in pressure takes place immediately ahead of the shock, especially in regions where the outer fast shock is propagating obliquely to the mesh. Other practitioners have also reported similarly in private communications. Such an unphysical drop in pressure is not seen in Fig. 6 where the divergence-free reconstruction strategy was used to obtain the volume-averaged magnetic fields. We see, therefore, that using the divergence-free reconstruction from Section III.1 produces a substantial improvement in the simulation of low-$\beta$ plasmas. We also point out that both the fast TVD limiter and the genuinely multidimensional limiter when used along with an HLLE-type Riemann solver produce solutions with velocities and transverse magnetic fields that are symmetrical up to machine accuracy. This can be a very desirable feature in certain kinds of physical problems where we want symmetries to be preserved in the flow unless those symmetries are explicitly broken by an external perturbation.

It is worthwhile to mention the performance of several variant algorithms that can be constructed by using the basic building blocks presented in this paper. Since the present problem is a stringent test problem it is interesting to examine how the varying algorithms perform on this test problem. Our first algorithmic variant consists of solving just one Riemann problem at each face-center and using eqns. (7) to (10) from Balsara and Spicer (1999b) to obtain the electric field instead of eqns. (2.10) to (2.12) from this paper. On applying such a scheme to this problem with the same r=3 WENO limiting described above we find that the pressure is not evolved properly. We, therefore, conclude that the four Riemann solver steps that are needed in eqns. (2.10) to (2.12) from this paper do help produce a more robust scheme. A second algorithmic variant consists of using point 2c instead of point 2b in Section V to obtain the face-centered slopes of the transverse magnetic fields. When that is used with eqns. (2.10) to (2.12) from this paper we do obtain pressure fields that are comparable to Fig. 6e. We, therefore, conclude that reducing the computational cost by simplifying the evaluation of the face-centered magnetic field slopes is indeed a good option. Our third algorithmic variant consists of using point 2c along with eqns. (7) to (10) from Balsara and Spicer (1999b). This variant also has problems with the pressure evolution. The present study of variant algorithms suggests that for stringent problems it is better to reduce the cost by simplifying the slope evaluation at the face centers rather than to try and reduce the cost by reducing the number of Riemann problems that are evaluated per zone.

**VII.d) Disk-Magnetosphere Interaction Problem**

This test problem is meant to illustrate the performance of our scheme on stringent physical problems in spherical geometry. Spherical geometry is one of the complex metrical geometries treated in Section IV. It is a good test problem because it brings together all the ingredients that make MHD simulations difficult: 1) a complex geometry, 2) the presence of high velocity flows in low plasma-$\beta$ environments and 3) challenging boundary conditions. The problem deals with the development of an accretion disk as it interacts with the magnetic field around a neutron star. The interaction of the accretion disk with the magnetic field of the neutron star leads to accretion on to the pole-caps of the neutron star. It has also been anticipated that it produces outflows. These results have long been anticipated in theoretical astrophysics. We show that both outflows as well as pole-cap accretion can be reproduced via time-dependent simulations.

An accretion disk is set up around a neutron star. The radius of the neutron star is $R_* = 10$ Km and sets the inner radial boundary of the computational domain in the r-$\theta$ plane. The accretion disk extends out to a radius of 90 Km. The outer radial boundary of the computational domain is set at 90 Km. The azimuthal coordinate, $\theta$, extends from 0 to $\pi/2$. The problem is set up in 2.5-dimensional geometry, i.e. the toroidal direction suppressed but the toroidal velocity and its self-consistent evolution are not. The gravitational potential is given by :

$$\Phi(r) = -\frac{GM}{r - r_g} \qquad (7.6)$$

Here we take $G = 6.67 \times 10$ gm$^{-1}$ cm$^3$ sec$^{-1}$, $M = 1.4$ times the solar mass $= 2.8 \times 10^{33}$ gm and $r_g = 0$, i.e. the Newtonian limit. The specific angular momentum (i.e. the angular momentum per unit mass) of the accretion disk is parametrized as :

$$L = r \, v_\phi = L_0 \, r^a \quad \text{(Set } L_0 = \sqrt{G\,M} \text{ and } a = \frac{1}{2} \text{ for Keplerian flow)} \tag{7.7}$$

Here we choose to set up a Keplerian disk. The pressure in the accretion disk is related to its density by the following polytropic relation :

$$P = K \rho^{1+1/n} \tag{7.8}$$

Here we choose n = 3. The gas in this problem has a ratio of specific heats, $\gamma$, given by 4/3. The density at any point in the disk can be obtained by specifying the density $\rho_0$ = 474.24 gm/cm$^3$ and temperature $T_0 = 3 \times 10^8$ K ~ 30 keV in the mid-plane of the disk at a fixed radius $r_0$ = 30 Km . This choice also specifies K in eqn. (7.8). Dynamical balance in the disk along with the polytropic relation requires that the following relation be a constant in the disk :

$$(1 + n) \frac{P}{\rho} + \Phi(r) + \frac{1}{2(1-a)} L_0^2 \, r^{2a-2} = \Phi_0 \tag{7.9}$$

Equns. (7.8) and (7.9) enable us to specify the pressure and density anywhere in the disk. The disk is set up so that its upper surface is in pressure balance with the halo. The halo is initially static and has a constant temperature $T_h = 2000 \, T_0$ . The high temperature of the halo gas simply serves to make the mass of the halo very small so that the initial halo is dynamically unimportant, except for providing pressure balance at the boundary of the disk. A reference density $\rho_h$ = 0.11442008 gm/cm$^3$ then specifies the halo entirely because dynamical balance in the radial direction gives us the run of the density with radius as follows :

$$\rho = \rho_h \, \exp\left[ - \frac{\mu}{R \, T_h} \left( \Phi(r) - \Phi(r_0) \right) \right] \tag{7.10}$$

Here $\mu$ = 1.217 and R = 8.31$\times 10^7$ erg deg$^{-1}$ mole$^{-1}$ is the gas constant. At each zone in the computational domain the pressure from the disk and the pressure from the halo are evaluated. The zone is taken to be part of the disk if the disk's pressure is greater, otherwise it is set to be part of the halo. Such a model for building a disk and halo that are in dynamical balance seems to have been initially explored by Suchkov et al (1994) and a little later in its present form by Matsumoto et al (1996). The magnetic field is initially dipolar and is specified all over by specifying the field $B_* = 10^{10}$ G at the neutron star's pole-cap. The magnetic field can be generated by specifying the vector potential of a magnetic dipole as:

$$A_\phi = \frac{B_* \, R_*^3 \, \sin \theta}{2 \, r^2} \tag{7.11}$$

Using the magnetic vector potential to initialize the magnetic field is in fact the preferred method of initializing the dipolar magnetic field on the computational mesh in divergence-free fashion. This completes the specification of the interior of the computational domain. The problem was normalized by taking : a) 1 Km to be one code unit of length, b) $10^{-2}$ sec to be one code unit of time and c) $10^{17}$ gm to be one code unit of mass. Other normalizations are possible and the reader can pick them as s/he wishes.

The boundary conditions for the base of a magnetosphere are indeed quite complicated and we describe that next. The boundary conditions at the surface of the star are especially important because inappropriate boundary conditions can cause the magnetic flux through each zone face at $r = R_*$ to change with time. We want to permit the possibility that the magnetic field can be buried if a strong enough accretion flow develops. However, we do not want the foot points of the magnetic field on the star's surface to wander in time. This is equivalent to requiring that the magnetic flux through each zone face at $r = R_*$ remains constant with time. It can be achieved in each fractional time step by enforcing the boundary conditions at $r = R_*$ via the following three steps: 1) Set the density and pressure immediately outside of the boundary to the same values as are obtained inside the boundary. 2) The three components of the velocity are made to flip signs across the boundary. 3) The transverse components of the magnetic field remain continuous across the boundary. With this arrangement of the states across the boundary it is easy to show that the HLLE Riemann solver returns a zero value for the transverse components of the electric field, i.e. $F_7$ and $F_8$ are zero in eqns. (2.1) and (2.2). Our construction, therefore, ensures that *all* the fluxes returned by the Riemann solver are physically consistent with a radial magnetic field that is non-evolving at the inner boundary. $E_y$ and $E_z$ can then be kept zero at the $r = R_*$ boundary by replacing eqns. (2.11) and (2.12) at the boundary by:

$$E_{y,\ 1/2,j,k+1/2}\left(\mathbf{U}^n\right) = \frac{1}{2c}\left(F^+_{8,\ 1/2,j,k+1/2}\left(\mathbf{U}^n\right) + F^-_{8,\ 1/2,j,k+1/2}\left(\mathbf{U}^n\right)\right) \tag{7.12}$$

$$E_{z,\ 1/2,j+1/2,k}\left(\mathbf{U}^n\right) = \frac{1}{2c}\left(-F^+_{7,\ 1/2,j+1/2,k}\left(\mathbf{U}^n\right) - F^-_{7,\ 1/2,j+1/2,k}\left(\mathbf{U}^n\right)\right) \tag{7.13}$$

respectively. This ensures that the radial component of the magnetic field at $r = R_*$ does not change with time, see eqn. (2.7). One could always have zeroed out the electric fields in the surface of the star by fiat. But the present constructive strategy gives us a way of setting the ghost zones and the Godunov fluxes at the star's surface in a way that is totally physically consistent with the fact that we want the foot points of the magnetic field to stay anchored to the neutron star's surface.

The boundary conditions at the outer radial boundary do not play as critical a role as the inner boundary conditions. This is because the fluid flows of maximal interest to us arise from the interaction of the inner part of the accretion disk with the magnetosphere. However, an inappropriate choice of outer boundary conditions can trigger a runaway infall in either the outer halo or the outer disk regions. The portion of the outer radial boundary that abuts the halo should transparently permit any wind solution that develops to leave the mesh. For that reason, if the radial velocity at the halo's boundary is directed outwards, we let the flow leave the mesh without experiencing any fluid stresses. This is done by continuously extending the flow variables as well as the magnetic field variables past the outer boundary. However, if the radial velocity at the halo's boundary is directed inwards, then we pick a boundary treatment that ensures that there is zero mass flux outside the outer radial boundary while letting the transverse velocities and magnetic fields be continuous across the boundary. Zero mass flux is ensured by reflecting the radial velocity at the outer boundary while retaining the same values of the density and pressure. The portion of the outer radial boundary that abuts the disk should just bring in new fluid as the disk slowly sinks inward by its interaction with the magnetosphere. In nature that boundary condition is set by the rate at which accreting fluid is brought into the disk – a highly time-variable phenomenon. Hence that boundary condition cannot be set without knowledge of the physical conditions that lie outside the computational domain. We do this by applying the outer boundary condition that is used for the halo to the disk too. It represents one possible physical situation where there is no new matter being accreted from the outside to the disk. It is physically justified by the fact that the interaction of the inner disk with the magnetosphere takes place much faster than the evolution of the disk at the outer boundary.

The mesh is a logically rectangular tensor product mesh in spherical geometry with 300 ratioed zones in the radial direction and 100 ratioed zones in the azimuthal direction. The innermost radial zone has a radial size of 0.078644 Km and each successive zone outwards is larger than the previous one by a ratio of 1.00701. The zoning in the $\theta$-direction is such that the zones are smallest at the equator and increase in angular size as they approach either of the two poles. The smallest zone at the equator has $\Delta\theta = 0.0174767$ radians. The zones in the $\theta$-direction increase by a ratio of 1.03557 as they go further away from the equator. The second order accurate Runge-Kutta time-stepping scheme described in Section II was used with a Courant number of 0.4. For the spatial interpolation we used the r=2 WENO limiting described in sub-section III.3. The volume-averaged magnetic fields were obtained by using eqns. (4.14) to (4.16) from sub-section IV.2. An HLLE Riemann solver was used. Divergence-free reconstruction of the magnetic field in spherical geometry, as described in sub-section IV.2, was used to obtain the left and right states of the magnetic field for the Riemann solver. The flow was evolved in an angular momentum conserving fashion, as shown in sub-section IV.2.

Figure 7a shows the logarithm (base 10) of the density at a simulation time of 0.24 , which corresponds to $0.24 \times 10^{-2}$ sec of physical time. This time is also equal to about 1 orbital period of the disk at the fiducial radius, $r_0$ . Figures 7b and 7c show the logarithm of the velocity and magnetic field respectively in the azimuthal plane at the same time. Vectors, with their lengths scaled logarithmically, are also shown to indicate the direction of the field. Since the dipolar magnetic field threads the disk and halo material the disk and its initial halo are magnetically coupled. Since the disk and halo have very different rotation speeds, some of the angular momentum in the upper, more tenuous, disk material is readily transferred to the halo, causing the upper layers of the outer disk to fall radially towards the star. The core of the disk, where most of its mass is concentrated, retains its Keplerian motion. The infalling material does cause the magnetic field to kink at the upper surface of the disk. The kinking of the field is maximal at the inner edge of the disk and an outflow develops from the inner part of the disk's upper surface. The infalling material cannot, however, keep falling radially all the way to the surface of the star because the magnetic field near the star is strong enough to arrest the radial infall. Thus the infalling matter begins to travel along the magnetic field lines in the region nearer to the star. It is this material that will eventually form the star's magnetosphere. The velocity field clearly shows the flow developing along dipolar field lines in the magnetospheric region. The accretion stream splatters when it hits the neutron star's surface, as can be seen from the velocity plot (i.e. the velocity at the pole-cap is close to zero). Under suitable conditions the splatter spot could be the site for rp-process nucleosynthesis, a process not included here. In this problem the accretion stream impacts the splatter spot with only 20% of the Keplerian velocity. However, as the strength of the magnetic field is increased we anticipate that the magnetosphere will grow in size. As a result, the accretion stream will experience larger amounts of radial motion along the inner field lines resulting in larger impact velocities with the surface of the star. These larger velocities would further enhance the rp-process yields.

Figure 7d shows the logarithm (base 10) of the density at a simulation time of 0.48 , which corresponds to $0.48 \times 10^{-2}$ sec of physical time. This time is also equal to about 2 orbital periods of the disk at the fiducial radius, $r_0$ . Figures 7e and 7f show the logarithm of the velocity and magnetic field respectively in the azimuthal plane at the same time. We see that the outflow has become larger and more prominent as expected. However, we also see a very interesting phenomenon in the velocity plot. The pole-cap accretion flow is not present at this instant of time. Watching a movie of the flow as it develops shows that the pole-cap accretion is very episodic. Thus in Figs 7d,e and f we show the flow at an epoch when the pole-cap accretion is in the process of re-establishing itself but has not fully re-established itself. We understand, therefore, that accretion onto pole-caps of magnetized neutron stars can be a highly time-dependent process. We also see that pole-cap accretion can establish itself simultaneously with an outflow. The highly time-dependent aspect of the accretion causes the magnetosphere (that is in the process of forming) to be highly turbulent. Notice from Fig. 7f that the magnetic field has acquired a quadrupolar morphology. In this simulation we have not included physical resistivity. However, the turbulence could cause the region where the disk rubs against the magnetosphere to develop anomalously high diffusivities, thereby further enhancing the pole-cap accretion. Such processes have not been included in this round of simulations and will be explored in subsequent work.

## VIII) Conclusions

1) We have shown that it is possible to arrive at well-designed schemes for numerical MHD.

2) The schemes designed here overcome the four key unresolved issues that were spelled out in the Introduction.

3) We have shown that both second and third order accurate Runge-Kutta time-stepping can be used with a divergence-free, staggered mesh formulation. This is true on structured and unstructured meshes.

4) We have shown that the divergence-free reconstruction of magnetic fields that was invented in Balsara (2001b) is very useful in designing schemes for numerical MHD.

5) We have seen, via the blast test problem, that the divergence-free reconstruction helps overcome some of the problems that arise in maintaining the positivity of the pressure variable in low-$\beta$ plasma simulations.

6) We have shown that the divergence-free reconstruction can even be extended to more complicated geometries on logically rectangular meshes, such as cylindrical and spherical geometry. The disk-magnetosphere interaction problem provides us with an example of an extremely stringent test problem in spherical geometry for which our algorithm can reliably calculate a very useful result. This extension also opens the door to doing divergence-free AMR on logically rectangular meshes.

7) Furthermore, we have explicited the divergence-free reconstruction on unstructured meshes. In doing so we have established important points of similarity between divergence-free MHD on structured and unstructured meshes.

8) We have shown that the schemes designed here do meet their design accuracies. More refined interpolation yields better accuracy, as shown in Section VI. The quality of the Riemann solver, on the other hand, does not have a significant influence on the accuracy of the scheme though it does influence the dissipation characteristics of the scheme. It is shown that schemes that use very refined interpolation strategies are not influenced much by the choice of Riemann solver.

9) The schemes presented here in fact have all the essential building blocks for designing schemes for numerical MHD that are better than second order accurate. We have shown here that designing such third and higher order accurate schemes for numerical MHD basically consists of extending the building blocks to higher orders. Earlier higher order Godunov schemes for numerical MHD did not have these building blocks and, therefore, could not have been naturally extended to higher orders.

10) We have shown that the schemes presented here perform well on several stringent test problems.

**Acknowledgements** : Balsara acknowledges support via NSF grants 005569-001 and DMS-0204640. Some portion of this work was done when the author was visiting NAOJ Japan and the kind hospitality of K. Tomisaka is a pleasure to acknowledge.

**Appendix A**

Consider a zone in cylindrical geometry extending from $(r_1, \theta_1, z_1)$ at its lower coordinate limits to $(r_2, \theta_2, z_2)$ at its upper coordinate limits. There is a freedom in where we choose to collocate $B_r$ and $B_z$ so that they can be collocated at the facial centroids if one wishes. There is less freedom in where we choose to collocate $B_\theta$ and $r_{coll} = (r_1 + r_2)/2$ is an optimal choice. We also choose the collocation points in the $\theta$ and z-directions as $\theta_{coll} = (\theta_1 + \theta_2)/2$ and $z_{coll} = (z_1 + z_2)/2$. It is assumed that whenever a zone is subdivided in AMR it is bisected along these optimal collocation points. One can, of course, choose not to subdivide a zone in a given direction. Alternatively, one may choose to recursively subdivide a zone twice to obtain refinement by a factor of four in any or all directions.

In the rest of this appendix we describe the extension of the divergence-free prolongation strategy from Section IV of Balsara (2001) to cylindrical AMR-MHD. Many of the prolongation formulae are unchanged and are not repeated here. We only describe here the prolongation formulae that do change in cylindrical geometry. The rest of the steps in carrying out cylindrical AMR-MHD are as described in Balsara (2001) and do not require any further changes.

When carrying out AMR on the zone described above it helps to work in the local volumetric coordinates for that zone. The local volumetric coordinates for a zone are given by applying the transformation

$$x \leftarrow x - x_{coll} \; ; \; y \leftarrow y - y_{coll} \; ; \; z \leftarrow z - z_{coll} \tag{A.1}$$

to the global volumetric coordinates given in Section IV. The local volumetric coordinates $(x_1, y_1, z_1)$ and $(x_2, y_2, z_2)$ correspond to $(r_1, \theta_1, z_1)$ and $(r_2, \theta_2, z_2)$ respectively. If the $x = x_2$ boundary does not abut a refined mesh boundary then reconstruct the magnetic field there using eqn. (3.2). If it does abut a refined zone then write $B_x$ at that zone face as

$$B_x(x = x_2, y, z) = B_x^+ + \frac{\Delta_y B_x^+}{\Delta y} y + \frac{\Delta_z B_x^+}{\Delta z} z + \frac{\Delta_{yz} B_x^+}{\Delta y \, \Delta z} y z \tag{A.2}$$

We want closed form expressions that relate the coefficients in (A.2) to the fine mesh values:

$$\begin{aligned}
b_{r,+,+}^+ &= B_r \left( r = r_2, \theta = (\theta_{coll} + \theta_2)/2, z = (z_{coll} + z_2)/2 \right) ; \\
b_{r,+,-}^+ &= B_r \left( r = r_2, \theta = (\theta_{coll} + \theta_2)/2, z = (z_1 + z_{coll})/2 \right) ; \\
b_{r,-,+}^+ &= B_r \left( r = r_2, \theta = (\theta_1 + \theta_{coll})/2, z = (z_{coll} + z_2)/2 \right) ; \\
b_{r,-,-}^+ &= B_r \left( r = r_2, \theta = (\theta_1 + \theta_{coll})/2, z = (z_1 + z_{coll})/2 \right)
\end{aligned} \tag{A.3}$$

Such expressions are given by:

$$B_x^+ = \frac{r_2}{4}\left(b_{r,+,+}^+ + b_{r,-,-}^+ + b_{r,+,-}^+ + b_{r,-,+}^+\right)$$

$$\Delta_{yz}B_x^+ = 4\,r_2\left(b_{r,+,+}^+ + b_{r,-,-}^+ - b_{r,+,-}^+ - b_{r,-,+}^+\right)$$

$$\Delta_y B_x^+ = r_2\left(b_{r,+,+}^+ - b_{r,-,-}^+ + b_{r,+,-}^+ - b_{r,-,+}^+\right)$$

$$\Delta_z B_x^+ = r_2\left(b_{r,+,+}^+ - b_{r,-,-}^+ - b_{r,+,-}^+ + b_{r,-,+}^+\right)$$

(A.4)

A similar strategy can be carried out at $x = x_1$. If the $y = y_2$ boundary does not abut a refined mesh boundary then reconstruct the magnetic field there using eqn. (3.3). If it does abut a refined zone then write $B_y$ at that zone face as

$$B_y(x, y = y_2, z) = B_y^+ + \frac{\Delta_x B_y^+}{\Delta x}\,x + \frac{\Delta_z B_y^+}{\Delta z}\,z + \frac{\Delta_{xz} B_y^+}{\Delta x\,\Delta z}\,x\,z \qquad (A.5)$$

Writing expressions analogous to eqn. (A.3) for the $\theta$-components of the magnetic field on the fine mesh we obtain closed form expressions for the coefficients in eqn. (A.5) as follows:

$$B_y^+ = \frac{1}{4}\left(\frac{b_{\theta,+,+}^+}{r_{coll,+}} + \frac{b_{\theta,-,-}^+}{r_{coll,-}} + \frac{b_{\theta,+,-}^+}{r_{coll,+}} + \frac{b_{\theta,-,+}^+}{r_{coll,-}}\right)$$

$$\Delta_{xz}B_y^+ = 4\left(\frac{b_{\theta,+,+}^+}{r_{coll,+}} + \frac{b_{\theta,-,-}^+}{r_{coll,-}} - \frac{b_{\theta,+,-}^+}{r_{coll,+}} - \frac{b_{\theta,-,+}^+}{r_{coll,-}}\right)$$

$$\Delta_x B_y^+ = \left(\frac{b_{\theta,+,+}^+}{r_{coll,+}} - \frac{b_{\theta,-,-}^+}{r_{coll,-}} + \frac{b_{\theta,+,-}^+}{r_{coll,+}} - \frac{b_{\theta,-,+}^+}{r_{coll,-}}\right)$$

$$\Delta_z B_y^+ = \left(\frac{b_{\theta,+,+}^+}{r_{coll,+}} - \frac{b_{\theta,-,-}^+}{r_{coll,-}} - \frac{b_{\theta,+,-}^+}{r_{coll,+}} + \frac{b_{\theta,-,+}^+}{r_{coll,-}}\right)$$

(A.6)

where

$$r_{coll,-} = (r_1 + r_{coll})/2 \quad ; \quad r_{coll,+} = (r_{coll} + r_2)/2 \qquad (A.7)$$

A similar strategy can be carried out at $y = y_1$. If the $z = z_2$ boundary does not abut a refined mesh boundary then reconstruct the magnetic field there using eqn. (3.4). If it does abut a refined zone then write $B_z$ at that zone face as

$$B_z(x, y, z = z_2) = B_z^+ + \frac{\Delta_x B_z^+}{\Delta x}\,x + \frac{\Delta_y B_z^+}{\Delta y}\,y + \frac{\Delta_{xy} B_z^+}{\Delta x\,\Delta y}\,x\,y \qquad (A.8)$$

Writing expressions analogous to eqn. (A.3) for the z–components of the magnetic field on the fine mesh we obtain closed form expressions for the coefficients in eqn. (A.8) as follows:

$$B_z^+ = \frac{1}{4} \left( b_{z,+,+}^+ + b_{z,-,-}^+ + b_{z,+,-}^+ + b_{z,-,+}^+ \right)$$
$$\Delta_{xy} B_z^+ = 4 \left( b_{z,+,+}^+ + b_{z,-,-}^+ - b_{z,+,-}^+ - b_{z,-,+}^+ \right)$$
$$\Delta_x B_z^+ = \left( b_{z,+,+}^+ - b_{z,-,-}^+ + b_{z,+,-}^+ - b_{z,-,+}^+ \right) \tag{A.9}$$
$$\Delta_y B_z^+ = \left( b_{z,+,+}^+ - b_{z,-,-}^+ - b_{z,+,-}^+ + b_{z,-,+}^+ \right)$$

A similar strategy can be carried out at $z = z_1$.

The coefficients in eqns. (A.2), (A.5) and (A.8) now enable us to obtain the coefficients a… , b… and c… in eqns. (4.6) to (4.8) of Balsara (2001) and an efficient constructive strategy for obtaining the coefficients is presented in Section IV of that paper. As a result, divergence-free, directionally unsplit, TVD-preserving reconstruction of the magnetic field can be obtained even in cylindrical AMR. The last step consists of providing closed form expressions for the magnetic field components evaluated over areal elements of interest. These are just the components that one assigns to a newly-formed refined mesh in order to obtain a divergence-free representation of the magnetic field on adaptive mesh hierarchies. We provide such closed form expressions below for the case of cylindrical AMR-MHD. Say that at some general "r" we want the area-averaged value of $B_r$ over the range $[\theta_3, \theta_4] \times [z_3, z_4]$. This corresponds to the range $[y_3, y_4] \times [z_3, z_4]$ in the local volumetric coordinates. Here "x" is the local volumetric coordinate that corresponds to "r". Thus we obtain:

$$\langle B_r \rangle_{area-avg} = \frac{1}{r} A(x, y_3, y_4, z_3, z_4) \tag{A.10}$$

where

$$A(x, y_3, y_4, z_3, z_4) = \{ [a_0 + a_x x + a_{xx} x^2] + [a_y + a_{xy} x + a_{xxy} x^2] \frac{1}{2}(y_3 + y_4)$$
$$+ [a_z + a_{xz} x + a_{xxz} x^2] \frac{1}{2}(z_3 + z_4) + [a_{yz} + a_{xyz} x] \frac{1}{4}(y_3 + y_4)(z_3 + z_4) \}$$
$$\tag{A.11}$$

Say that at some general "θ" we want the area-averaged value of $B_\theta$ over the range $[r_3, r_4] \times [z_3, z_4]$. This corresponds to the range $[x_3, x_4] \times [z_3, z_4]$ in the local volumetric coordinates. Here "y" is the local volumetric coordinate that corresponds to "θ". Thus we obtain:

$$\langle B_\theta \rangle_{area-avg} = \frac{1}{2}\,(r_3 + r_4)\,B(x_3, x_4, y, z_3, z_4) \tag{A.12}$$

where

$$B(x_3, x_4, y, z_3, z_4) = \{\,[b_0 + b_y\,y + b_{yy}\,y^2] + [b_x + b_{xy}\,y + b_{xyy}\,y^2]\,\frac{1}{2}(x_3 + x_4)$$
$$+ [b_z + b_{yz}\,y + b_{yyz}\,y^2]\,\frac{1}{2}(z_3 + z_4) + [b_{xz} + b_{xyz}\,y]\,\frac{1}{4}(x_3 + x_4)(z_3 + z_4)\,\} \tag{A.13}$$

Say that at some general "z" we want the area-averaged value of $B_z$ over the range $[\,r_3\,,\,r_4\,] \times [\,\theta_3\,,\,\theta_4\,]$. This corresponds to the range $[\,x_3\,,\,x_4\,] \times [\,y_3\,,\,y_4\,]$ in the local volumetric coordinates. Thus we obtain:

$$\langle B_z \rangle_{area-avg} = C(x_3, x_4, y_3, y_4, z) \tag{A.14}$$

where

$$C(x_3, x_4, y_3, y_4, z) = \{\,[c_0 + c_z\,z + c_{zz}\,z^2] + [c_x + c_{xz}\,z + c_{xzz}\,z^2]\,\frac{1}{2}(x_3 + x_4)$$
$$+ [c_y + c_{yz}\,z + c_{yzz}\,z^2]\,\frac{1}{2}(y_3 + y_4) + [c_{xy} + c_{xyz}\,z]\,\frac{1}{4}(x_3 + x_4)(y_3 + y_4)\,\} \tag{A.15}$$

Eqns. (A.10), (A.12) and (A.14) give us the desired closed form expressions for the magnetic field components evaluated over areal elements of interest in cylindrical AMR-MHD.

**Appendix B**

Consider a zone in spherical geometry extending from $(r_1, \theta_1, \phi_1)$ at its lower coordinate limits to $(r_2, \theta_2, \phi_2)$ at its upper coordinate limits. There is a freedom in where we choose to collocate $B_r$ so that it can be collocated at the facial centroids if one wishes. There is less freedom in where we choose to collocate $B_\theta$ and $r_{coll} = \left[2\left(r_2^3 - r_1^3\right)\right]/\left[3\left(r_2^2 - r_1^2\right)\right]$ is an optimal choice. Likewise, there is less freedom in where we choose to collocate $B_\phi$ and $\theta_{coll} = \sin^{-1}\left[(\cos\theta_1 - \cos\theta_2)/(\theta_2 - \theta_1)\right]$ is an optimal choice. We also choose the collocation point in the $\phi$-direction as $\phi_{coll} = (\phi_1 + \phi_2)/2$. When carrying out adaptive mesh refinement it is best to refine along these optimal choices. It is, therefore, assumed that whenever a zone is subdivided in AMR it is bisected along these optimal collocation points. One can, of course, choose not to subdivide a zone in a given direction. Alternatively, one may choose to recursively subdivide a zone twice to obtain refinement by a factor of four in any or all directions.

In the rest of this appendix we describe the extension of the divergence-free prolongation strategy from Section IV of Balsara (2001) to spherical AMR-MHD. Many of the prolongation formulae are unchanged and are not repeated here. We only describe here the prolongation formulae that do change in spherical geometry. The rest of the steps in carrying out spherical AMR-MHD are as described in Balsara (2001) and do not require any further changes.

As in the cylindrical case above we transform to the local volumetric coordinates for spherical geometry using eqn. (A.1). The local volumetric coordinates ( $x_1$ , $y_1$ , $z_1$ ) and ( $x_2$ , $y_2$ , $z_2$ ) correspond to ( $r_1$ , $\theta_1$ , $\phi_1$ ) and ( $r_2$ , $\theta_2$ , $\phi_2$ ) respectively. If the x = $x_2$ boundary does not abut a refined mesh boundary then reconstruct the magnetic field there using eqn. (3.2). If it does abut a refined zone then write $B_x$ at that zone face using eqn. (A.2). Writing equations that are analogous to eqn. (A.3) at the x = $x_2$ boundary we obtain closed-form expressions for the coefficients in eqn. (A.2). Notice, though, that the explicit expressions in spherical geometry are different and so we get:

$$\begin{aligned}
B_x^+ &= \frac{r_2^2}{4} \left( b_{r,+,+}^+ + b_{r,-,-}^+ + b_{r,+,-}^+ + b_{r,-,+}^+ \right) \\
\Delta_{yz} B_x^+ &= 4\, r_2^2 \left( b_{r,+,+}^+ + b_{r,-,-}^+ - b_{r,+,-}^+ - b_{r,-,+}^+ \right) \\
\Delta_y B_x^+ &= r_2^2 \left( b_{r,+,+}^+ - b_{r,-,-}^+ + b_{r,+,-}^+ - b_{r,-,+}^+ \right) \\
\Delta_z B_x^+ &= r_2^2 \left( b_{r,+,+}^+ - b_{r,-,-}^+ - b_{r,+,-}^+ + b_{r,-,+}^+ \right)
\end{aligned} \qquad (B.1)$$

A similar strategy can be carried out at x = $x_1$ . If the y = $y_2$ boundary does not abut a refined mesh boundary then reconstruct the magnetic field there using eqn. (3.3). If it does abut a refined zone then write $B_y$ at that zone face using eqn. (A.5). Writing expressions analogous to eqn. (A.3) for the θ-components of the magnetic field on the fine mesh we obtain closed form expressions for the coefficients in eqn. (A.5) as follows:

$$\begin{aligned}
B_y^+ &= \frac{\sin \theta_2}{4} \left( \frac{b_{\theta,+,+}^+}{r_{coll,+}} + \frac{b_{\theta,-,-}^+}{r_{coll,-}} + \frac{b_{\theta,+,-}^+}{r_{coll,+}} + \frac{b_{\theta,-,+}^+}{r_{coll,-}} \right) \\
\Delta_{xz} B_y^+ &= 4 \sin \theta_2 \left( \frac{b_{\theta,+,+}^+}{r_{coll,+}} + \frac{b_{\theta,-,-}^+}{r_{coll,-}} - \frac{b_{\theta,+,-}^+}{r_{coll,+}} - \frac{b_{\theta,-,+}^+}{r_{coll,-}} \right) \\
\Delta_x B_y^+ &= \sin \theta_2 \left( \frac{b_{\theta,+,+}^+}{r_{coll,+}} - \frac{b_{\theta,-,-}^+}{r_{coll,-}} + \frac{b_{\theta,+,-}^+}{r_{coll,+}} - \frac{b_{\theta,-,+}^+}{r_{coll,-}} \right) \\
\Delta_z B_y^+ &= \sin \theta_2 \left( \frac{b_{\theta,+,+}^+}{r_{coll,+}} - \frac{b_{\theta,-,-}^+}{r_{coll,-}} - \frac{b_{\theta,+,-}^+}{r_{coll,+}} + \frac{b_{\theta,-,+}^+}{r_{coll,-}} \right)
\end{aligned} \qquad (B.2)$$

where

$$r_{coll,+} = \left[2\left(r_2^3 - r_{coll}^3\right)\right]/\left[3\left(r_2^2 - r_{coll}^2\right)\right]$$
$$r_{coll,-} = \left[2\left(r_{coll}^3 - r_1^3\right)\right]/\left[3\left(r_{coll}^2 - r_1^2\right)\right]$$
(B.3)

A similar strategy can be carried out at $y = y_1$. If the $z = z_2$ boundary does not abut a refined mesh boundary then reconstruct the magnetic field there using eqn. (3.4). If it does abut a refined zone then write $B_z$ at that zone face using eqn. (A.8). Writing expressions analogous to eqn. (A.3) for the $\phi$–components of the magnetic field on the fine mesh we obtain closed form expressions for the coefficients in eqn. (A.8) as follows:

$$B_z^+ = \frac{1}{4}\left(\frac{b_{\phi,+,+}^+}{r_{coll,+}\sin\theta_{coll,+}} + \frac{b_{\phi,-,-}^+}{r_{coll,-}\sin\theta_{coll,-}} + \frac{b_{\phi,+,-}^+}{r_{coll,+}\sin\theta_{coll,-}} + \frac{b_{\phi,-,+}^+}{r_{coll,-}\sin\theta_{coll,+}}\right)$$

$$\Delta_{xy}B_z^+ = 4\left(\frac{b_{\phi,+,+}^+}{r_{coll,+}\sin\theta_{coll,+}} + \frac{b_{\phi,-,-}^+}{r_{coll,-}\sin\theta_{coll,-}} - \frac{b_{\phi,+,-}^+}{r_{coll,+}\sin\theta_{coll,-}} - \frac{b_{\phi,-,+}^+}{r_{coll,-}\sin\theta_{coll,+}}\right)$$

$$\Delta_x B_z^+ = \left(\frac{b_{\phi,+,+}^+}{r_{coll,+}\sin\theta_{coll,+}} - \frac{b_{\phi,-,-}^+}{r_{coll,-}\sin\theta_{coll,-}} + \frac{b_{\phi,+,-}^+}{r_{coll,+}\sin\theta_{coll,-}} - \frac{b_{\phi,-,+}^+}{r_{coll,-}\sin\theta_{coll,+}}\right)$$

$$\Delta_y B_z^+ = \left(\frac{b_{\phi,+,+}^+}{r_{coll,+}\sin\theta_{coll,+}} - \frac{b_{\phi,-,-}^+}{r_{coll,-}\sin\theta_{coll,-}} - \frac{b_{\phi,+,-}^+}{r_{coll,+}\sin\theta_{coll,-}} + \frac{b_{\phi,-,+}^+}{r_{coll,-}\sin\theta_{coll,+}}\right)$$
(B.4)

where

$$\theta_{coll,+} = \sin^{-1}\left[\left(\cos\theta_{coll} - \cos\theta_2\right)/\left(\theta_2 - \theta_{coll}\right)\right]$$
$$\theta_{coll,-} = \sin^{-1}\left[\left(\cos\theta_1 - \cos\theta_{coll}\right)/\left(\theta_{coll} - \theta_1\right)\right]$$
(B.5)

A similar strategy can be carried out at $z = z_1$.

The coefficients in eqns. (A.2), (A.5) and (A.8) now enable us to obtain the coefficients a… , b… and c… in eqns. (4.6) to (4.8) of Balsara (2001) and an efficient constructive strategy for obtaining the coefficients is presented in Section IV of that paper. As a result, divergence-free, directionally unsplit, TVD-preserving reconstruction of the magnetic field can be obtained even in spherical AMR. The last step consists of providing closed form expressions for the magnetic field components evaluated over areal elements of interest. These are just the components that one assigns to a newly-formed refined mesh in order to obtain a divergence-free representation of the magnetic field on adaptive mesh hierarchies. We provide such closed form expressions below for the case of spherical AMR-MHD. Say that at some general "r" we want the area-averaged value of $B_r$ over the range $[\theta_3, \theta_4] \times [\phi_3, \phi_4]$. This corresponds to the range $[\, y_3 \,,\, y_4 \,]X[\, z_3 \,,\, z_4 \,]$ in the local volumetric coordinates. Here "x" is the local volumetric coordinate that corresponds to "r". Thus we obtain:

$$\langle B_r \rangle_{area-avg} = \frac{1}{r^2} A(\, x \,,\, y_3 \,,\, y_4 \,,\, z_3 \,,\, z_4 \,) \tag{B.6}$$

where the function "A" in the above equation is just eqn. (A.11) above. Say that at some general "θ" we want the area-averaged value of $B_\theta$ over the range $[\, r_3 \,,\, r_4 \,]X[\, \phi_3 \,,\, \phi_4 \,]$. This corresponds to the range $[\, x_3 \,,\, x_4 \,]X[\, z_3 \,,\, z_4 \,]$ in the local volumetric coordinates. Here "y" is the local volumetric coordinate that corresponds to "θ". Thus we obtain:

$$\langle B_\theta \rangle_{area-avg} = \frac{r_{coll,34}}{\sin \theta} B(\, x_3 \,,\, x_4 \,,\, y \,,\, z_3 \,,\, z_4 \,) \tag{B.7}$$

where

$$r_{coll,34} = \left[ 2\left(r_4^3 - r_3^3\right) \right] / \left[ 3\left(r_4^2 - r_3^2\right) \right] \tag{B.8}$$

and the function "B" in eqn. (B.7) is just eqn. (A.13) above. Say that at some general "$\phi$" we want the area-averaged value of $B_\phi$ over the range $[\, r_3 \,,\, r_4 \,]X[\, \theta_3 \,,\, \theta_4 \,]$. This corresponds to the range $[\, x_3 \,,\, x_4 \,]X[\, y_3 \,,\, y_4 \,]$ in the local volumetric coordinates. Here "z" is the local volumetric coordinate that corresponds to "$\phi$". Thus we obtain:

$$\langle B_\phi \rangle_{area-avg} = r_{coll,34} \sin \theta_{coll,34} \, C(\, x_3 \,,\, x_4 \,,\, y_3 \,,\, y_4 \,,\, z \,) \tag{B.9}$$

where

$$\theta_{coll,34} = \sin^{-1} \left[ (\cos \theta_3 - \cos \theta_4) / (\theta_4 - \theta_3) \right] \tag{B.10}$$

and the function "C" in eqn. (B.9) is just eqn. (A.15) above. Eqns. (B.6), (B.7) and (B.9) give us the desired closed form expressions for the magnetic field components evaluated over areal elements of interest in spherical AMR-MHD.

**Appendix C**

For the reference tetrahedron on the right of Fig. 3 let the vertices 1, 2, 3 and 4 have coordinates given by (0,0,0), (1,0,0), (0,1,0) and (0,0,1) respectively. Also let the vertices 1, 2, 3 and 4 have magnetic field vectors given by $\vec{B}_1$, $\vec{B}_2$, $\vec{B}_3$ and $\vec{B}_4$ respectively. (These are the same magnetic field vectors that are shown in Fig. 3 where we explicitly show that the magnetic field vector on each vertex of the given triangle is the same as the magnetic field vector on the corresponding vertex of the reference triangle.) Letting $(\tilde{x}, \tilde{y}, \tilde{z})$ be the coordinates within the reference tetrahedron, the reconstructed magnetic field in the reference tetrahedron is given by:

$$\begin{aligned}
B_x\left(\tilde{x}, \tilde{y}, \tilde{z}\right) &= \tilde{a}_0 + \tilde{a}_x\, \tilde{x} + \tilde{a}_y\, \tilde{y} + \tilde{a}_z\, \tilde{z} \\
B_y\left(\tilde{x}, \tilde{y}, \tilde{z}\right) &= \tilde{b}_0 + \tilde{b}_x\, \tilde{x} + \tilde{b}_y\, \tilde{y} + \tilde{b}_z\, \tilde{z} \\
B_z\left(\tilde{x}, \tilde{y}, \tilde{z}\right) &= \tilde{c}_0 + \tilde{c}_x\, \tilde{x} + \tilde{c}_y\, \tilde{y} + \tilde{c}_z\, \tilde{z}
\end{aligned} \quad (C.1)$$

with

$$\begin{aligned}
\tilde{a}_0 &= B_{1,x} \,;\, \tilde{b}_0 = B_{1,y} \,;\, \tilde{c}_0 = B_{1,z} \,; \\
\tilde{a}_x &= B_{2,x} - B_{1,x} \,;\, \tilde{b}_x = B_{2,y} - B_{1,y} \,;\, \tilde{c}_x = B_{2,z} - B_{1,z} \,; \\
\tilde{a}_y &= B_{3,x} - B_{1,x} \,;\, \tilde{b}_y = B_{3,y} - B_{1,y} \,;\, \tilde{c}_y = B_{3,z} - B_{1,z} \,; \\
\tilde{a}_z &= B_{4,x} - B_{1,x} \,;\, \tilde{b}_z = B_{4,y} - B_{1,y} \,;\, \tilde{c}_z = B_{4,z} - B_{1,z}
\end{aligned} \quad (C.2)$$

Let the vertices 1, 2, 3 and 4 of the given tetrahedron on the left of Fig. 3 have coordinates given by ( $x_1$ , $y_1$ , $z_1$ ), ( $x_2$ , $y_2$ , $z_2$ ), ( $x_3$ , $y_3$ , $z_3$ ) and ( $x_4$ , $y_4$ , $z_4$ ) respectively. The coordinates (x,y,z) of the given tetrahedron can be related to the coordinates $(\tilde{x}, \tilde{y}, \tilde{z})$ of the reference tetrahedron by the linear relationship:

$$\begin{aligned}
x &= x_1 + (x_2 - x_1)\, \tilde{x} + (x_3 - x_1)\, \tilde{y} + (x_4 - x_1)\, \tilde{z} \\
y &= y_1 + (y_2 - y_1)\, \tilde{x} + (y_3 - y_1)\, \tilde{y} + (y_4 - y_1)\, \tilde{z} \\
z &= z_1 + (z_2 - z_1)\, \tilde{x} + (z_3 - z_1)\, \tilde{y} + (z_4 - z_1)\, \tilde{z}
\end{aligned} \quad (C.3)$$

The inverse of eqn. (C.3) is also easily obtained. Using the inverse in eqn. (C.1) we can obtain the reconstructed, divergence-free magnetic field in the given tetrahedron. We also point out that for the above strategy to work the given and reference tetrahedra should have their vertices numbered according to a right-handed convention.

**Figure Captions**

Figure 1 applies to logically rectangular meshes. Figure 1a shows the collocation of the magnetic fields at the control volume's faces and the collocation of electric fields at the control volume's edges. The edge centered quadrature points at which the electric field components are desired have been shown by solid dots. Figure 1b shows the facial quadrature points at which the Riemann problems are evaluated. The components of the fluxes from the Riemann problems that contribute to the edge centered quadrature points are also shown.

Figure 2 applies to tetrahedral meshes. Figure 2a shows a tetrahedron with a locally orthogonal coordinate system constructed on face 1 that opposes vertex 1. The magnetic field component that is collocated on face 1 is also shown. The edge centered quadrature points at which the electric field components are desired have been shown by solid dots. Figure 1b shows the facial quadrature points at which the Riemann problems are evaluated where the red points lie in face 1, the green points in face 2, the blue points in face 3 and the black points in face 4.

Figure 3 shows the given tetrahedron on the left and its mapping to the reference tetrahedron on the right. The tetrahedron of interest is shown on the left side of Fig. 3. Taking face 1 in Fig. 3 as an example, we know $B_{\eta_1} = \vec{B} \bullet \hat{\eta}_1$ at the centroid of face 1 as well as its linear variation along $\hat{\xi}_1$ and $\hat{\phi}_1$. The magnetic field vectors are obtained at the vertices of the given tetrahedron and are shown as thick dashed vectors on the left side of Fig. 3. The *same* magnetic field vectors are assigned to the corresponding vertices of the reference tetrahedron on the right side of Fig. 3.

Figure 4 shows the decay of torsional Alfven waves that are propagating obliquely to a 2d computational mesh. The waves propagate at an angle of $9.462^0$ with respect to the y-axis. "split" refers to a directionally split scheme. "us" stands for unsplit scheme. "Roe" refers to the use of a Roe-type linearized Riemann solver. "HLL" refers to the use of the HLLE-type Riemann solver. "TVD" denotes TVD interpolation applied to the characteristic variables. "r=2WENO" refers to the r=2 WENO interpolation from sub-section III.3. "r=3WENO" refers to the r=3 WENO interpolation from sub-section III.4 being applied to all the characteristic fields. "b-WENO" refers to the blended WENO interpolation from sub-section III.4. Figure 4a shows the evolution of the maximum value of the z-component of the velocity as a function of time on a log-linear plot. ( The schemes labeled "e" and "f" almost overlie each other for the z-velocity plot.) Figure 4b shows the evolution of the maximum value of the z-component of the magnetic field as a function of time again on a log-linear plot. Several schemes are shown and are labeled from "a" to "g". The actual algorithms corresponding to these labels are specified in the text of sub-section VII.a.

Figures 5a, 5b, 5c and 5d show the density, pressure, Mach number and the magnitude of the magnetic field respectively at a time of 0.29 for the rotor test problem. Genuinely multidimensional limiting from sub-section III.5 was used.

Figures 6a, 6b, 6c and 6d show the logarithm (base 10) of the density, the logarithm (base 10) of the pressure, the magnitude of the velocity and the magnitude of the magnetic field respectively at a time of 0.01 for the blast test problem. Fast TVD limiting was used and the volume-averaged magnetic fields were obtained by using eqn. (3.17) from Sub-section III.1. Figure 6e shows the logarithm (base 10) of the pressure variable when r=3 WENO slopes from Sub-section III.4 were kept within physical bounds by using the multidimensional limiter in Sub-section III.5. Figure 6f shows the logarithm (base 10) of the pressure variable when the slopes from the genuinely multidimensional limiter were used by themselves.

Figure 7a shows the logarithm (base 10) of the density at a simulation time of 0.24 , which corresponds to $0.24 \times 10^{-2}$ sec of physical time. This time is also equal to about 1 orbital period at the fiducial radius. Figures 7b and 7c show the logarithm of the velocity and magnetic field respectively in the azimuthal plane at the same time. Vectors, with their lengths scaled logarithmically, are also shown to indicate the direction of the field. Variables are displayed in scaled units, not cgs units. Figures 7d, e and f show information that is analogous to figures 7a, b and c respectively but at a later time of $0.48 \times 10^{-2}$ sec of physical time.

Table I (r=2 WENO slopes with Roe Riemann Solver and RK-3 time stepping)

| #of zones | $L_1$ error (momentum) | Order (momentum) | $L_1$ error (B-field) | Order (B-field) |
|---|---|---|---|---|
| 50X50 | $5.5995 \times 10^{-3}$ | | $1.4351 \times 10^{-2}$ | |
| 100X100 | $1.8944 \times 10^{-3}$ | 1.56 | $6.1119 \times 10^{-3}$ | 1.23 |
| 200X200 | $5.0303 \times 10^{-4}$ | 1.91 | $1.6746 \times 10^{-3}$ | 1.87 |
| 400X400 | $1.2086 \times 10^{-4}$ | 2.06 | $4.1782 \times 10^{-4}$ | 2.01 |

Table II (r=3 WENO slopes with Roe Riemann Solver and RK-3 time stepping)

| #of zones | $L_1$ error (momentum) | Order (momentum) | $L_1$ error (B-field) | Order (B-field) |
|---|---|---|---|---|
| 50X50 | $2.9487 \times 10^{-3}$ | | $1.0782 \times 10^{-2}$ | |
| 100X100 | $8.3775 \times 10^{-4}$ | 1.82 | $2.9789 \times 10^{-3}$ | 1.86 |
| 200X200 | $2.1239 \times 10^{-4}$ | 1.98 | $7.5473 \times 10^{-4}$ | 1.98 |
| 400X400 | $5.3302 \times 10^{-5}$ | 1.99 | $1.8932 \times 10^{-4}$ | 2.00 |

Table III (r=2 WENO slopes with HLLE Riemann Solver and RK-3 time stepping)

| #of zones | $L_1$ error (momentum) | Order (momentum) | $L_1$ error (B-field) | Order (B-field) |
|---|---|---|---|---|
| 50X50 | $6.6865 \times 10^{-3}$ | | $1.6856 \times 10^{-2}$ | |
| 100X100 | $2.1524 \times 10^{-3}$ | 1.64 | $6.6692 \times 10^{-3}$ | 1.34 |
| 200X200 | $5.9867 \times 10^{-4}$ | 1.85 | $1.9290 \times 10^{-3}$ | 1.79 |
| 400X400 | $1.4499 \times 10^{-4}$ | 2.05 | $4.8205 \times 10^{-4}$ | 2.00 |

Table IV (Blended-WENO slopes with HLLE Riemann Solver and RK-3 time stepping)

| #of zones | $L_1$ error (momentum) | Order (momentum) | $L_1$ error (B-field) | Order (B-field) |
|---|---|---|---|---|
| 50X50 | $3.7699 \times 10^{-3}$ | | $1.2723 \times 10^{-2}$ | |
| 100X100 | $1.1648 \times 10^{-3}$ | 1.69 | $4.0570 \times 10^{-3}$ | 1.65 |
| 200X200 | $3.3608 \times 10^{-4}$ | 1.79 | $1.1739 \times 10^{-3}$ | 1.79 |
| 400X400 | $8.9955 \times 10^{-5}$ | 1.90 | $3.1544 \times 10^{-4}$ | 1.90 |

Table V (Fast TVD with HLLE Riemann Solver and RK-3 time stepping)

| #of zones | $L_1$ error (momentum) | Order (momentum) | $L_1$ error (B-field) | Order (B-field) |
|---|---|---|---|---|

| 50X50 | 8.6817X10$^{-3}$ | | 1.2025X10$^{-2}$ | |
| 100X100 | 2.8120X10$^{-3}$ | 1.63 | 5.1786X10$^{-3}$ | 1.22 |
| 200X200 | 8.8480X10$^{-4}$ | 1.67 | 2.0507X10$^{-3}$ | 1.34 |
| 400X400 | 2.4946X10$^{-4}$ | 1.83 | 6.1975X10$^{-4}$ | 1.73 |
| 800X800 | 6.6402X10$^{-5}$ | 1.91 | 1.7457X10$^{-4}$ | 1.83 |

Table VI (Multi-d limiter $\psi = 1.0$ with HLLE Riemann Solver and RK-3 time stepping)

| #of zones | L$_1$ error (momentum) | Order (momentum) | L$_1$ error (B-field) | Order (B-field) |
|---|---|---|---|---|
| 50X50 | 3.4250X10$^{-3}$ | | 1.0142X10$^{-2}$ | |
| 100X100 | 8.6663X10$^{-4}$ | 1.98 | 2.9426X10$^{-3}$ | 1.79 |
| 200X200 | 2.2289X10$^{-4}$ | 1.96 | 7.7723X10$^{-4}$ | 1.92 |
| 400X400 | 5.6268X10$^{-5}$ | 1.99 | 1.9857X10$^{-4}$ | 1.97 |
| 800X800 | 1.4167X10$^{-5}$ | 1.99 | 5.0075X10$^{-5}$ | 1.99 |

Table VII (Multi-d slopes $\psi = 1.0$ ($\rho$, **B**)/0.5 (P,**v**) with HLLE Riemann Solver and RK-3 time stepping)

| #of zones | L$_1$ error (momentum) | Order (momentum) | L$_1$ error (B-field) | Order (B-field) |
|---|---|---|---|---|
| 50X50 | 8.2735X10$^{-3}$ | | 9.9802X10$^{-3}$ | |
| 100X100 | 2.6163X10$^{-3}$ | 1.66 | 3.4637X10$^{-3}$ | 1.53 |
| 200X200 | 8.1804X10$^{-4}$ | 1.68 | 1.1019X10$^{-3}$ | 1.65 |
| 400X400 | 2.3563X10$^{-4}$ | 1.80 | 3.3516X10$^{-4}$ | 1.72 |
| 800X800 | 6.1940X10$^{-5}$ | 1.93 | 9.4423X10$^{-5}$ | 1.83 |

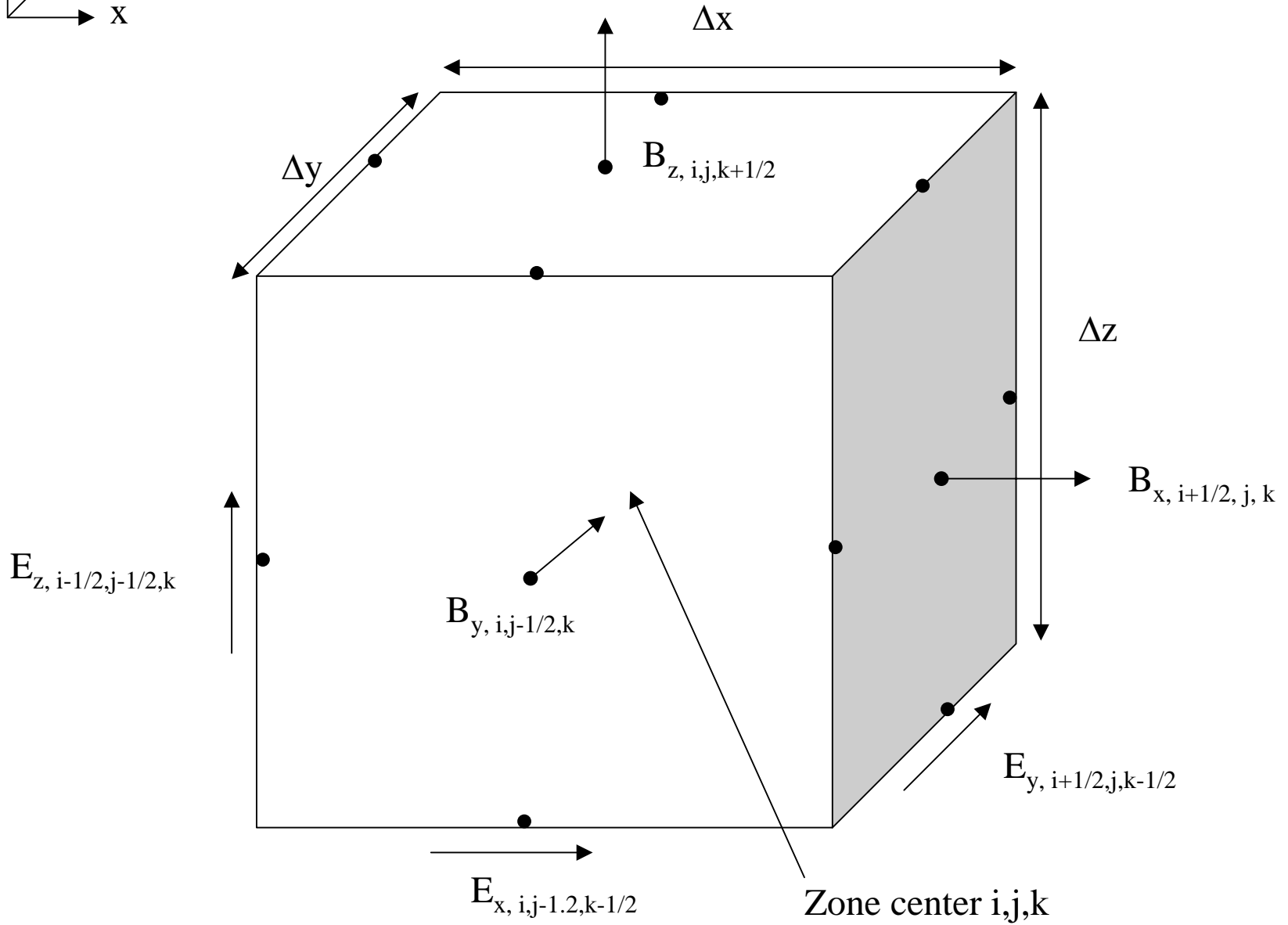

Fig 1a

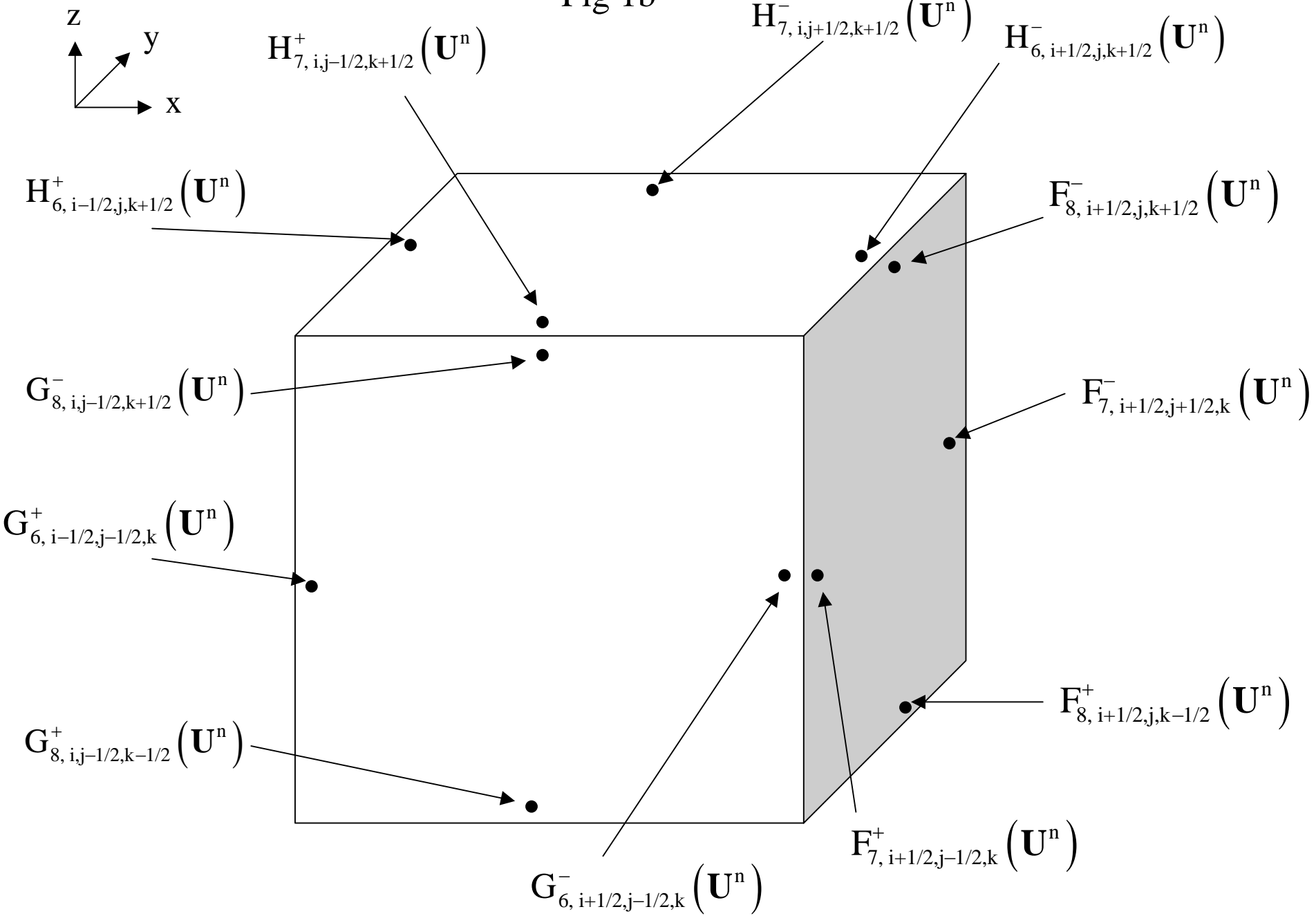

Fig 1b

Fig 2a

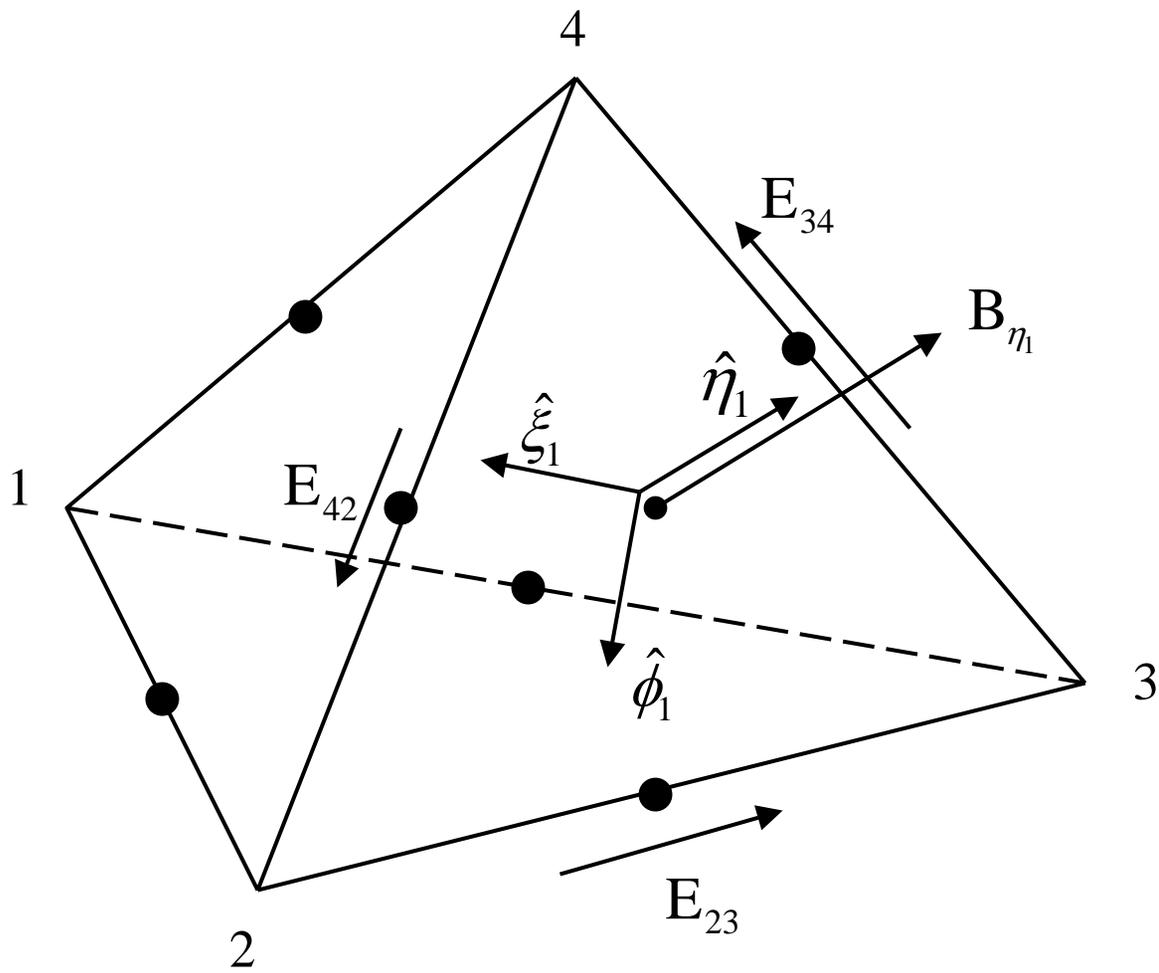

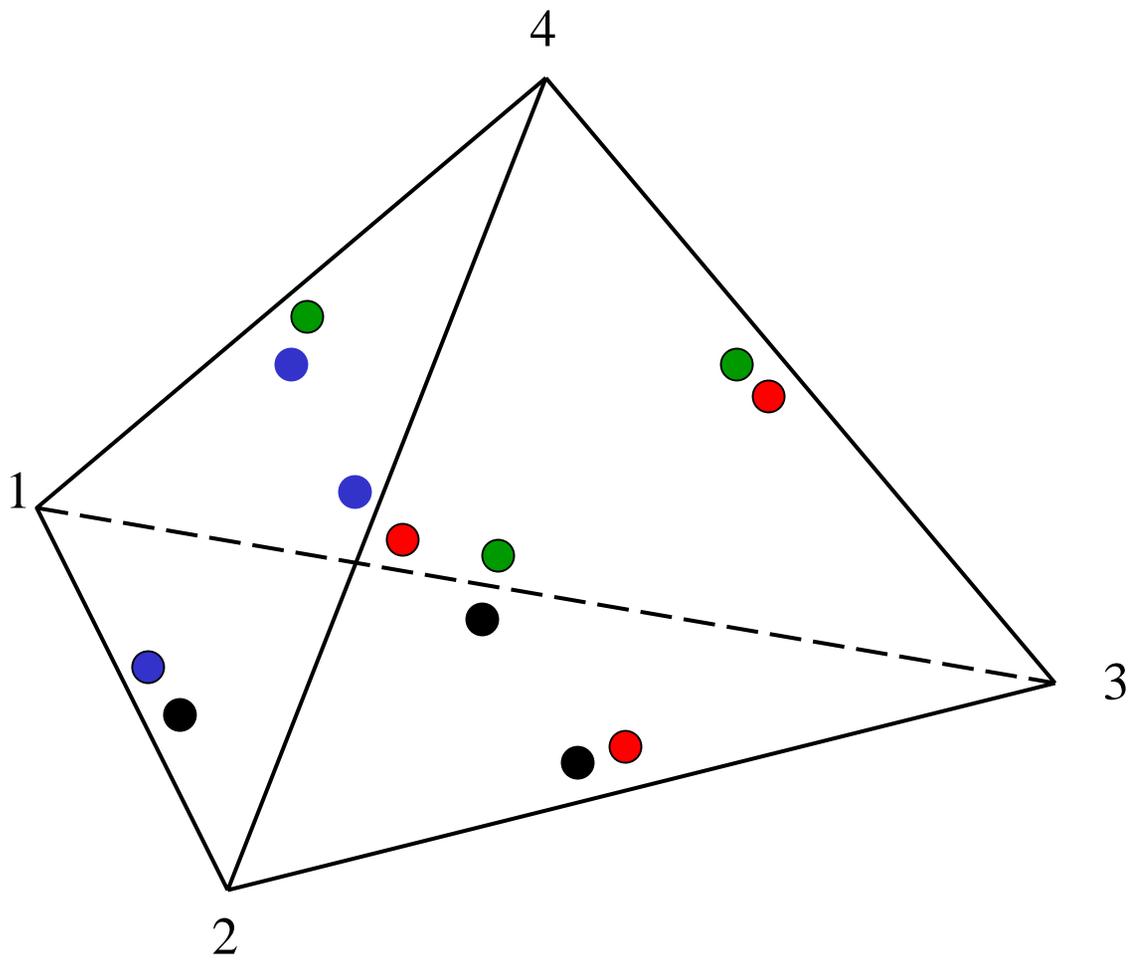

Fig 2b

Fig 3

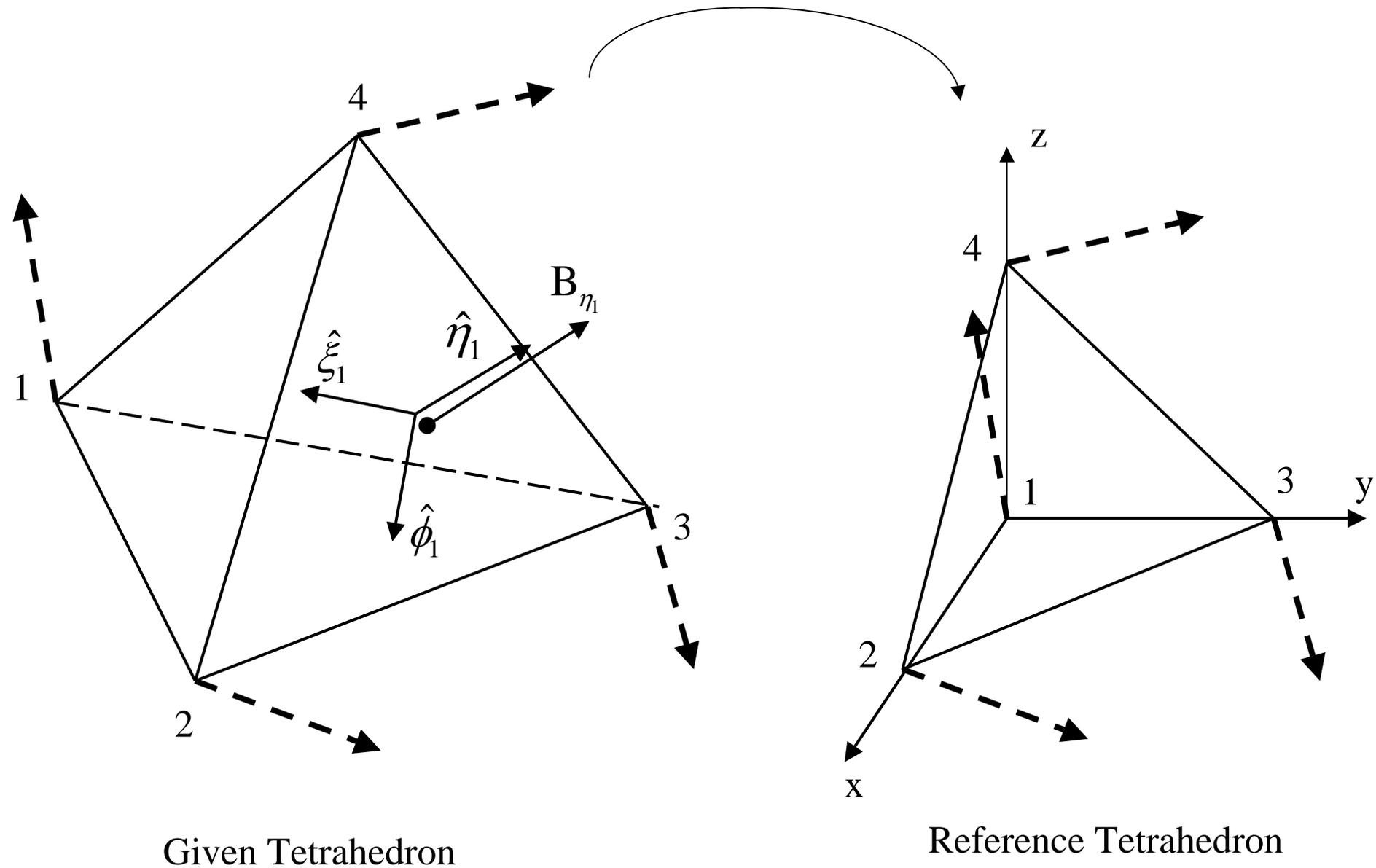

Given Tetrahedron

Reference Tetrahedron

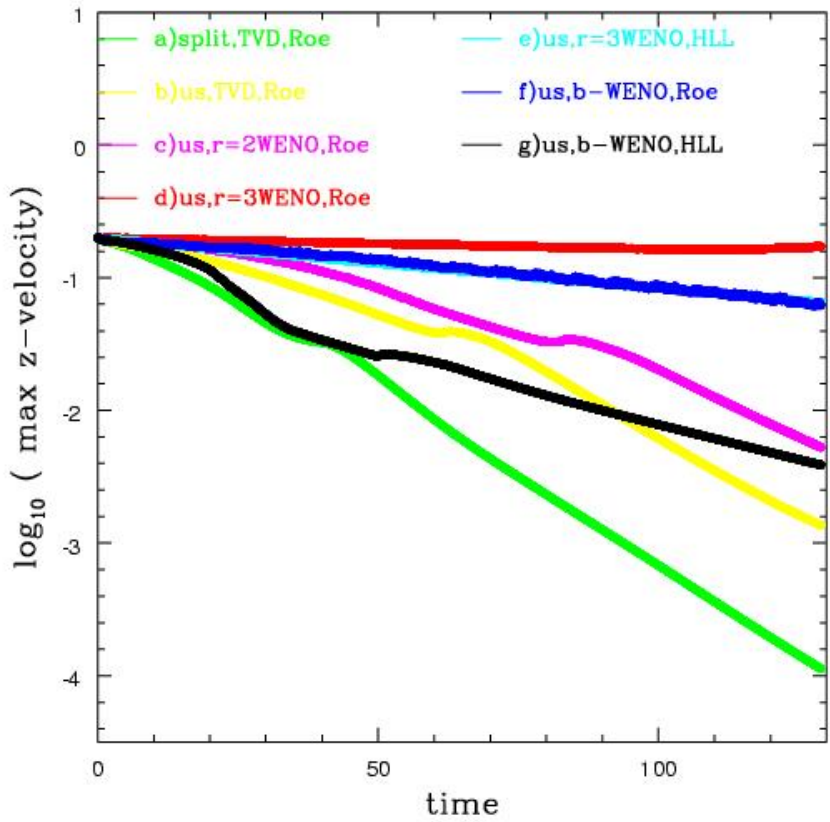

Fig 4a

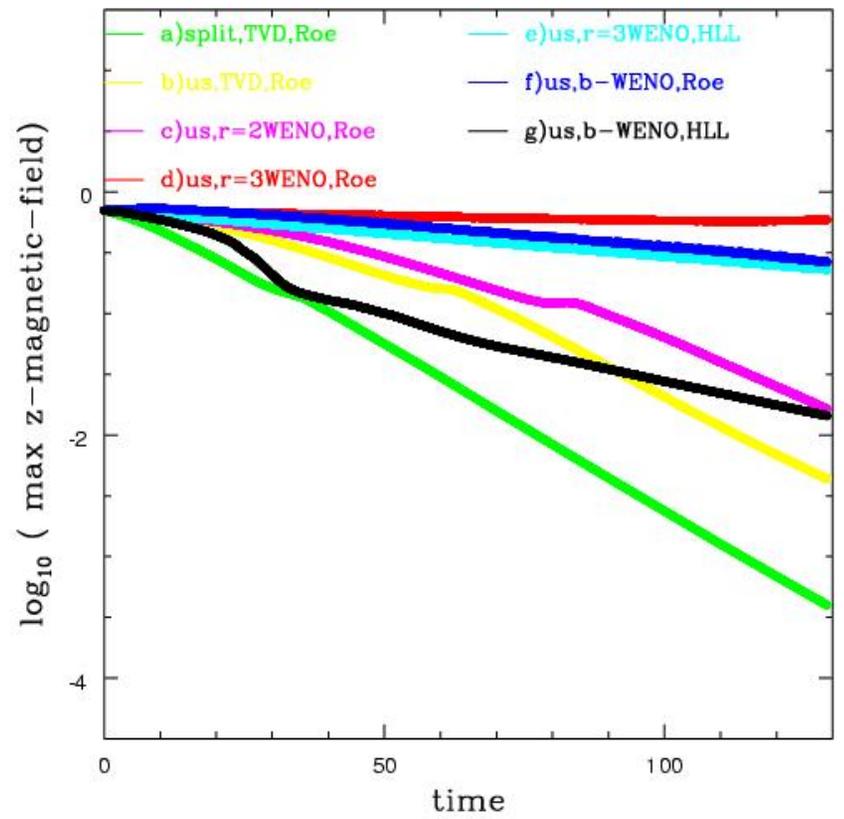

Fig 4b

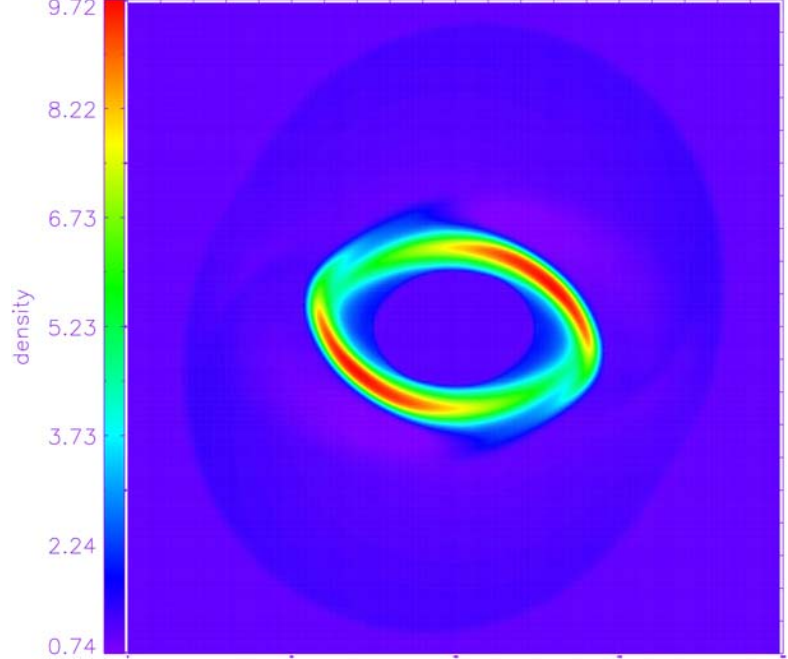
Fig 5a

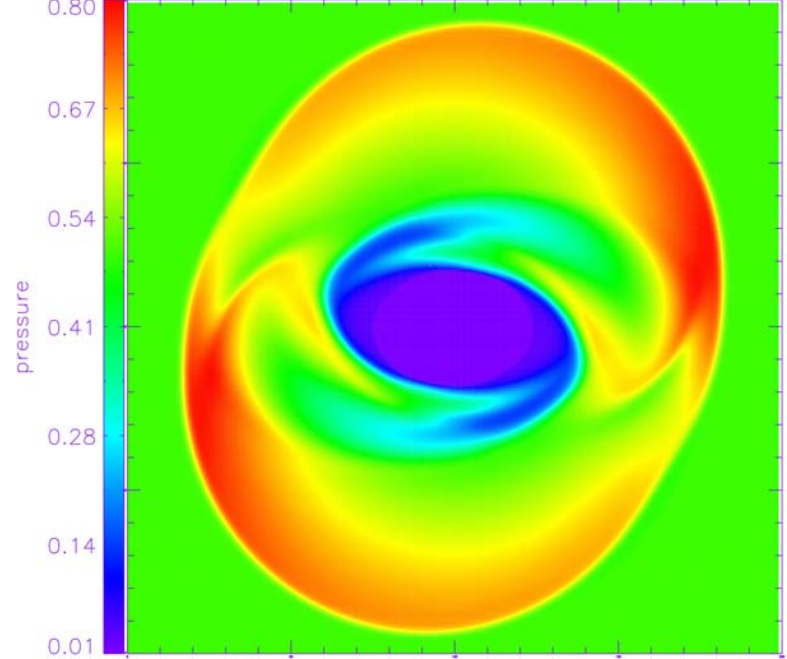
Fig 5b

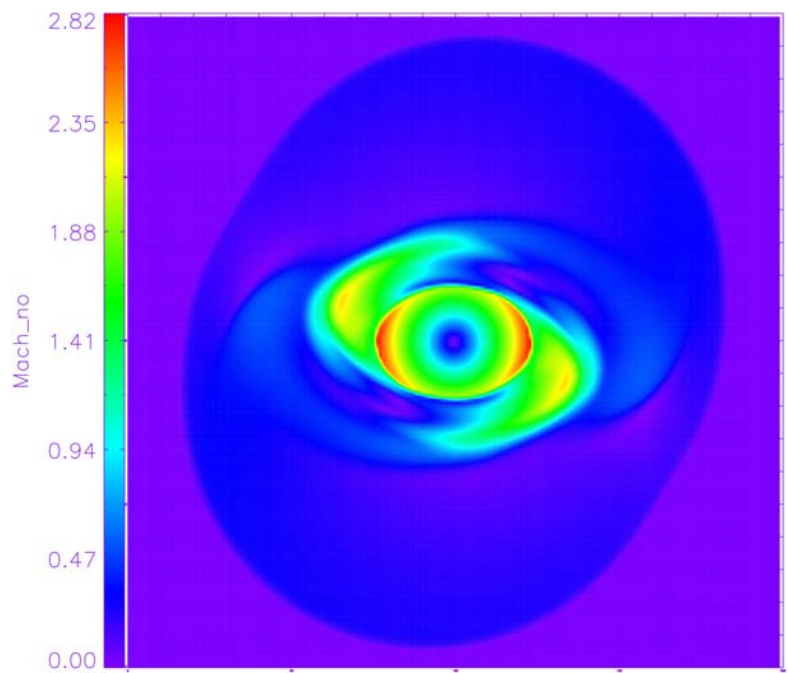
Fig 5c

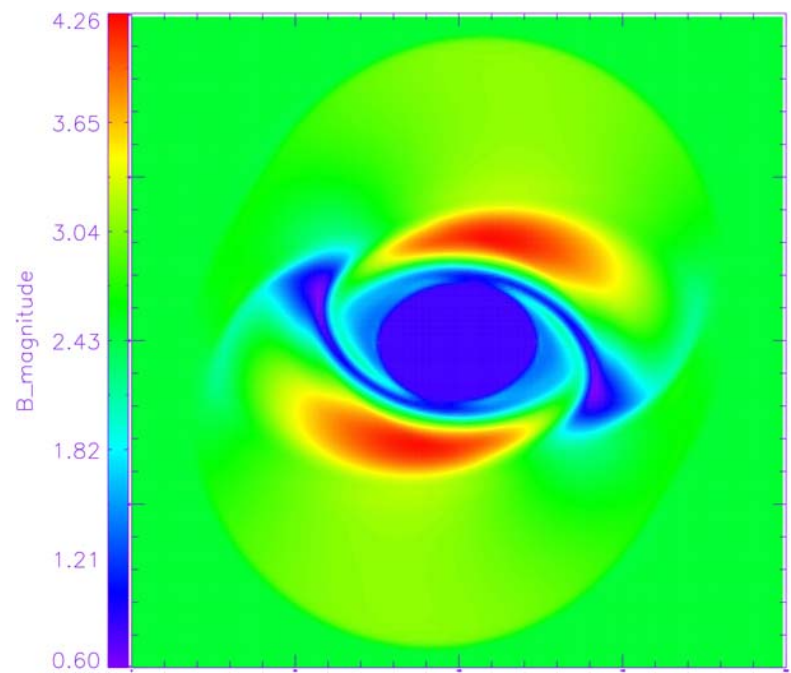
Fig 5d

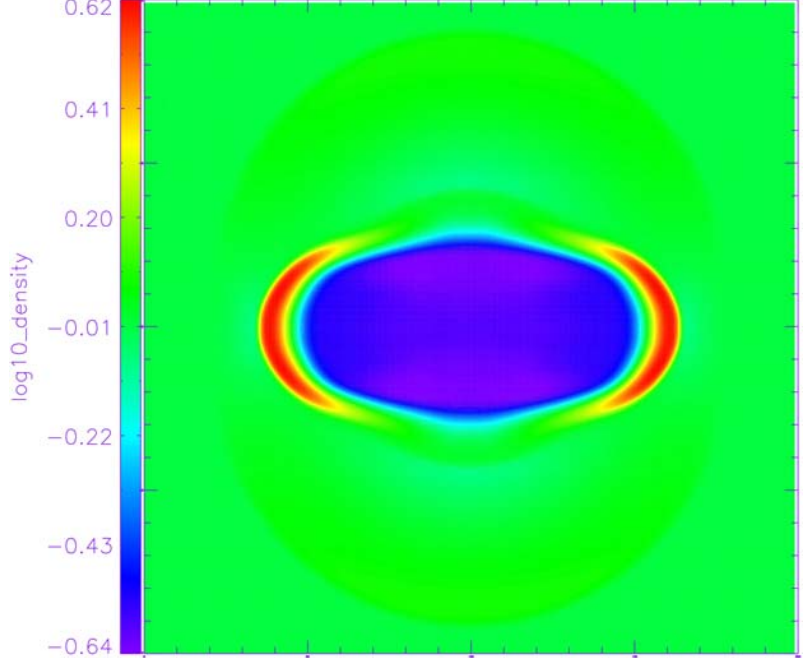
Fig 6a

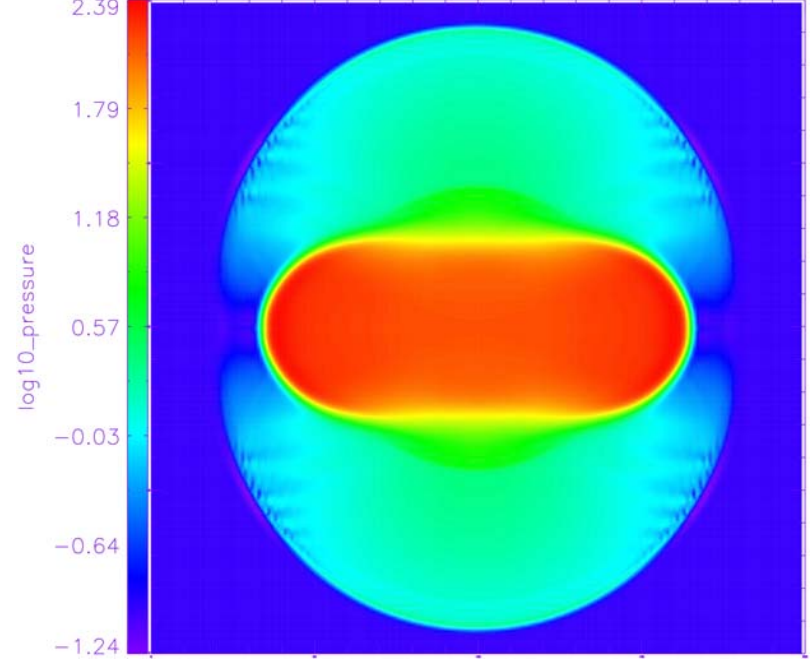
Fig 6b

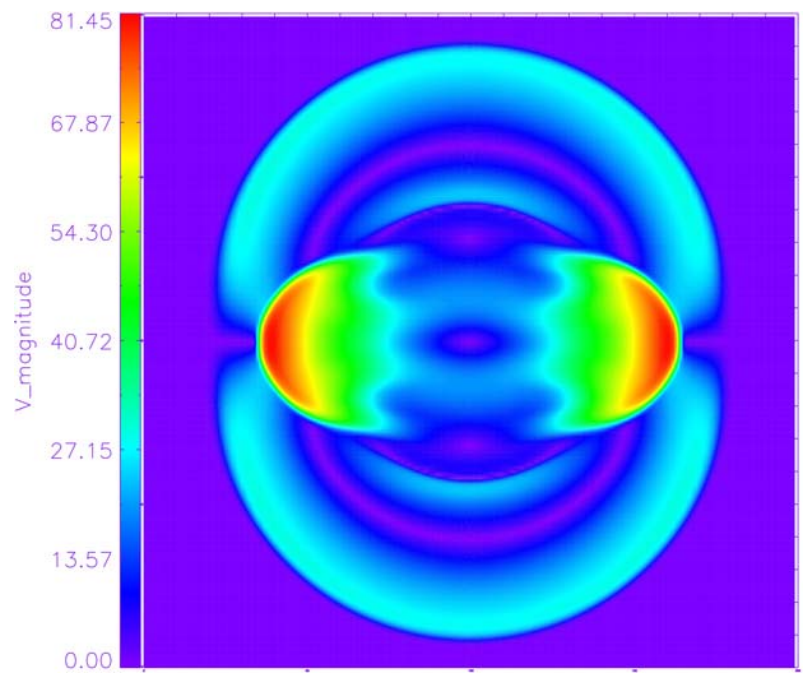
Fig 6c

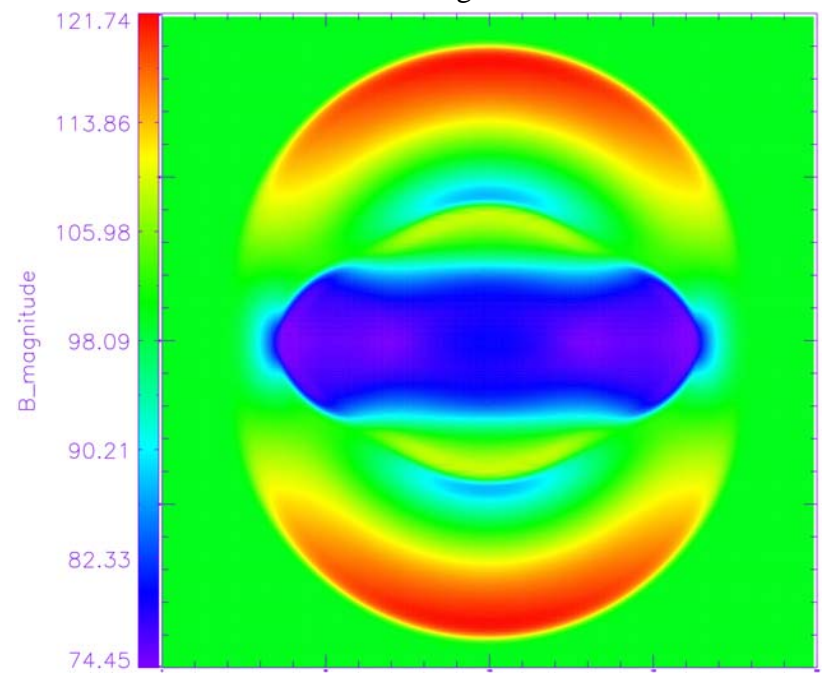
Fig 6d

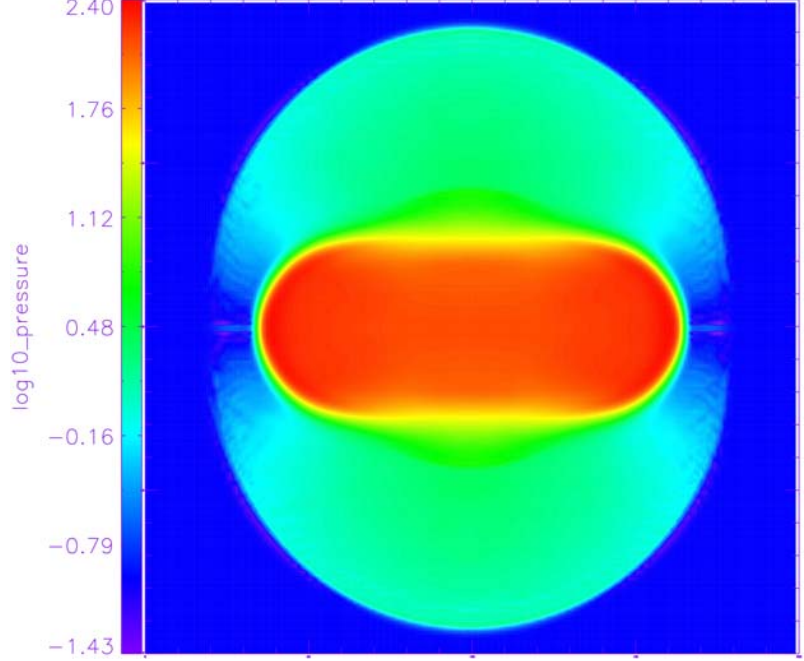 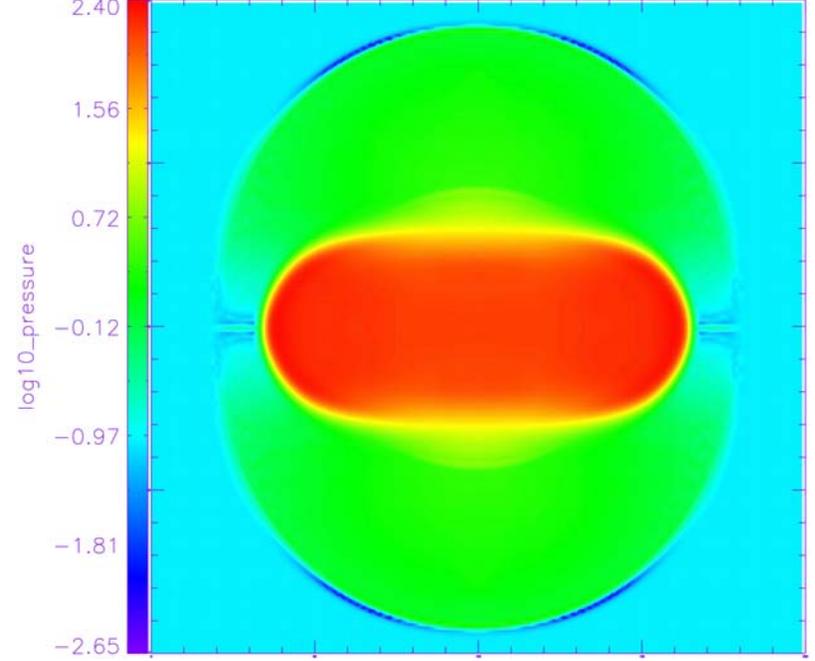

Fig 6e  Fig 6f

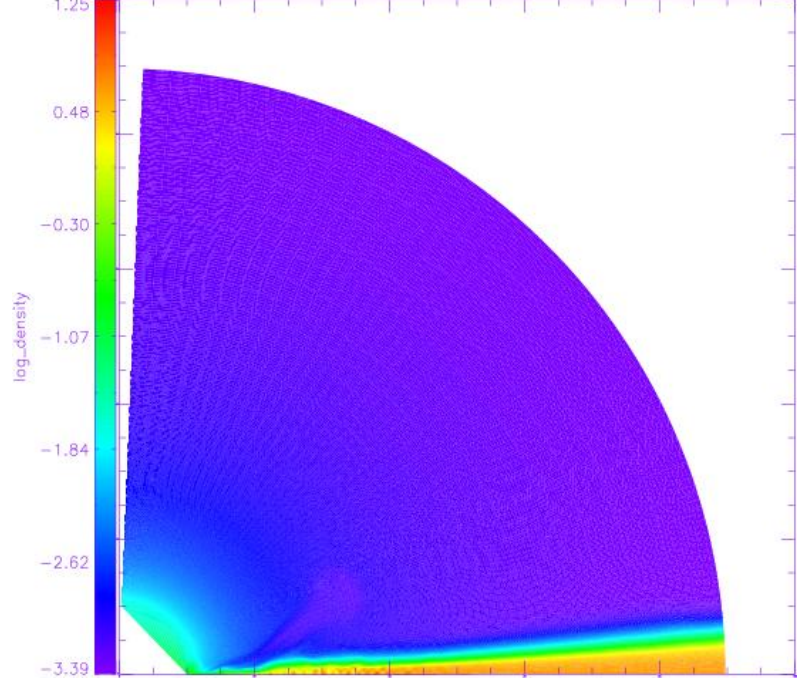
Fig 7a

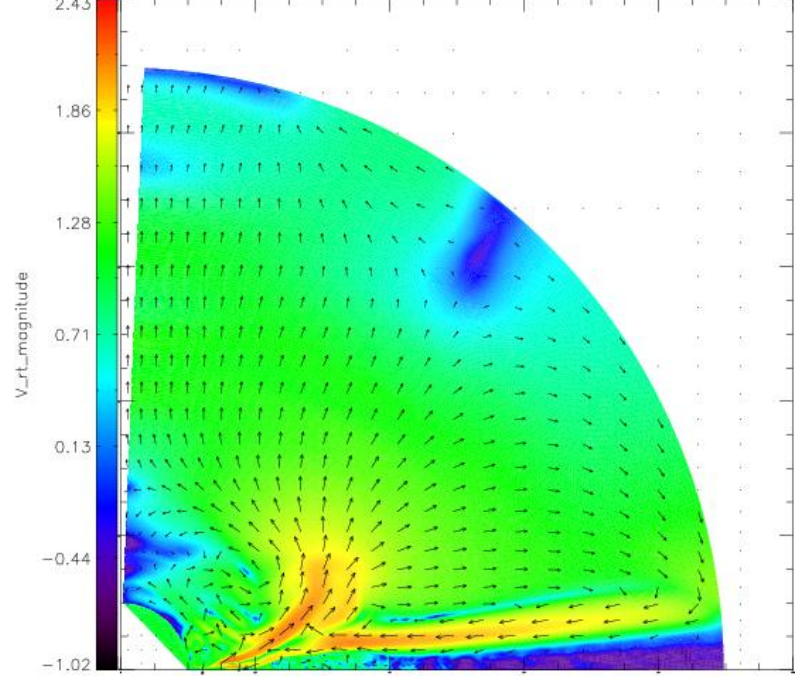
Fig 7b

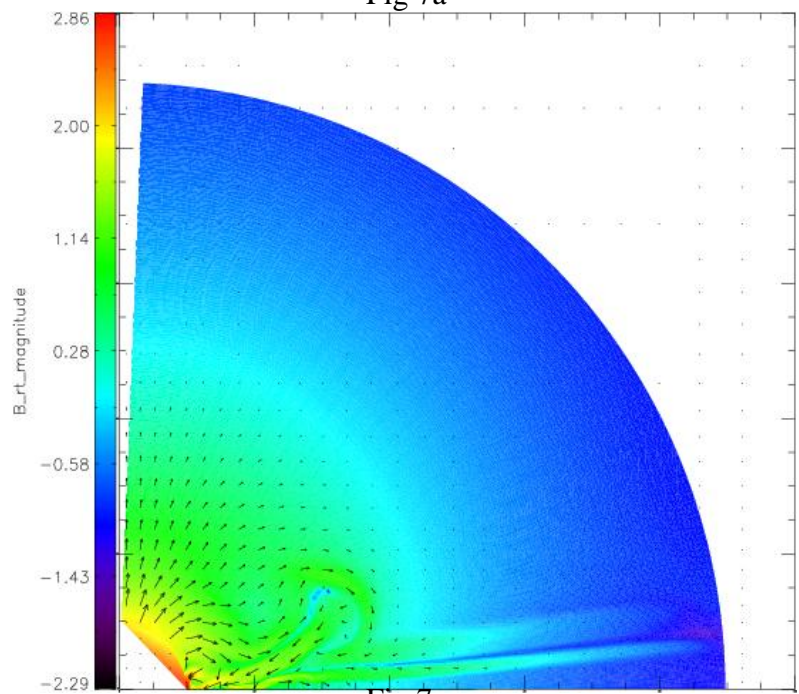
Fig 7c

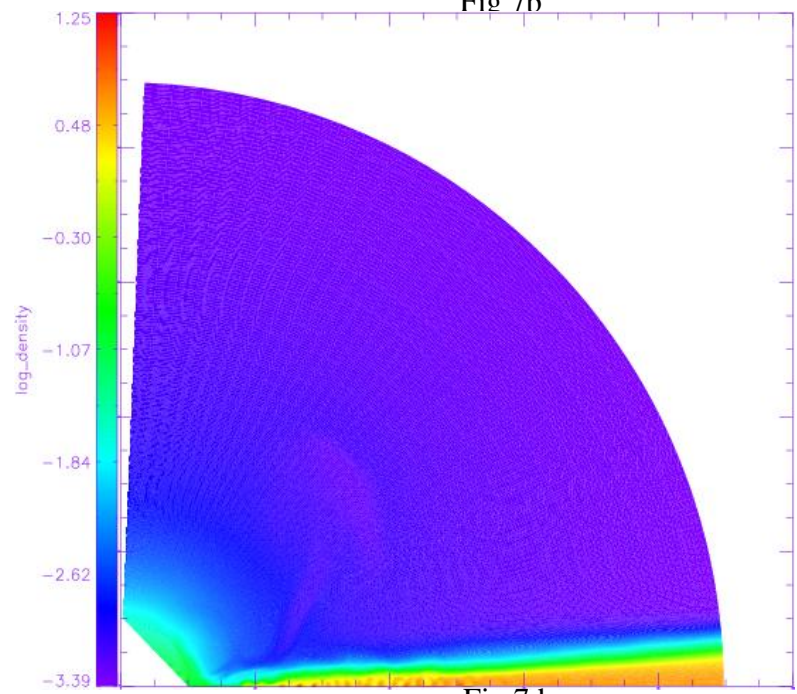
Fig 7d

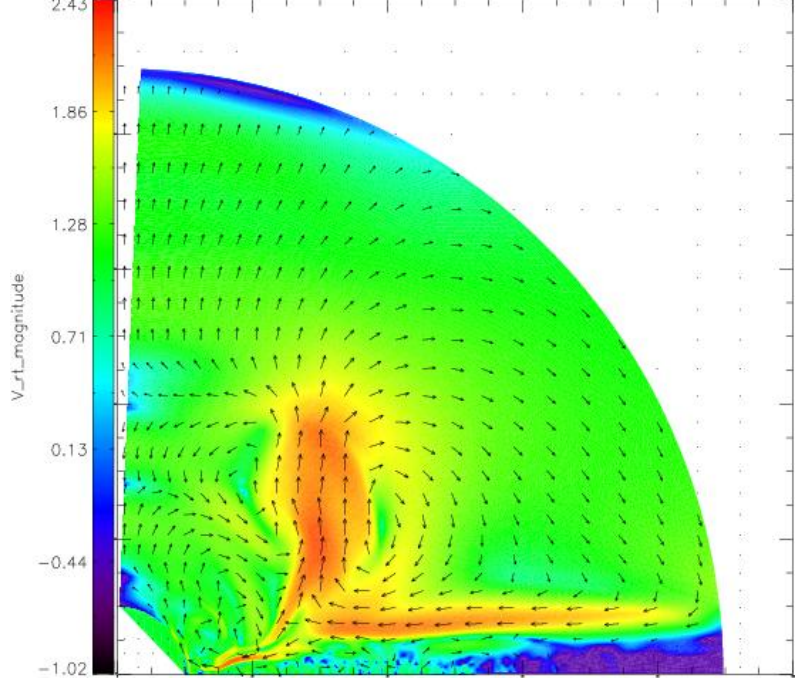 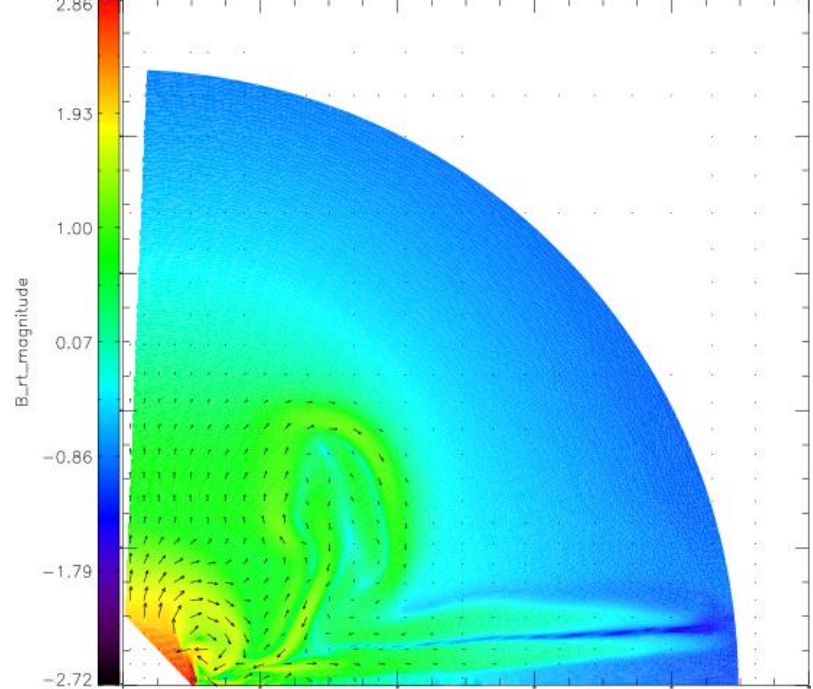

Fig 7e | Fig 7f